\documentclass[journal]{IEEEtran}
\usepackage{tipa}
\usepackage{bbm}
\usepackage{mathrsfs}
\usepackage{cite,url,subfigure,epsfig,graphicx}
\usepackage{amssymb,amsmath}
\usepackage{indentfirst}
\usepackage{algorithmic}
\usepackage{algorithm}
\usepackage{pdfpages}
\usepackage{times,verbatim,amsfonts,amsmath,color}
\usepackage{pifont}
\usepackage{multirow}
\usepackage{booktabs}
\usepackage{adjustbox}

\IEEEoverridecommandlockouts
\makeatletter
\def\ps@headings{\def\@oddhead{\mbox{}\scriptsize\rightmark \hfil \thepage}\def\@evenhead{\scriptsize\thepage \hfil \leftmark\mbox{}}\def\@oddfoot{}\def\@evenfoot{}}
\makeatother 

\newcommand{\tabincell}[2]{\begin{tabular}{@{}#1@{}}#2\end{tabular}}
\usepackage{geometry}
\usepackage[switch,pagewise]{lineno}

\begin{document}

\title{A Survey on Metaverse: Fundamentals, Security, \\and Privacy}

\author{
\IEEEauthorblockN{Yuntao~Wang\IEEEauthorrefmark{2}, Zhou~Su\IEEEauthorrefmark{2}\IEEEauthorrefmark{1}, Ning~Zhang\IEEEauthorrefmark{3}, Rui~Xing\IEEEauthorrefmark{2}, Dongxiao~Liu\IEEEauthorrefmark{4}, Tom~H.~Luan\IEEEauthorrefmark{2}, and Xuemin~Shen\IEEEauthorrefmark{4}}\\
\IEEEauthorblockA{
\IEEEauthorrefmark{2}School of Cyber Science and Engineering, Xi'an Jiaotong University, Xi'an, China\\
\IEEEauthorrefmark{3}Department of Electrical and Computer Engineering, University of Windsor, Windsor, ON, Canada\\
\IEEEauthorrefmark{4}Department of Electrical and Computer Engineering, University of Waterloo, Waterloo, ON, Canada\\
\IEEEauthorrefmark{1}Corresponding author: Zhou~Su (zhousu@ieee.org)
}}

\maketitle

\begin{abstract}
Metaverse, as an evolving paradigm of the next-generation Internet, aims to build a fully immersive, hyper spatiotemporal, and self-sustaining virtual shared space for humans to play, work, and socialize.
Driven by recent advances in emerging technologies such as extended reality, artificial intelligence, and blockchain, metaverse is stepping from science fiction to an upcoming reality.
However, severe privacy invasions and security breaches (inherited from underlying technologies or emerged in the new digital ecology) of metaverse can impede its wide deployment.
At the same time, a series of fundamental challenges (e.g., scalability and interoperability) can arise in metaverse security provisioning owing to the intrinsic characteristics of metaverse, such as immersive realism, hyper spatiotemporality, sustainability, and heterogeneity.
In this paper, we present a comprehensive survey of the fundamentals, security, and privacy of metaverse.
Specifically, we first investigate a novel distributed metaverse architecture and its key characteristics with ternary-world interactions.
Then, we discuss the security and privacy threats, present the critical challenges of metaverse systems, and review the state-of-the-art countermeasures.
Finally, we draw open research directions for building future metaverse systems.
\end{abstract}

\begin{IEEEkeywords}
Metaverse, security, privacy, distributed virtual worlds, extended reality, artificial intelligence, and blockchain.
\end{IEEEkeywords}

\IEEEpeerreviewmaketitle
\section{Introduction}
The metaverse, literally a combination of the prefix ``meta'' (meaning transcendence) and the suffix ``verse'' (shorthand for universe), is a computer-generated world with a consistent value system and an independent economic system linked to the physical world. 
The term metaverse was created by Neal Stephenson in his science fiction novel named \emph{Snow Crash} in 1992. In this novel, humans in the physical world enter and live in the metaverse (a parallel virtual world) through digital avatars (in analogy to user's physical self) via virtual reality (VR) equipment. Since its first appearance, the concept of metaverse is still evolving with various descriptions, such as a second life \cite{sanchez2007second}, 3D virtual worlds \cite{dionisio20133d}, and life-logging \cite{2019Lifelogging}.
Commonly, the metaverse is regarded as a fully immersive, hyper spatiotemporal, and self-sustaining virtual shared space blending the ternary physical, human, and digital worlds \cite{ning2021survey}.
Metaverse is recognized as an evolving paradigm of the next-generation Internet after the web and the mobile Internet revolutions \cite{MetaverseReport}, where users can live as digital natives and experience an alternative life in virtuality. 

The metaverse integrates a variety of emerging technologies \cite{lee2021all,yang2022fusing,Metaverse2021Duan}. 
In particular, digital twin produces a mirror image of the real world, VR and augmented reality (AR) provide immersive 3D experience, 5G and beyond offer ultra-high reliable and ultra-low latency connections for massive metaverse devices, wearable sensors and brain-computer interface (BCI) enable user/avatar interaction in the metaverse, artificial intelligence (AI) enables the large-scale metaverse creation and rendering, and blockchain and non-fungible token (NFT) play an important role in determining authentic rights for metaverse assets \cite{Wei2022Realizing}.
Currently, with the popularity of smart devices and the maturity of enabling technologies, the metaverse is stepping out of its infancy into an upcoming reality in the near future.
Furthermore, significant innovations and advances in the above emerging technologies are giving birth to a new information ecology and new demands for applications, as well as the metaverse for becoming a platform of the new ecology and applications \cite{Metaverse2021Duan}.
Driven by realistic demands and the prospect of feasibility of metaverse construction, metaverse recently has attracted increasing attention from around the world and many tech giants such as Facebook, Microsoft, Tencent, and NVIDIA have announced their ventures into Metaverse. Particularly, Facebook rebranded itself as ``Meta'' to dedicate itself to building the future metaverse \cite{facebookrename}. 

\begin{figure}[!t]\setlength{\abovecaptionskip}{-0.0cm}
\centering
  \includegraphics[width=9cm]{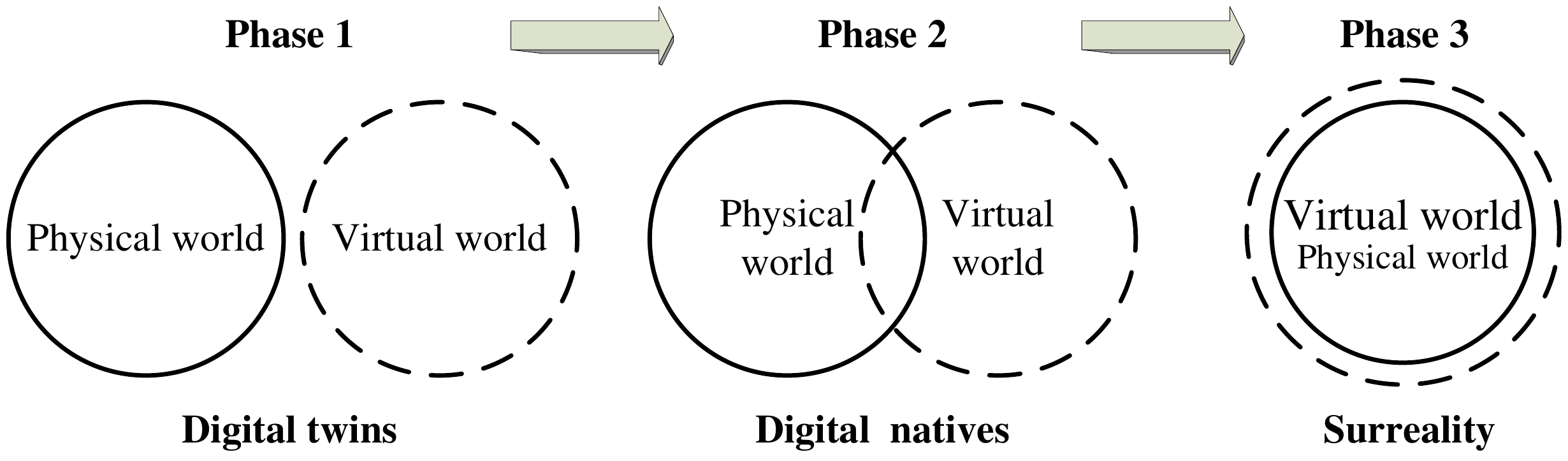}
  \caption{Three phases of the development of the metaverse.}\label{fig:1}\vspace{-0.3cm}
\end{figure}

Generally, the development of metaverse consists of three successive phases from a macro perspective \cite{lee2021all}: (i) \emph{digital twins}, (ii) \emph{digital natives}, and eventually (iii) \emph{surreality}, as depicted in Fig.~\ref{fig:1}.
The first phase produces a mirror world consisting of large-scale and high-fidelity digital twins of humans and things in virtual environments, aimed for a vivid digital representation of the physical reality. In this phase, virtual activities and properties such as user emotion and movement are imitations of their physical counterparts, where reality and virtuality are two parallel spaces. The second phase mainly focuses on the native content creation, where digital natives represented by avatars can produce innovations and insights inside the digital worlds and these digital creations may only exist in the virtual spaces. In this phase, the massively created contents in the digital world become equal with their physical counterparts, and the digital world has the ability to transform and innovate the production process of the physical world, thereby creating more intersections between these two worlds.
The metaverse grows to its maturity in the last phase and turns into a persistent and self-sustaining surreality world which assimilates the reality into itself. The seamless integration and mutual symbiosis of physical and virtual worlds will be realized in this phase, where the scope of the virtual world will be larger than that of the real world and more scenes and lives that do not exist in reality can exist in virtual realms.

\subsection{Challenges for Securing Metaverse}\label{sec:Challenges}
In spite of the promising sign of metaverse, security and privacy issues are the prime concerns that hinder its further development. 
A wide range of security breaches and privacy invasions may arise in the metaverse from the management of massive data streams, pervasive user profiling activities, unfair outcomes of AI algorithms, to the safety of physical infrastructures and human bodies.
Firstly, since metaverse integrates a variety of latest technologies and systems built on them as its basis, their vulnerabilities and intrinsic flaws may also be inherited by the metaverse. 
There have been incidents of emerging technologies, such as hijacking of wearable devices or cloud storage, theft of virtual currencies, and the misconduct of AI to produce fake news. 
Secondly, driven by the interweaving of various technologies, the effects of existing threats can be amplified and become more severe in virtual worlds, while new threats nonexistent in physical and cyber spaces can breed such as virtual stalking and virtual spying \cite{2008Privacy}.
Particularly, the personal data involved in the metaverse can be more granular and unprecedentedly ubiquitous to build a digital copy of the real world, which opens new horizons for crimes on private big data \cite{Falchuk8371577}. For example, to build a virtual scene using AI algorithms, users will inevitably wear wearable AR/VR devices with built-in sensors to comprehensively collect brain wave patterns, facial expressions, eye movements, hand movements, speech and biometric features, as well as the surrounding environment.
Besides, as users need to be uniquely identified in the metaverse, it means that headsets, VR glasses, or other devices can be used for tracking users' real locations illegally \cite{9134928}.
Lastly, hackers can exploit system vulnerabilities and compromise devices as entry points to invade real-world equipments such as household appliances to threaten personal safety, and even threaten critical infrastructures such as power grid systems, high-speed rail systems, and water supply systems via advanced persistent threat (APT) attacks \cite{7218444}.

Nevertheless, existing security countermeasures can still be ineffective and lack adaptability for metaverse applications.
Particularly, the intrinsic characteristics of metaverse including \emph{immersiveness}, \emph{hyper spatiotemporality}, \emph{sustainability}, \emph{interoperability}, \emph{scalability}, and \emph{heterogeneity} may bring about a series of challenges for efficient security provision. 
1) The real-time fully immersive experience in the metaverse brings not only sensual pleasures of the flawless virtual environment, but also challenges in the secure fusion of massive multimodal user-sensitive big data for interactions between users and avatars/environments.
2) The integration of the ternary world contributes to the hyper spatiotemporality in the metaverse  \cite{nevelsteen2018virtual}, which greatly increases the complexity and difficulty of trust management.
Due to the deepening blurring of the boundary between the real and the virtual, the metaverse will make the fact and fiction more confusing such as Deepfake events, especially for regulations and digital forensics. 
3) To avoid the single point of failure (SPoF) and the control by a few powerful entities, the metaverse should be built on a decentralized architecture to be self-sustaining and persistent \cite{nguyen2021metachain}, which raises severe challenges in reaching unambiguous consensus among massive entities in the time-varying metaverse. 
4) The interoperability and scalability in the metaverse indicate users can freely shuttle across various sub-metaverses concurrently under different scenes and interaction modes, which also poses challenges to ensure fast service authorization, compliance auditing, and accountability enforcement in seamless service mitigation and multi-source data fusion. 
5) The virtual worlds in the large-scale metaverse can be highly heterogeneous in terms of hardware implementation, communication interfaces, and softwares, which poses huge interoperability difficulties.

\subsection{Related Works}\label{subsec:Contributions}
The topic of metaverse has attracted various research attention. Until now, there have been several survey papers from different aspects of the metaverse.
For example, Dionisio \emph{et al}. \cite{dionisio20133d} specify four characteristics of viable 3D virtual worlds (or metaverse) including ubiquity, realism, scalability, and interoperability, and discuss ongoing improvements of the underlying virtual world technology.
Lee \emph{et al}. \cite{lee2021all} review and examine eight fundamental technologies to build up the metaverse as well as its opportunities from six user-centric factors.
{Huynh-The \emph{et al}. \cite{Thien2022AI} study the role of AI approaches in the foundation and development of the metaverse.}
Yang \emph{et al}. \cite{yang2022fusing} investigate the potential of AI and blockchain technologies for future metaverse construction.
Ning \emph{et al}. \cite{ning2021survey} present a survey of the development status of metaverse in terms of national policies, industrial projects, infrastructures, supporting technologies, VR, and social metaverse.
Park \emph{et al}. \cite{9667507} discuss three components (i.e., hardware, software, and content) of metaverse and review the user interaction, implementation, and representative applications in the metaverse.
{Xu \emph{et al}. \cite{Xu2022EdgeMeta} present an in-depth survey on the edge-enabled metaverse from communication, networking, computation, and blockchain perspectives.}
Leenes \cite{2008Privacy} investigate potential privacy risks in the online game \emph{Second Life} from both social and legal perspectives.
Different from the above existing surveys on the general metaverse \cite{dionisio20133d,lee2021all,ning2021survey,9667507,2008Privacy}, {AI-empowered metaverse \cite{Thien2022AI,yang2022fusing}, edge-enabled metaverse \cite{Xu2022EdgeMeta},} or the potential in service provisioning in social VR/AR games \cite{Falchuk8371577}, retailing \cite{Bourlakis2009Retail}, education \cite{JEM2020Virtual}, social goods \cite{Metaverse2021Duan}, and computational arts \cite{Lee2021When}, we focus on the perspective of metaverse security and privacy such as potential security/privacy threats, critical security/privacy challenges, and state-of-the-art defenses, etc. 

In this paper, we present a comprehensive survey on the fundamentals of metaverse, as well as the key challenges and solutions to build the secure and privacy-preserving metaverse. 
{By discussing existing/potential solutions for the challenges facing the metaverse, our survey offers critical insights and useful guidelines for readers to better understand how these security/privacy threats could arise and be prevented in the metaverse.}
The contributions of this survey are four-fold.
\begin{itemize}
  \item We discuss the fundamentals of metaverse including the general architecture, key characteristics, and enabling technologies, as well as existing modern prototypes of metaverse applications.
  \item We investigate the security and privacy threats in the metaverse from seven aspects (i.e., authentication \& access control, data management, privacy, network, economy, governance, and physical/social effects) and discuss the critical challenges to address them. 
  \item We survey the state-of-the-art security and privacy countermeasures in both academic and industry and discuss their feasibility toward building the secure and privacy-preserving metaverse paradigm. 
  \item We outline open future research directions in building the secure, privacy-preserving, and efficient metaverse realm.
\end{itemize}

Table~\ref{contribution} summarizes the contribution of our work in comparison to previous relevant surveys in the metaverse. 

\begin{table}[!t]
   \centering \setlength{\abovecaptionskip}{0cm}
    \caption{{A Comparison of Contribution Between Our Survey and Relevant Surveys}}\label{contribution}
    \resizebox{1.01\linewidth}{!}{
        \begin{tabular}{|c|c|l|}
        \hline
        \textbf{Year.} &\textbf{Refs.} &\textbf{Contribution} \\ \hline 
        {2008} &\cite{2008Privacy} &\tabincell{l}{Discussions on privacy risks in the game \emph{Second Life} \\from both social and legal perspectives.} \\ \hline

        {2009} &\cite{Bourlakis2009Retail} &\tabincell{l}{Survey on metaverse applications in terms of retailing.} \\ \hline

        {2013} &\cite{dionisio20133d}  &\tabincell{l}{Discussions on key features of metaverse and ongoing\\ improvements of the underlying virtual world technology.} \\ \hline

        {2018} &\cite{Falchuk8371577} &\tabincell{l}{Survey on privacy issues and countermeasures related to\\digital footprints in social metaverse games.} \\ \hline

        {2020} &\cite{JEM2020Virtual} &\tabincell{l}{Survey on metaverse applications in terms of education.} \\ \hline

        {2021} &\cite{Metaverse2021Duan} &\tabincell{l}{Survey on metaverse applications in terms of social goods.} \\ \hline

        {2021} &\cite{lee2021all} &\tabincell{l}{Review on eight fundamental technologies to build up the \\ metaverse and its opportunities from six user-centric factors.} \\ \hline

        {2021} &\cite{ning2021survey} &\tabincell{l}{Overview of metaverse development in terms of national\\ policies, industrial projects, infrastructures, supporting \\technologies, VR, and social metaverse.} \\ \hline

        {2021} &\cite{Lee2021When} &\tabincell{l}{Survey on metaverse applications in terms of digital arts.} \\ \hline

        {2022} &\cite{yang2022fusing} &\tabincell{l}{Discuss the potential of AI and blockchain technologies \\in future metaverse construction.} \\ \hline

        {{2022}} &{\cite{Thien2022AI}} &{\tabincell{l}{Discuss the role of AI from six technical aspects \\in the development of the metaverse.}} \\ \hline

        {2022} &\cite{9667507} &\tabincell{l}{Discuss the hardware, software, and content components\\of metaverse and review user interaction, implementation,\\and representative applications in the metaverse.} \\ \hline

        {{2022}} &{\cite{Xu2022EdgeMeta}} &{\tabincell{l}{An in-depth survey on the edge-enabled metaverse in terms \\of communication, networking, and computation.}} \\ \hline

        {Now} &\textbf{Ours} &\tabincell{l}{Comprehensive survey of the fundamentals, security, and \\privacy of metaverse, discussions on the general architecture, \\characteristics, and security/privacy threats of the metaverse, \\discussions on critical challenges, state-of-the-art solutions, \\and future research directions in building secure metaverse.} \\ \hline
        \end{tabular}}
\end{table}

\begin{figure}[!t]
\centering \setlength{\abovecaptionskip}{-0.1cm}
  \includegraphics[width=8.0cm,height=19cm]{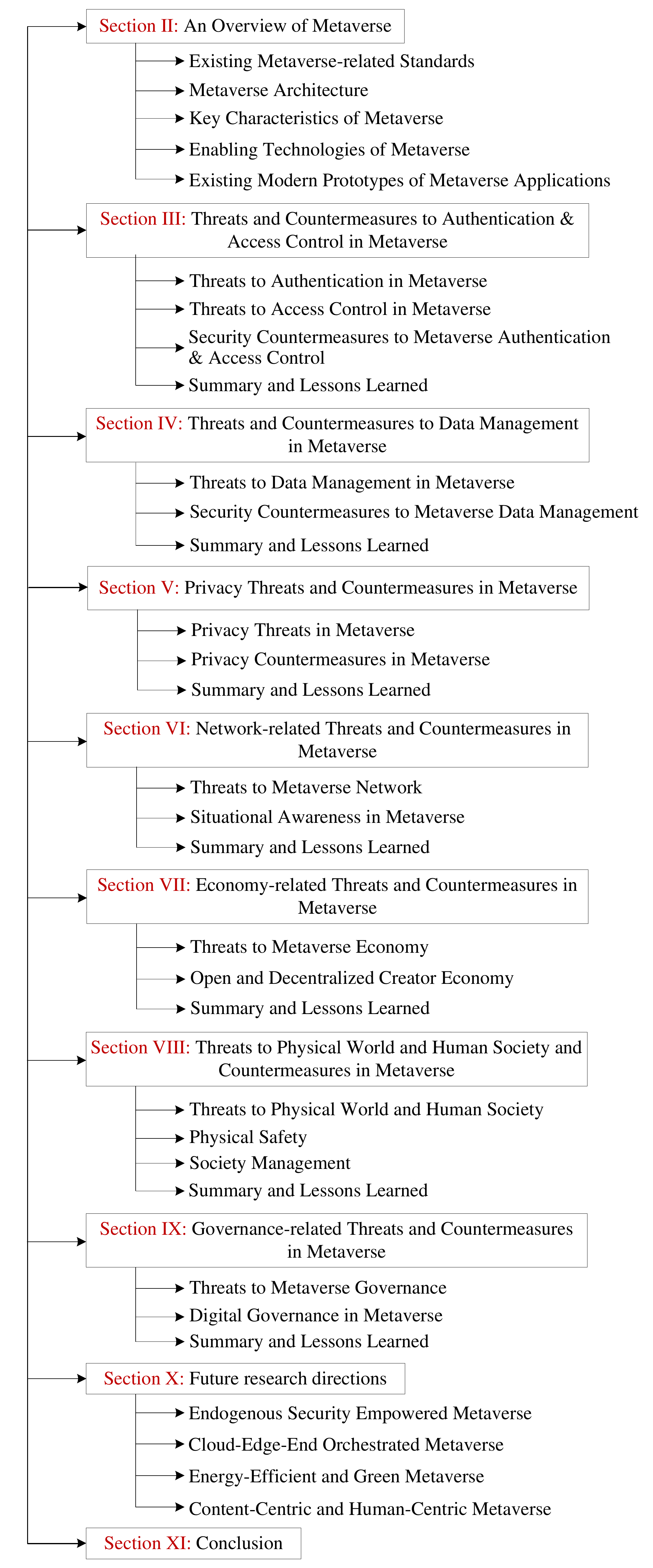}
  {\caption{Organization structure of this paper.}\label{fig:organization}}\vspace{-5mm}
\end{figure}

The remainder of this paper is organized as follows. Section~\ref{sec:METAVERSE} presents the standards, architecture, characteristics, supporting technologies, and current prototypes of the metaverse.
Sections~\ref{sec:Threat1}--\ref{sec:Threat7} present the taxonomy of security and privacy threats in the metaverse and discuss critical challenges and existing/potential solutions to resolve them from seven aspects.
Then, we discuss open research issues in Section~\ref{sec:FUTUREWORK}. Finally, we draw the conclusions in Section~\ref{sec:CONSLUSION}. {Fig.~\ref{fig:organization} illustrates the organization of this survey.} The key acronyms are listed in Table~\ref{table-abbr}.

\begin{table*}[!t]
   \begin{center}\setlength{\abovecaptionskip}{0cm}
    \caption{Summary of Important Abbreviations in Alphabetical Order}\label{table-abbr}
        \begin{tabular}{ll|ll|ll}
        \hline
        \textbf{Abbr.}  &\textbf{Definition}            & \textbf{Abbr.} &\textbf{Definition}           &\textbf{Abbr.} &\textbf{Definition} \\ \hline 
        ABE   &Attribute-Based Encryption               &AR  &Augmented Reality       &AI   &Artificial Intelligence   \\ \hline
        APT   &Advanced Persistent Threat               &BCI &Brain-Computer Interface      &B5G &Beyond 5G \\ \hline
        CA   &Certificate Authority                     &CPSS    &Cyber-Physical-Social System  &DL &Deep Learning \\ \hline
        DP   &Differential Privacy                      &ECG    &Electrocardiogram         &FL &Federated Learning \\ \hline
        GDPR  &General Data Protection Regulation       &HCI   &Human-Computer Interaction        &HE &Homomorphic Encryption \\ \hline
        IoT  &Internet of Things                        &MMO    &Massive Multi-player Online    &MR &Mixed Reality \\ \hline
        NFT  &Non-Fungible Token                &NPC   &Non-Player Character   &OSN &Online Social Network \\ \hline
        PUGC  &Professional- and User-Generated Content           &PGC &Professional-Generated Content &PKI &Public Key Infrastructure \\ \hline
        PPG  &Photoplethysmography                 &SDN &Software-Defined Network &SSI &Self-Sovereign Identity \\ \hline
        SMC &Secure Multi-party Computation  &SPoF &Single Point of Failure  &SVM&Support Vector Machine  \\ \hline
        QoE &Quality-of-Experience                   &QoS &Quality-of-Service      &UGC &User-Generated Content  \\ \hline
        VR   &Virtual Reality                     &XR  &Extended Reality           &ZKP&Zero-Knowledge Proof \\ \hline
        \end{tabular}
    \end{center}
\end{table*}

\section{An Overview of Metaverse}\label{sec:METAVERSE}
In this section, we introduce the metaverse from the following aspects: {existing standards,} the general architecture, key characteristics, enabling technologies, potential applications, and existing prototypes.

\subsection{{Existing Metaverse-Related Standards}}\label{subsec:Standards}
\begin{figure}[!t]\setlength{\abovecaptionskip}{-0.0cm}
\centering
  \includegraphics[width=5.2cm,height=4.2cm]{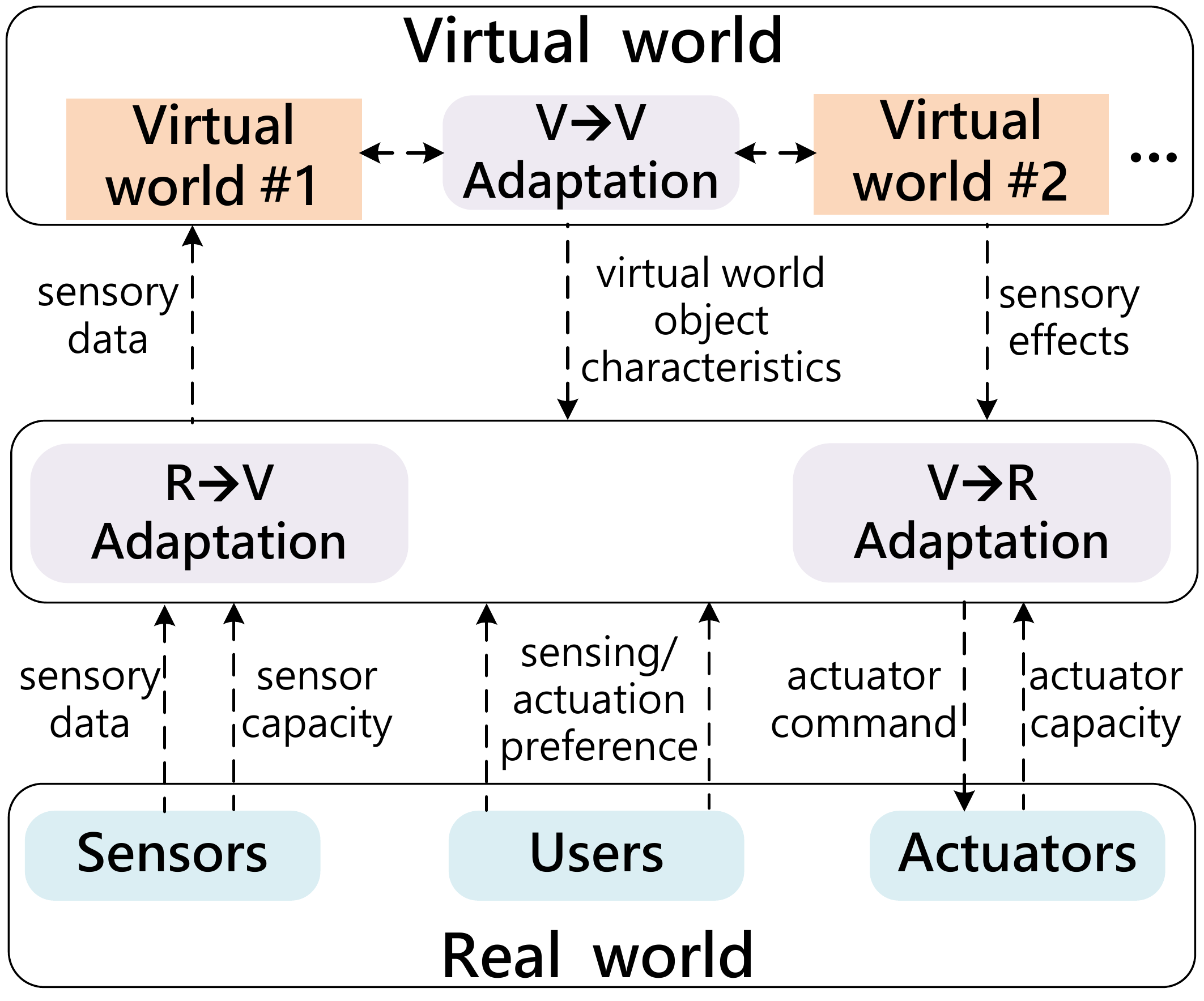}
  \caption{{The architecture of ISO/IEC 23005 (MPEG-V) standards \cite{MPEG-V}. $R\!\rightarrow\! V$ adaptation means the conversion of sensory data from the real world (RW) to virtual world (VW) object characteristics. $V\!\rightarrow\! R$ adaptation means the conversion of sensory effects from VW into actuator commands to RW. $V\!\rightarrow\! V$ adaptation means the conversion of native representations of information in a VW to the standard format.}}\label{fig:MPEG-V}\vspace{-1mm}
\end{figure}

\begin{figure}[!t]\setlength{\abovecaptionskip}{-0.0cm}
\centering
  \includegraphics[width=6.1cm]{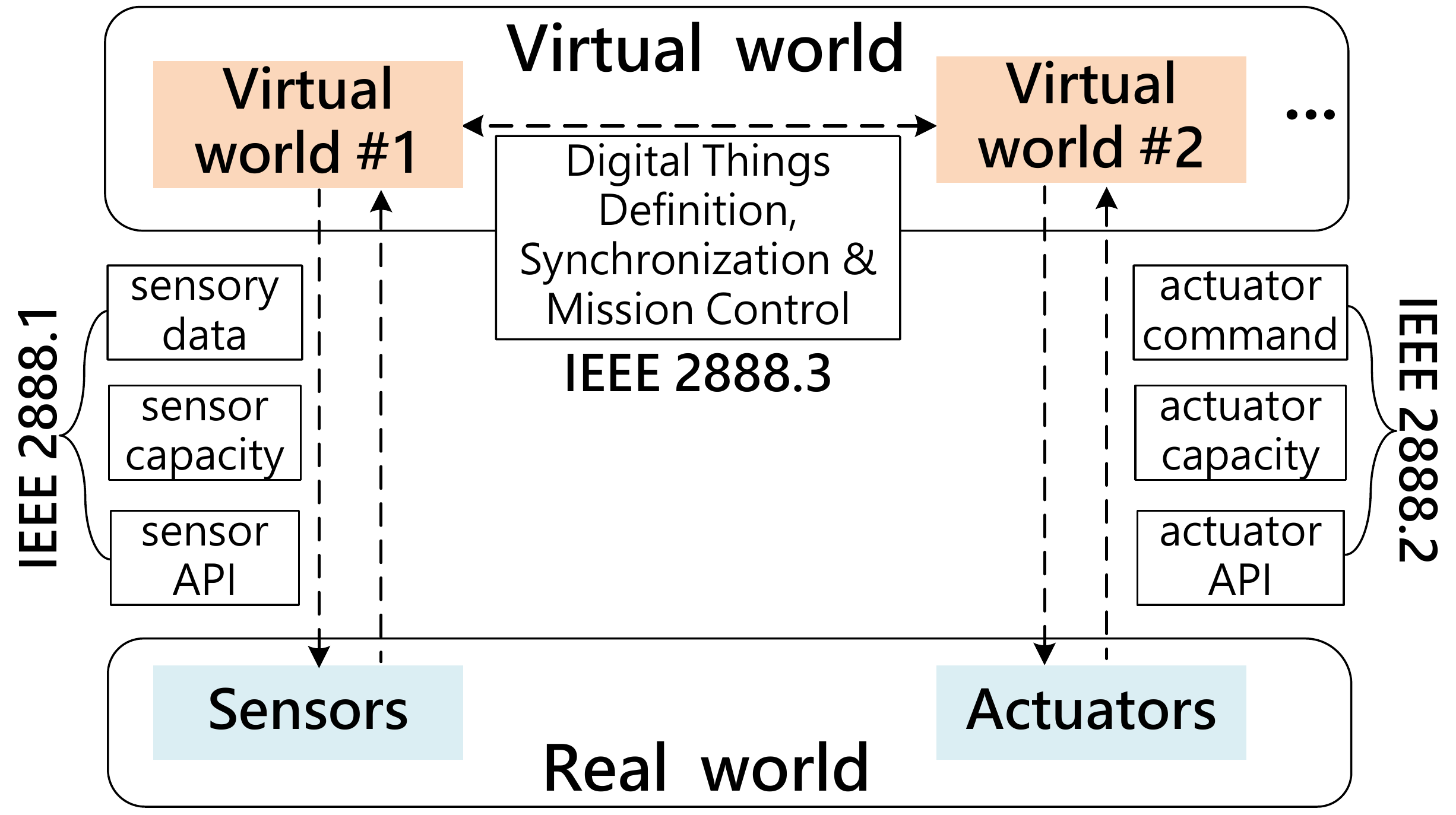}
  \caption{The architecture of IEEE 2888 standards \cite{IEEE-2888}. IEEE 2888.1, IEEE 2888.2, and IEEE 2888.3 specify the standards on sensor interface, actuator interface, and orchestration of digital synchronization, respectively. }\label{fig:IEEE-2888}\vspace{-4mm}
\end{figure}

In what follows, we briefly introduce two existing metaverse-related standards: ISO/IEC 23005 \cite{MPEG-V} and IEEE 2888 \cite{IEEE-2888}. 

\emph{1)} As the first standardized framework for networked virtual environments (NVEs) in the metaverse, ISO/IEC 23005 (MPEG-V) aims to standardize the interfaces between the real world and the virtual world, and among virtual worlds, to realize seamless information exchange, simultaneous reactions, and interoperability \cite{MPEG-V}. Its first version was published in 2011, and the latest 4th edition was released in 2020. ISO/IEC 23005 standards are applicable for a variety of metaverse-related business services, where the association of audiovisual information, rendered sensory effects, and characteristics of virtual objects (e.g., avatars and virtual items) can benefit the interactions between virtual and real worlds. Fig.~\ref{fig:MPEG-V} illustrates the general architecture of ISO/IEC 23005 standards. 

\emph{2)} ISO/IEC 23005 standards mainly focus on the sensory effects and lack capability in offering general-purpose interfaces between virtual and real worlds. As a supplement to ISO/IEC 23005 standards, IEEE 2888 project launched in 2019 aims to define standardized interfaces for synchronization of cyber and physical worlds \cite{IEEE-2888}. By specifying information formats and application program interfaces (APIs) to control actuators and obtain sensory information, IEEE 2888 standards offer foundations for building metaverse systems, where both virtual and real worlds can affect each other. 
Fig.~\ref{fig:IEEE-2888} illustrates the general architecture of IEEE 2888 standards. In Fig.~\ref{fig:IEEE-2888}, the sensory information and actuator-related information are exchanged between virtual and real worlds via IEEE 2888.1 and IEEE 2888.2 standards, respectively. Besides, the definition, synchronization, and mission control data are defined by the IEEE 2888.3 standard for digital things (i.e., virtual objects).

\begin{figure}[!t]\setlength{\abovecaptionskip}{-0.0cm}\vspace{-3mm}
\centering
  \includegraphics[width=8.25cm]{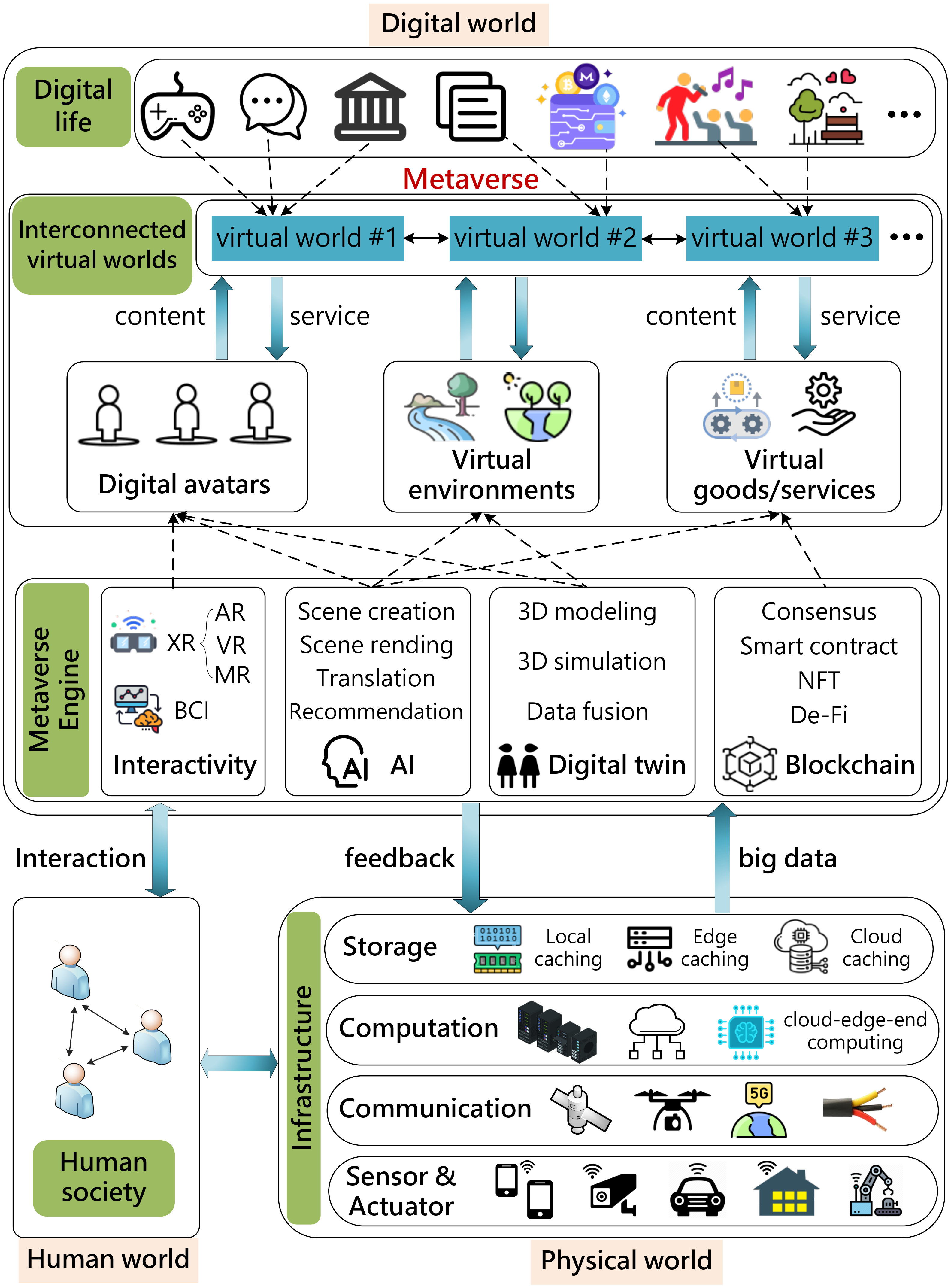}
  \caption{The architecture of metaverse in integration of the human, physical, and digital worlds.}\label{fig:architecture}\vspace{-2mm}
\end{figure}

\subsection{Metaverse Architecture}\label{subsec:Architecture}
Metaverse is a self-sustaining, hyper spatiotemporal, and 3D immersive virtual shared space, created by the convergence of physically persistent virtual space and virtually enhanced physical reality. {In other words, the metaverse is a synthesized world which is composed of user-controlled avatars, digital things, virtual environments, and other computer-generated elements, where humans (represented by avatars) can use their virtual identity through any smart device to communicate, collaborate, and socialize with each other.}
The construction of metaverse blends the ternary physical, human, and digital worlds. Fig.~\ref{fig:architecture} shows the general architecture of the metaverse with consideration of its intrinsic ternicity. {In the following, we elaborate on the relationships between the three worlds, the components in the metaverse, and the information flow of the metaverse in detail.}

\subsubsection{{Human Society}}
{The metaverse is regarded as human-centric \cite{7863165}.
Human users along with their inner psychologies and social interactions constitute the human world.
Equipped with smart wearable devices (e.g., VR/AR helmets), humans can interact and control their digital avatars to play, work, socialize, and interact with other avatars or virtual entities in the metaverse via human-computer interaction (HCI) and extended reality (XR) technologies \cite{9495125} (as depicted in the film \emph{Ready Player One}).}

\subsubsection{{Physical Infrastructures}}
The physical world offers supporting infrastructures (including sensing/control, communication, computation, and storage infrastructures) to the metaverse to support multi-sensory data perception, transmission, processing, and caching, as well as physical control, thereby enabling efficient interactions with both the digital and human worlds.
Specifically, pervasive smart objects, sensors, and actuators constitute the sensing/control infrastructure to enable all-around and multimodal data perception from the environment and human bodies and high-accuracy device control. Networking is provided via the communication infrastructure consisting of various heterogeneous wireless or wired networks (e.g., cellular communications, unmanned aerial
vehicle (UAV) communications, and satellite communications). Besides, powerful computation and storage capacities are provisioned via the computation and storage infrastructure assisted by cloud-edge-end computing \cite{9171865}.
For instance, a virtual world runs at a minimum rate of 30 frames per second \cite{4623222}, posing huge computational demands and latency constraints (e.g., within 1/30th of a second at most) in rendering high-quality graphics for each avatar.

\subsubsection{{Interconnected Virtual Worlds}}
According to ISO/IEC 23005 and IEEE 2888 standards \cite{MPEG-V,IEEE-2888}, the digital world can be composed of a series of interconnected distributed virtual worlds (i.e., sub-metaverses), and each sub-metaverse can offer certain kinds of virtual goods/services (e.g., gaming, social dating, online museum, and online concert) and virtual environments (e.g., game scenes and virtual cities) to users represented as digital avatars.
\begin{itemize}
  \item \emph{Digital avatars}. Avatars refer to the digital representation of human users in the metaverse. A user can create various avatars in different metaverse applications, and the produced avatars can be like a human shape, animals, imaginary creatures, etc.
  \item \emph{Virtual environments}. Virtual environments refer to the simulated real or imaginary environments (consisting of 3D digital things and their attributes) in the metaverse. Besides, the virtual environments in the metaverse can have distinct spatiotemporal dimensions (e.g., in ancient times or future worlds) for users to experience an alternate life.
  \item \emph{Virtual goods/services}. Virtual goods refer to the tradeable commodities (e.g., skins, digital arts, and land parcels) produced by virtual service providers (VSPs) or the users in the metaverse. Virtual services in the metaverse have a broad of scopes including digital market, digital currency, digital regulation, social service, etc.
\end{itemize}
There are two main sources of information in the metaverse: one is the input of the real world (i.e., the captured information and obtained knowledge from the real space digitally displayed in the virtual space), and the other is the output of virtual worlds (i.e., the information generated by avatars, digital objects, and metaverse services in the virtual space).
{For the massive fine-grained metaverse data collected/generated in real time, efficient authentication and access control should be enforced, as well as the data reliability, traceability, and privacy protection in the life-cycle of metaverse services.}

\subsubsection{{Metaverse Engine}}
The metaverse engine \cite{Xu2022EdgeMeta} uses the big data from the real world as inputs to generate, maintain, and update the virtual world via the interactivity, AI, digital twin, and blockchain technologies.
Particularly, with the assistance of XR and HCI (especially brain-computer interaction (BCI)) techniques, users situated in physical environments are able to immersively control their digital avatars in the metaverse via their senses and bodies for diverse collective and social activities such as car racing, dating, and virtual item trading. The virtual economy as a spontaneous derivative of such digital creation activities of avatars can be built in the metaverse.
AI algorithms perform personalized avatar/content creation, large-scale metaverse rending, and intelligent service offering to enrich the metaverse ecology.
Besides, the knowledge derived via AI-based big data analytics can be beneficial to perform simulating, digitalizing, and mirroring the real world via digital twin technology to produce vivid virtual environments for users to experience.
Finally, the created digital twins, as well as native contents created by avatars, can be transparently managed, uniquely tokenized, and monetized by the blockchain technology to enable trust-free trading and service offering, towards building the economic system and value system in the metaverse.
More details of these enabling technologies are elaborated at Sect.~\ref{subsec:Technologies}.

In summary, information is the core resource of the metaverse and the free data flow in the ternary world makes the digital ecology, which eventually promotes the integration of virtual and actual worlds.
Next, we discuss the information flow in a single world and across different worlds, respectively.

\begin{figure}[!t]\setlength{\abovecaptionskip}{-0.0cm}
\centering
  \includegraphics[width=9.5cm]{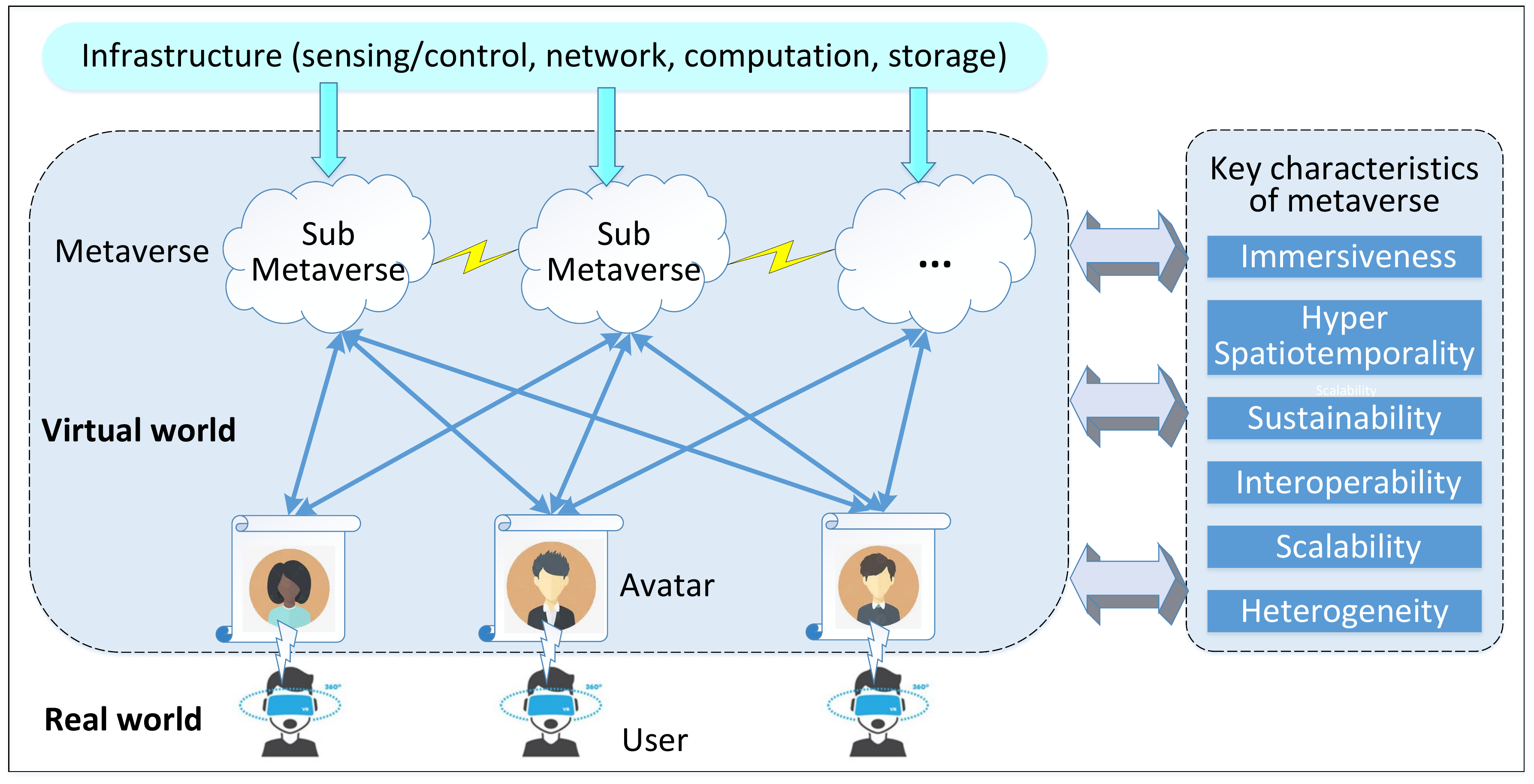}
  \caption{General network architecture and key characteristics of the metaverse.}\label{fig:networkarchit}\vspace{-2mm}
\end{figure}

\subsubsection{In-World Information Flow}
The human society or human world is interconnected by social networks and formed based on common activities and mutual interactions among human beings.

In the physical world, {the IoT-enabled sensing/control infrastructure} plays an important role in digitalizing{/transforming} the physical world via pervasive sensors {and actuators}, and the generated IoT big data is transmitted and processed {via network and computation infrastructures}.

In the digital world, the produced digital information of physical and human worlds are processed and managed via {the metaverse engine} to support large-scale metaverse {creation/rendering and various metaverse services}. Besides, users, represented as avatars, can produce and distribute digital {creations across} various sub-metaverses to promote the creativity of metaverse.

\subsubsection{Information Flow Across Worlds}
As depicted in Fig.~\ref{fig:architecture}, the subjective consciousness, the Internet, and the IoT are the main media among the three worlds.
(i) Humans can interact with physical objects via HCI technology and experience virtually augmented reality (e.g., holographic telepresence) via XR technology.
(ii) The human world and the digital world are connected through the Internet, i.e., the largest computer network in the world. Users can interact with the digital world via smart devices such as smartphones, wearable sensors, and VR helmets, for the creation, sharing, and acquisition of knowledge.
(iii) The IoT infrastructure bridges the physical world and the digital world by using inter-connected smart devices for digitalization, and thereby information can flow freely between the two worlds \cite{8364607}.
Besides, the feedback information from the digital world (e.g., processed results of big data and intelligent decisions) can guide the transformation (e.g., manufacturing process) of the physical world. 
As the metaverse blends physical systems, human society, and cyber worlds, threats in virtual worlds can be amplified and severely affect physical infrastructures and personal safety, which also raises huge governance demands and challenges.

\subsection{Key Characteristics of Metaverse}\label{subsec:Characteristics}
In web 1.0, Internet users are just content consumers, where contents are provided by the websites. 
In web 2.0 (i.e., mobile Internet), users are both content producers and consumers, and the websites turn into platforms for service offering. Typical such platforms include Wikipedia, WeChat, and TikTok.
Metaverse is recognized as the evolving paradigm of web 3.0.
In metaverse, as shown in Fig.~\ref{fig:networkarchit}, users represented as digital avatars can seamlessly shuttle across various virtual worlds (i.e., sub-metaverses) to experience a digital life, as well as make digital creations and economic interactions, supported by physical infrastructures and the metaverse engine.
Specifically, metaverse exhibits unique features from the following perspectives.

\subsubsection{Immersiveness}
The immersiveness means that the computer-generated virtual space is sufficiently realistic to allow users to feel psychologically and emotionally immersed \cite{5606335}. It can be also called \emph{immersive realism} \cite{dionisio20133d}. According to the perspective of realism, human beings interact with the environment through their senses and their bodies. The immersive realism can be approached through the structure of sensory perception (e.g., sight, sound, touch, temperature, and balance) and expression (e.g., gestures).

\subsubsection{Hyper Spatiotemporality}
The real world is restricted by the finiteness of space and the irreversibility of time.
As metaverse is a virtual space-time continuum parallel to the real one, the hyper spatiotemporality refers to the break of limitations of time and space \cite{ning2021survey}. As such, users can freely shuttle across various worlds with different spatiotemporal dimensions to experience an alternate life with seamless scene transformation. 

\subsubsection{Sustainability}
The sustainability indicates that the metaverse maintains a closed economic loop and a consistent value system with a high level of independence. On the one hand, it should be \emph{open}, i.e., continuously arousing users' enthusiasm in digital content creation as well as open innovations.
On the other hand, to remain persistent, it should be built on a \emph{decentralized} architecture to get rid of SPoF risks and prevent from being controlled by a few powerful entities.

\subsubsection{Interoperability}
The interoperability in the metaverse represents that (i) users can seamlessly move across virtual worlds (i.e., sub-metaverses) without interruption of the immersive experience \cite{lee2021all}; and (ii) digital assets for rendering or reconstruction of virtual worlds are interchangeable across distinct platforms \cite{dionisio20133d}.

\subsubsection{Scalability}
The scalability refers to the capacity of metaverse to remain efficient with the number of concurrent users/avatars, the level of scene complexity, and the mode of user/avatar interactions (in terms of type, scope, and range) \cite{dionisio20133d}.

\subsubsection{Heterogeneity}
The heterogeneity of metaverse includes heterogeneous virtual spaces (e.g., with distinct implementations), heterogeneous physical devices (e.g., with distinct interfaces), heterogeneous data types (e.g., unstructured and structured), heterogeneous communication modes (e.g., cellular and satellite communications), as well as the diversity of human psychology. It also entails the poor interoperability of metaverse systems. 

\subsection{Enabling Technologies of Metaverse}\label{subsec:Technologies}
\begin{figure}[!t]
\centering\setlength{\abovecaptionskip}{-0.0cm}
  \includegraphics[width=6.4cm]{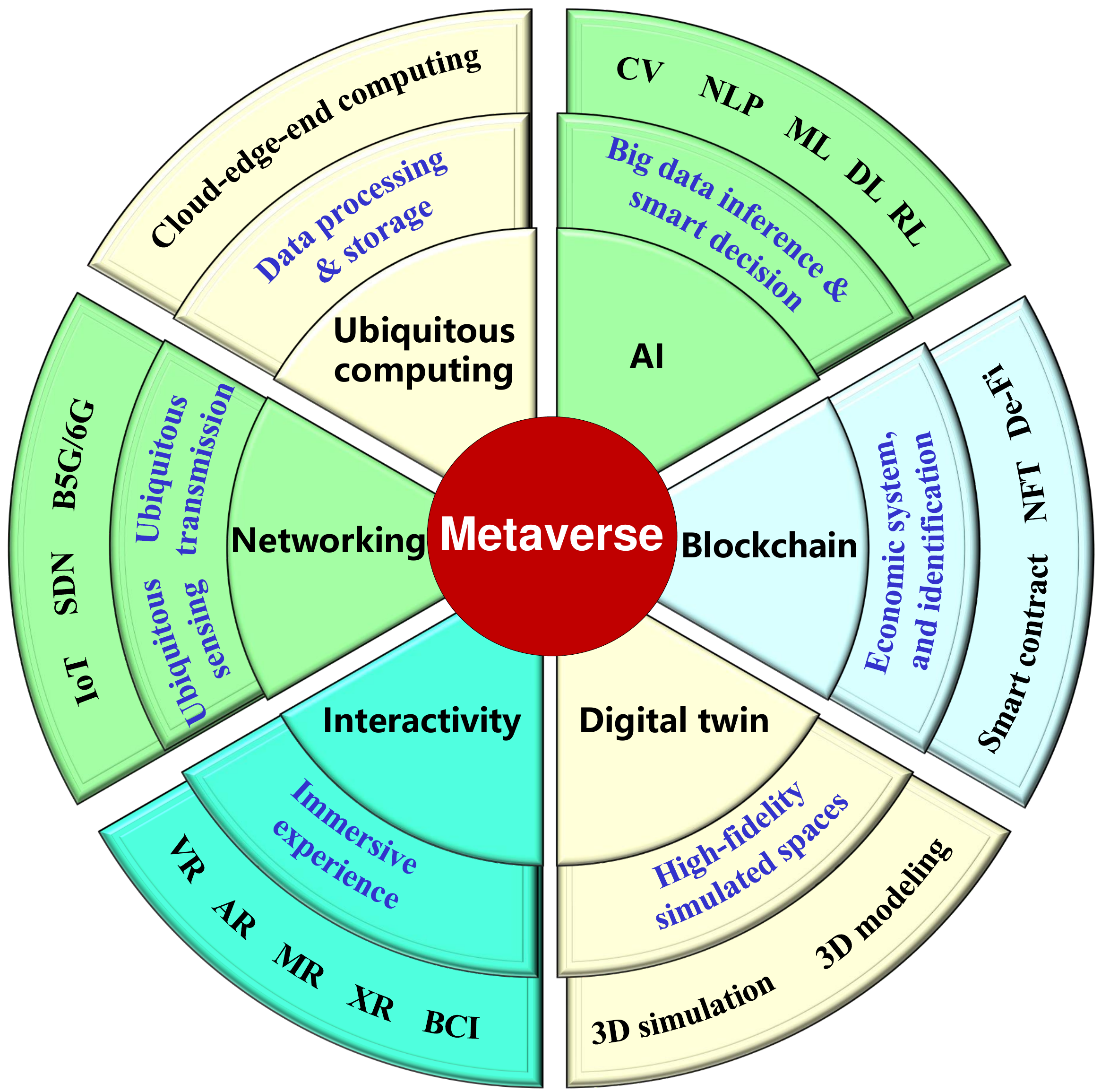}
  \caption{The illustration of six underlying technologies including its roles and key components in the metaverse.}\label{fig:enablingTech}\vspace{-2mm}
\end{figure}

As shown in Fig.~\ref{fig:enablingTech}, there are the following six enabling technologies underlying the metaverse.

\subsubsection{Interactivity}
With the maturity of miniaturized sensors, embedded technology, and XR technology, XR devices such as helmet-mounted displays (HMDs) are expected to be the main terminal for entering the metaverse \cite{sugimoto2021extended}.
The XR deeply incorporates virtual reality/augmented reality/mixed reality (VR/AR/MR) technologies to offer multi-sensory immersiveness, augmented experience, and real-time user/avatar/environment interaction via front-projected holographic display, HCI (especially BCI), and large-scale 3D modeling \cite{10.1145/769953.769967}.
{Particularly, VR provides immersive experiences in a virtual world, AR delivers true presence experiences of virtual holograms, graphics, and videos in the real world, and MR offers a transition experience between VR and AR.}
The wearable XR devices perform fine-grained human-specific information perception, {as well as ubiquitous sensing for objects and surroundings, with the assistance of indoor smart devices (e.g., cameras)}.
In this manner, the user/avatar interactivity will no longer be limited to mobile inputs (e.g., hand-held phones and laptops), but all kinds of interactive devices connected to the metaverse.
Besides, negative experiences such as dizziness in wearing XR helmets can be resolved by low-latency edge computing systems and AI-empowered real-time rendering.

\subsubsection{Digital Twin}
Digital twin represents the digital clone of objects and systems in the real world with high fidelity and consciousness \cite{9429703}. 
It enables the mirroring of physical entities, as well as prediction and optimization of their virtual bodies, by analyzing real-time streams of sensory data, physical models, and historical information.
In digital twin, data fed back from physical entities can be used for self-learning and self-adaption in the mirrored space.
Moreover, digital twins can provide precise digital models of the expected objects with intended attributes in the metaverse with high accuracy through the simulation of complex physical processes and the assistance of AI technologies, which is beneficial for large-scale metaverse creation and rendering.
Besides, digital twin enables predictive maintenance and accident traceability for physical safety, due to the bidirectional connection between physical entities and their virtual counterparts, thereby improving efficiency and reducing risks in the physical world.

\subsubsection{Networking}
In the metaverse, networking technologies such as 6G, software-defined network (SDN), and IoT empower the ubiquitous network access and real-time massive data transmission between real and virtual worlds, as well as between sub-metaverses. 
Beyond 5G (B5G) and 6G offer possibilities for ubiquitous, real-time, and ultra-reliable communications for massive metaverse devices with enhanced mobility support \cite{du2021optimal}. {In 6G, space-air-ground integrated network (SAGIN) \cite{9631953} is a promising trend for seamless and ubiquitous network access to metaverse services.}
SDN enables the flexible and scalable management of large-scale metaverse networks via the separation of the control plane and data plane. In SDN-based metaverse, the physical devices and resources are managed by a logically centralized controller using a standardized interface such as OpenFlow, thereby virtualized computation, storage, and bandwidth resources can be dynamically allocated according to real-time demands of various sub-metaverses \cite{9258002}.
Besides, IoT is a network of numerous physical objects that are embedded with sensors, softwares, communication components, and other technologies with the aim to connect, exchange, and process data between things, systems, clouds, and users over the Internet. In the metaverse, IoT sensors are extensions of human senses. 

\subsubsection{Ubiquitous Computing}
Ubiquitous computing, or ubicomp aims to create an environment where computing appears anytime and everywhere for users \cite{6157577}.
Through pervasive (often mobile) smart objects embedded in the environment or carried on the human body, ubiquitous computing enables smooth adaptation to the interactions between human users and the physical space.
With ubicomp, instead of using specific equipment (e.g., laptop), human users can freely interact with their avatars and experience real-time immersive metaverse services via ubiquitous smart objects and network access in the environment.
For improved users' quality-of-experience (QoE) in ubicomp, the cloud-edge-end computing \cite{9171865} orchestrates the highly scalable cloud infrastructures (with powerful computation and storage capacity) and heterogeneous edge computing infrastructures (closer to end users/devices) {via complex inner/inter-layer cooperation paradigms. As such, it allows} flexible and on-demand resource allocation to satisfy various requirements of end users/devices in different metaverse applications.

\subsubsection{AI}
AI technology acts as the ``brain'' of metaverse which empowers personalized metaverse services (e.g., vivid and customized avatar creation), massive metaverse scene creation and rendering, multilingual support in the metaverse by learning from massive multimodal input via big data inference \cite{Thien2022AI}. 
Moreover, AI enables smart interactions (e.g., smart shopping guide and user movement prediction) between user and avatar/NPC (non-player character) via intelligent decision-making.
For example, by continuously learning users' facial expressions, emotions, hairstyles, and so on, AI algorithms can create vivid and personalized avatars and intelligently recommend interested goods or information to users in the metaverse.
More details of AI in the metaverse can refer to the survey \cite{Thien2022AI}.

\subsubsection{Blockchain}
\begin{figure}[!t]\setlength{\abovecaptionskip}{-0.0cm}
\centering
  \includegraphics[width=9.0cm,height=5cm]{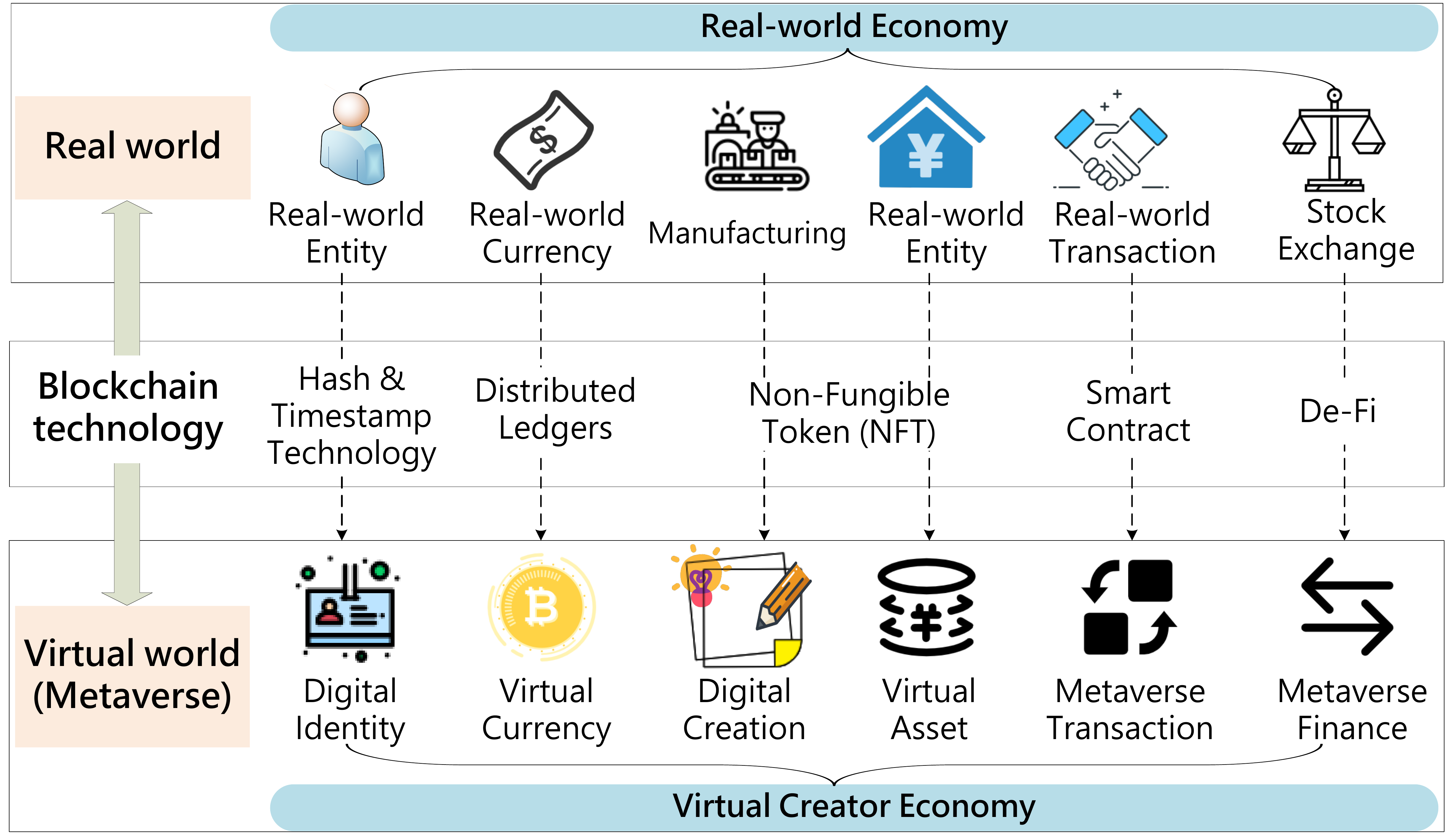}
  \caption{The role of blockchain technologies in bridging the conventional economy and metaverse economy.}\label{fig:blockchainMeta}\vspace{-2mm}
\end{figure}

To be persistent, the metaverse should be constructed on a decentralized architecture to avoid centralization risks such as SPoF, low transparency, and control by a few entities \cite{nguyen2021metachain}. Besides, the virtual economy and value system provided by the blockchain are essential components of the metaverse.
As shown in Fig.~\ref{fig:blockchainMeta}, blockchain technologies offer an open and decentralized solution for building the sustainable virtual economy, as well as constructing the value system in the metaverse.
Blockchain is a distributed ledger, in which data is structured into hash-chained blocks and featured with decentralization, immutability, transparency, and auditability \cite{9631953}. The blockchain can be classified into three categories, i.e., public, consortium, and private, based on the decentralization degree \cite{9631953}. The consensus protocols are the key component of blockchain, which determines the ledger consistency and system scalability. 
Besides, smart contracts can be deployed atop the blockchain to allow automatic function execution among distrustful parties in a prescribed fashion.
NFT represents irreplaceable and indivisible tokens \cite{wang2021NFT}, which can help asset identification and ownership provenance in the blockchain.
De-Fi stands for decentralized finance, which aims to deliver secure, transparent, and complex financial services (e.g., stock/currency exchange) in the metaverse.

\subsection{Existing Modern Prototypes of Metaverse Applications}\label{subsec:Applications}
In this subsection, we introduce existing representative prototypes in the following metaverse applications.
\subsubsection{Game}
Game is the current hottest metaverse application. 
Considering the technological maturity, user matching, and content adaptability, games are an excellent way to explore the metaverse.
We list some representative examples of metaverse games.
The sandbox game \emph{Second Life}\footnotemark[1]\footnotetext[1]{https://secondlife.com/} offers a modifiable 3D virtual world where players can join in as avatars and create their virtual architectures and sell them, as well as participate in social activities such as art shows and even political gatherings and visiting embassy.
\emph{Roblox}\footnotemark[2]\footnotetext[2]{https://developer.roblox.com/en-us/} is a global user-created game platform, in which players can create games and design items such as skins and clothes. It proposes eight key features of the metaverse: identity, friends, immersion, anywhere, diversity, low latency, economy, and civilization \cite{han2021analysis}. \emph{Fortnite}\footnotemark[3]\footnotetext[3]{https://www.epicgames.com/fortnite/en-US/home} is a massive multi-player online (MMO) shooter game designed by Epic Games, where players can build buildings and bunkers as well as construct islands, while the in-game items such as skins can only be designed by the platform.

\subsubsection{Social Experience}
Metaverse can revolutionize our society and enable a series of immersive social applications such as virtual lives, virtual shopping, virtual dating, virtual chatting, global travel, and even space/time travel. For example, Lil Nas X held a virtual concert on Roblox in 2020, with over 30 million fans participating. Players can unlock special Lil Nas X goods in the digital store, e.g., commemorative items and emotions.
Due to the COVID-19 situation, UC Berkeley celebrated graduation festivities virtually in Minecraft in 2020 by digitally copying the campus scenery. 
Besides, Tencent developed a \emph{Digital Palace Museum}\footnotemark[4] in 2018 which allows tourists to freely visit the palace museum and its exhibitions with a panoramic and immersive view by wearing VR helmets in their homes. 
\footnotetext[4]{https://en.dpm.org.cn/about/news/2019-09-18/3089.html}

\subsubsection{Online Collaboration}
Metaverse also opens new possibilities for immersive virtual collaboration in terms of telecommuting in virtual workplaces, studying and learning in virtual classrooms, and panel discussion and meeting in virtual conference rooms. For example, \emph{Horizon Workroom}\footnotemark[5] is an office collaboration software (run in Oculus Quest 2 helmet) released by {Meta (parent company of Facebook)}, which allows people in any physical location to work and meet together in the same virtual room. 
{\emph{Microsoft Mesh}\footnotemark[6] is an MR platform supported by Azure, which enables users working from multiple sites to cooperate virtually via holographic presence and shared experience from anywhere in a digital copy of their office.}
\footnotetext[5]{https://www.theverge.com/2021/8/19/22629942/facebook-workrooms-horizon-oculus-vr}
\footnotetext[6]{https://www.microsoft.com/en-us/mesh}

\subsubsection{Simulation \& Design}
Another promising application is 3D simulation, modeling, and architectural design on metaverse. For example, NVIDIA has built its open platform named \emph{Omniverse}\footnotemark[7]\footnotetext[7]{https://www.nvidia.com/en-us/omniverse/} to support multi-user real-time 3D simulation and visualization of physical objects and attributes in a shared virtual space for industrial applications, e.g., automotive design. Besides, Omniverse can be compatible with Disney Pixart's open-source platform Universal Scene Description (USD).

\subsubsection{Creator Economy}
\begin{table}[!t]
   \centering
    \caption{A Summary of Content Creation Modes in The Metaverse}\label{contentmode} \resizebox{1.0\linewidth}{!}{
        \begin{tabular}{c|ccc}\hline
              \textbf{Mode} & \textbf{Description}
              & \textbf{Feature}                                                                                     & \textbf{Instance}                                             \\ \hline
\textbf{PGC}  & \begin{tabular}[c]{@{}c@{}}Contents are produced \\ by professionals\end{tabular}
& \begin{tabular}[c]{@{}c@{}}Centralization, \\low diversification,\\high quality \& cost\end{tabular}           & \begin{tabular}[c]{@{}c@{}}GTA,\\Unity\end{tabular}                                                           \\\hline
\textbf{PUGC} & \begin{tabular}[c]{@{}c@{}}Contents are produced by\\professionals and users\end{tabular} & \begin{tabular}[c]{@{}c@{}}Semi-centralization,\\medium diversification,\\medium cost\end{tabular} & \begin{tabular}[c]{@{}c@{}} Second Life,\\Minecraft,\\Fortnite\end{tabular} \\\hline
\textbf{UGC}  & \begin{tabular}[c]{@{}c@{}}Contents are produced \\and traded among users\end{tabular}                   & \begin{tabular}[c]{@{}c@{}}Decentralization, \\high diversification,\\uneven quality\,\&\,low cost\end{tabular}         & \begin{tabular}[c]{@{}c@{}}Roblox,\\Decentraland,\\Cryptovoxels\end{tabular} \\\hline
{\textbf{AIGC}}  & {\begin{tabular}[c]{@{}c@{}}Contents are produced or\\partially produced by AI\end{tabular}}                   & {\begin{tabular}[c]{@{}c@{}}High efficiency, \\low cost\,\&\,fast\end{tabular}}         & {\begin{tabular}[c]{@{}c@{}}MetaHuman\end{tabular}} \\\hline
\end{tabular}}
\end{table}

\begin{table*}[]
\centering \setlength{\abovecaptionskip}{0cm}
    \caption{Summary of Existing Metaverse Prototypes In Different Applications}\label{Prototypes}
    \resizebox{1.01\linewidth}{!}{
\begin{tabular}{ccccccccc}\hline
\multirow{2}{*}{\textbf{Prototype}} & \multirow{2}{*}{\textbf{Application}} & \multirow{2}{*}{\textbf{Immersive}} & \multirow{2}{*}{\textbf{Hyper Spatiotemporal}} & \multicolumn{2}{c}{\textbf{Sustainable~~~}} & \multirow{2}{*}{\textbf{Interoperable}} & \multirow{2}{*}{\textbf{Scalable}} & \multirow{2}{*}{\textbf{Heterogeneous}} \\\cline{5-6}
                               &                               &                                     &                                                & \textbf{Open}  & \textbf{Decentralized}  &                                         &                                    &                                         \\\hline
Second Life  &MMO Game &Partly & $\checkmark$ & Partly &$\times$ & $\times$ &$\checkmark$ &N/A\\ \hline
Roblox &MMO Game &$\checkmark$ & $\checkmark$ & $\checkmark$ &$\times$ & Partly &$\checkmark$ &N/A\\ \hline
Fortnite &MMO Game &$\checkmark$ & $\checkmark$ & Partly &$\times$ & Partly &$\checkmark$ &N/A\\ \hline
Digital Palace Museum &Travelling &$\checkmark$ &$\times$ & $\times$ &$\times$ & $\times$ &Partly &N/A\\ \hline
Horizon Workroom &Working &$\checkmark$ & $\times$ & $\times$ &$\times$ & $\times$ &Partly &N/A\\ \hline
Omniverse &Simulation &$\checkmark$ & $\checkmark$ & $\checkmark$ &$\times$ & Partly &$\checkmark$ &$\checkmark$\\ \hline
Decentraland &Game &$\checkmark$ & $\checkmark$ & $\checkmark$ &$\checkmark$ & $\times$ &$\checkmark$ &Partly\\ \hline
Cryptovoxels &Game &$\checkmark$ & $\checkmark$ & $\checkmark$ &$\checkmark$ & $\times$ &$\checkmark$ &Partly\\ \hline
\end{tabular} }
\end{table*}

\begin{figure*}[!t]\setlength{\abovecaptionskip}{-0.0cm}
\centering
  \includegraphics[width=19.2cm,height=6.725cm]{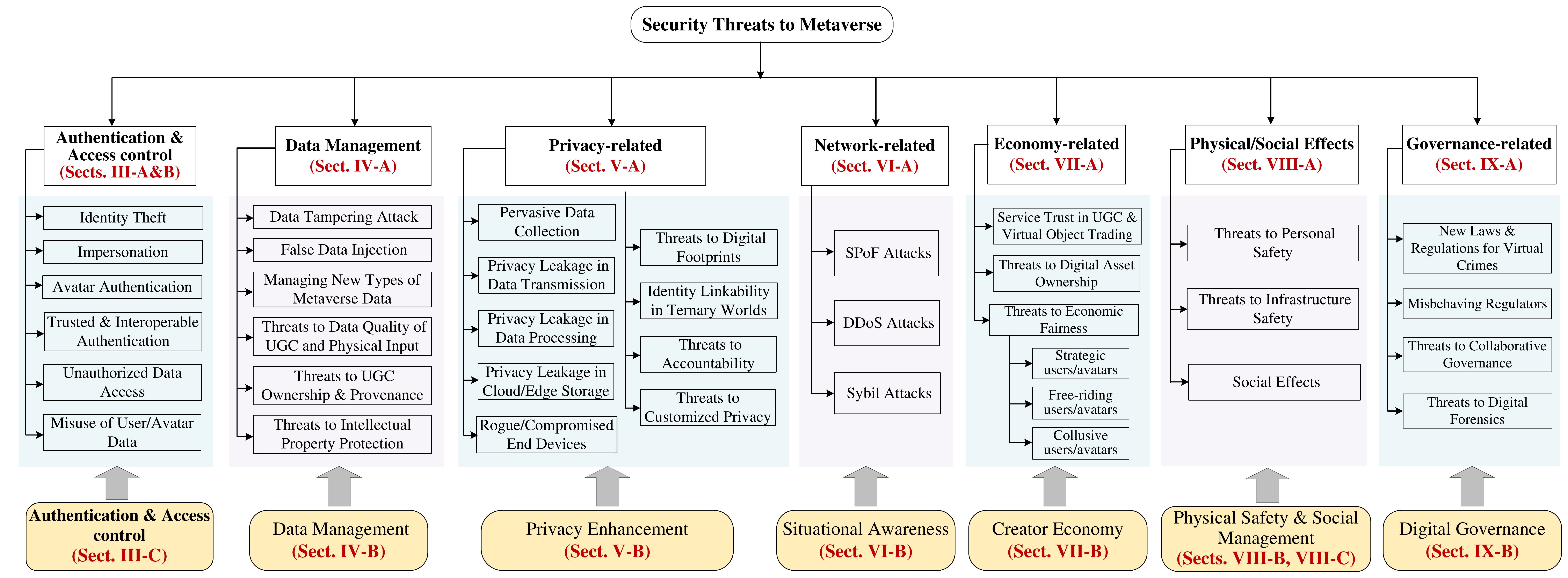}
  \caption{{The taxonomy of security threats and corresponding security countermeasures in the metaverse.}}\label{fig:threatstaxonomy}\vspace{-2mm}
\end{figure*}

The metaverse mainly includes four modes of content creation: professional-generated content (PGC), professional- and user-generated content (PUGC), user-generated content (UGC), and AI-generated content (AIGC), as illustrated in Table~\ref{contentmode}.
In PGC mode, contents (e.g., games) are created by professional content producers on the platform, and ordinary users are just participants and content viewers/experiencers. 
In UGC mode, all users produce contents and trade them freely in the marketplace provided by the platform, which is featured with high freedom degree, low cost, high diversification, and decentralization \cite{10.1145/3161570}. Users are dominant in the content production process under the UGC mode. 
For example, creators of game scenes, skins, and items in Roblox can earn a certain percentage of Robux (i.e., virtual tokens exchangeable with real-world currency) paid by their experiencers, leading to a virtuous cycle.
The PUGC mode is the combination of PGC and UGC modes, in which contents are jointly produced by professionals and ordinary users. 
{In the metaverse, as the number of content consumers can be far greater than the number of content producers, the AIGC mode can help VSPs to create massive qualified and personalized contents with much-improved efficiency and much-reduced cost. In AIGC, there exist two types of content creation: (i) AI fully replaces users for content production; and (ii) AI assists users to produce contents.
An example is that in the MetaHuman project \cite{MetaHuman}, Epic Games leverages AI algorithms to quickly create life-like virtual characters such as conversational virtual assistants.}

There are existing decentralized virtual worlds with a built-in creator economy supported by the Ethereum blockchain such as Decentraland\footnotemark[8] and Cryptovoxels\footnotemark[9]. In \emph{Decentraland}, users can trade the land parcel and equipments in the marketplace and build their own buildings as well as social games by calling the builder function, where the trading details are immutably recorded in Ethereum for auditablility. In \emph{Cryptovoxels}, players can trade the lands and build virtual stores and art galleries in the virtual world ``Origin City''. Besides, users can display and trade their digital assets such as artwork inside buildings.
\footnotetext[8]{https://decentraland.org/}\footnotetext[9]{https://www.cryptovoxels.com/}

Table~\ref{Prototypes} summarizes existing modern prototypes in different metaverse applications in terms of the six key characteristics of the metaverse.

In the next sections (i.e., from Sect.~\ref{sec:Threat1} to Sect.~\ref{sec:Threat7}), based on existing surveys \cite{9163078,9146364}, we classify a broad scope of security threats in the metaverse from the following seven dimensions: authentication \& access control, data management, privacy, network, economy, physical/social effects, and governance. Moreover, we review existing/potential defense mechanisms for the above security and privacy threats in the metaverse. Fig.~\ref{fig:threatstaxonomy} depicts the proposed taxonomy of security threats and the corresponding security countermeasures in the metaverse.

\section{Threats and Countermeasures to Authentication \& Access Control in Metaverse}\label{sec:Threat1}
In metaverse, identity authentication and access control play a vital role for massive users/avatars in metaverse service offering.

\subsection{Threats to Authentication in Metaverse}\label{subsec:Authenthreat}
The identities of users/avatars in the metaverse can be illegally stolen, impersonated, and interoperability issues can be encountered in authentication across virtual worlds.

\emph{1) Identity Theft}. If the identity of a user is stolen in the metaverse, his/her avatars, digital assets, social relationships, and even the digital life can be leaked and lost, which can be more severe than that in traditional information systems. For example, hackers can steal users' personal information (e.g., full names, secret keys of digital assets, and banking details) {in Roblox} through hacked personal {VR glasses}, phishing email scams, and authentication loopholes to commit fraud and crimes (e.g., steal the victim's avatar and digital assets) {in Roblox. For example, in 2022, the accounts of 17 users in the Opensea NFT marketplace are hacked due to smart contract flaws and phishing attacks, causing a lost of \$1.7 million \cite{opensea2022nft}.}

\emph{2) Impersonation Attack}. An attacker can carry out the impersonation attack by pretending to be another authorized entity to gain access to a service or system in the metaverse \cite{7218444}.
For example, hackers can invade the Oculus helmet and exploit the stolen behavioral and biological data gathered by the in-built motion-tracking system to create digital replicas of the user and impersonate the victim to facilitate social engineering attacks. The hackers can also create a fake avatar using digital replicas of the victim to deceive, fraud, and even commit a crime against the victim's friends in the metaverse.
Another example is that attackers can exploit Bluetooth impersonation threats \cite{9152758} to impersonate trusted endpoints and illegally access metaverse services by inserting rogue wearable devices into the established Bluetooth pairing.

\emph{3) Avatar Authentication Issue}. Compared with real-world identity authentication, the authentication of avatars (e.g., the verification of their friends' avatars) for users in the metaverse can be more challenging through verifying facial features, voice, video footage, and so on. Besides, adversaries can create multiple AI bots (i.e., digital humans), which appear, hear, and behave identical to user's real avatar, in the virtual world (e.g., Roblox) by imitating user's appearance, voice, and behaviors \cite{Falchuk8371577}. As a consequence, more additional personal information might be required as evidence to ensure secure avatar authentication, which may also open new privacy breach issues.

\emph{4) Trusted and Interoperable Authentication}. For users/avatars in the metaverse, it is fundamental to ensure fast, efficient, and trusted cross-platform and cross-domain identity authentication, i.e., across various service domains and virtual worlds (built on distinct platforms such as blockchains) \cite{dionisio20133d}. For example, the trust-free and interoperable asset exchange and avatar transfer between Roblox and Fortnite, as well as among distinct administrative domains for offering different services in Roblox.

\subsection{Threats to Access Control in Metaverse}\label{subsec:Authenthreat}
\emph{1) Unauthorized Data Access}. Complex metaverse services will generate new types of personal profiling data (e.g., biometric information, daily routine, and user habits). To deliver seamless personalized services (e.g., customized avatar appearance) in the metaverse, different VSPs in distinct sub-metaverses need to access real-time user/avatar profiling activities \cite{Xu2021Wireless}. Malicious VSPs may carry out attacks for unauthorized data access to earn benefits. An example is that malicious VSPs may illegally elevate their rights in data access via attacks such as buffer overflow and tampering access control lists \cite{8249924}.
Besides, as such massive personal information is produced and transmitted in real time, it is complicated to decide exactly what personal information to be shared, with whom, under what condition, for what purpose, and when it is destroyed.

\emph{2) Misuse of User/Avatar Data}. In the life-cycle of data services in the metaverse, user/avatar-related data can be disclosed intentionally by attackers or unintentionally by VSPs to facilitate user profiling and targeted advertising activities. Besides, due to the potential non-interoperability of certain sub-metaverses, it is hard to trace the data misuse activities in the large-scale metaverse.

\subsection{Security Countermeasures to Metaverse Authentication \& Access Control}\label{subsec:defense1}
For the metaverse, secure and efficient identity management is the basis for user/avatar interaction and service provisioning. Generally, digital identities can be classified into the following three kinds.
\begin{itemize}
  \item \emph{Centralized identity}. Centralized identity refers to the digital identity authenticated and managed by a single institution, such as the Gmail account.
  \item \emph{Federated identity \cite{6336740}}. Federated identity refers to the digital identity managed by multiple institutions or federations. It can reduce the administrative cost in identity authentication for cross-platform and cross-domain operations, and alleviate the cumbersome process of typing personal information repeatedly for users.
  \item \emph{Self-sovereign identity (SSI) \cite{9536956}}. SSI refers to the digital identity which is fully controlled by individual users. It allows users to autonomously share and associate different personal information (e.g., username, education information, and career information) in performing cross-domain operations to enable identity interoperability with users' consent.
\end{itemize}

In the metaverse, centralized identity systems can be prone to SPoF risks and suffer potential leakage risks. Federated identity systems are semi-centralized and the management of identities is controlled by a few institutions or federations, which may also suffer potential centralization risks. The identity systems built on SSIs will be dominant in future metaverse construction \cite{MetaverseReport}.
According to \cite{de2019key}, identity management schemes in the metaverse should follow the following design principles: (i) \emph{scalability} to massive users/avatars, (ii) \emph{resilience} to node damage, and (iii) \emph{interoperability} across various sub-metaverse during authentication.

\begin{figure}[!t]\setlength{\abovecaptionskip}{-0.0cm}
\centering
  \includegraphics[width=5.3cm]{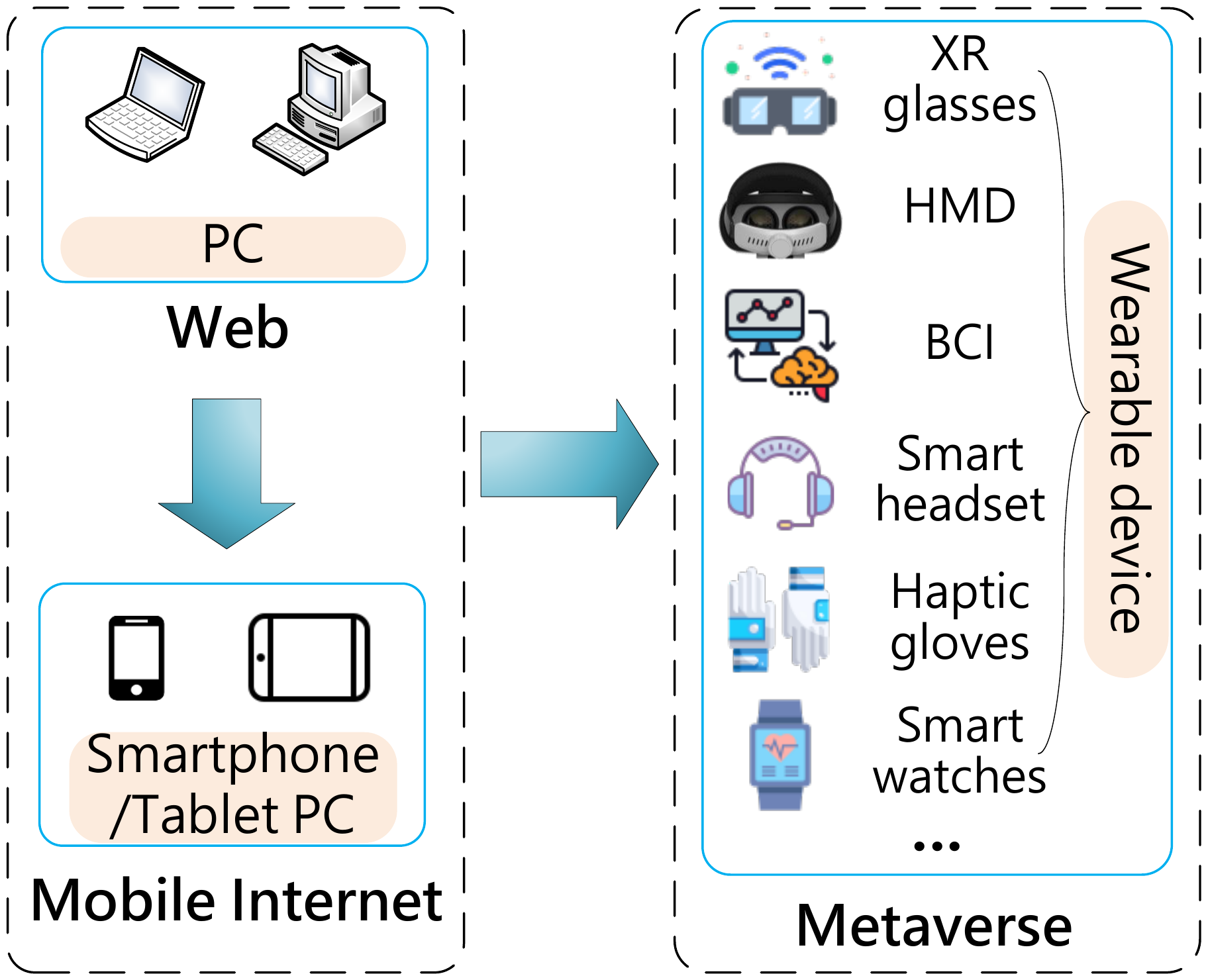}
  \caption{Comparison of hardware terminals for entering the web, mobile Internet, and the metaverse.}\label{fig:entrance}\vspace{-3mm}
\end{figure}

Fig.~\ref{fig:entrance} compares the hardware terminals for entering the web, mobile Internet, the metaverse.
As shown in Fig.~\ref{fig:entrance}, in the metaverse, empowered by XR and HCI technologies, wearable devices such as HMD and BCI enable user/avatar interactions and are expected as the major terminal to enter the metaverse \cite{lee2021all}. Besides, the metaverse usually includes various administrative domains and the sub-metaverses can be implemented on distinct blockchain platforms \cite{nguyen2021metachain}.
In the following, we first review existing works on the metaverse in terms of key management and identity authentication for wearable devices. Then, we give the literature review in cross-domain identity authentication in the metaverse.

\subsubsection{Key Management for Wearable Devices}
Wearable devices such as Oculus helmets and HoloLen headsets are anticipated to be the major terminal to enter the metaverse.
Key management (including generation, negotiation, distribution, update, revocation, and recovery) is essential for wearable devices to establish secure communication, deliver sensory data, receive immersive service, etc.
Conventional key management mechanisms are mainly built upon cryptographic systems such as Diffie-Hellman cryptosystem and public key infrastructure (PKI). These mechanisms usually require strict constraints on available resources (e.g., computation power, memory size, bandwidth, and transmit power) for sensor node operations, which are not applicable for battery-powered wearable devices with compact battery size and limited computational capacity.
In the literature, works \cite{8089389,sun2020accelerometer,chen2018lirek,zheng2018critical} take the intrinsic features of distinct wearable devices (e.g., wireless channel and gait signal) into account in designing efficient key management schemes, which can be beneficial for future metaverse construction.

Aimed to bridge the contactless secret key establishment among tiny wearable devices under wireless communication environments, Li \emph{et al}. \cite{8089389} design an innovative key establishment approach by utilizing unique wireless channel characteristics based on the positioning of wearable devices. The authors leverage the received signal strength (RSS) trajectories of two moving wearables to construct the secret key by moving or shaking the wearable devices.
Rigorous security analysis proves the defense of eavesdropping and experimental results validate its practicability for wearables with short-range communications and frequent movements. Apart from the RSS, the channel impulse response (CIR) is another typical unique physical-layer characteristic between communication parties.

To secure communications between wearable devices integrated with accelerometers, Sun \emph{et al}. \cite{sun2020accelerometer} exploit the gait-based biometric cryptography to design a group key generation and distribution scheme for wearable devices based on signed sliding window coding and fuzzy vault. The proposed acceleration-based key generation mechanism takes advantage of the randomness of noise signals imposed on the raw acceleration signals to produce a group key. Besides, it utilizes the common characteristic of gait signals sampled from distinct parts of the human body for key distribution to other sensors on the same body. Simulations prove that it can pass both the NIST and Dieharder statistical tests.

To further reduce system overheads and reduce response delay for resource-limited wearable devices, Chen \emph{et al}. \cite{chen2018lirek} introduce a lightweight and real-time key establishment model with gait regularity hiding functions for wearables by analyzing gestures and motions through the integrated accelerometer.
In their work, the shared key is established in real time based on user's motion (e.g., shaking and walking), and a lightweight bit-extraction method is devised based on the value difference of neighboring samples. Simulation results show that the generation rate of shake-to-generate key is 2.027 bit/sec and the matching rate can reach 91\%.

To protect patients from fatal cyber attacks, Zheng \emph{et al}. \cite{zheng2018critical} propose an electrocardiogram (ECG) signal based key distribution mechanism for wearable and implantable medical devices (WIMDs). In their work, two widely used cryptographic primitives, i.e., fuzzy commitment and fuzzy vault, are compared. Experimental results show that the solution built on fuzzy vault achieves a lower acceptable false reject rate (i.e., 5\%) and less energy cost of WIMDs, while the solution built on fuzzy commitment attains a higher false acceptance rate.

\subsubsection{Identity Authentication for Wearable Devices}
Identity authentication for wearable devices to guarantee device/user authenticity is also a promising topic in the metaverse.
To adapt to wearable devices with extremely low computing/storage capacity, Srinivas \emph{et al}. \cite{srinivas2018cloud} present a cloud-based mutual authentication model with low system cost for wearable medical devices to prevent device impersonation in healthcare monitoring systems with password change and smart card revocation functions. Rigorous security analysis {and formal security verification} prove the security of created session key in defense against active and passive attacks.
However, the one-time authentication in \cite{srinivas2018cloud} may cause friction such as unauthorized privileges.
To resolve this issue, Zhao \emph{et al}. \cite{zhao2020trueheart} propose a novel continuous authentication model to support seamless device authentication at a low cost. In \cite{zhao2020trueheart}, unique cardiac biometrics are extracted from photoplethysmography (PPG) sensors (embedded in wrist-worn wearables) for user authentication. Experimental results show that their proposed system obtains a high average continuous authentication accuracy rate of 90.73$\%$.
Jan \emph{et al}. \cite{9290438} design a privacy-aware mutual authentication mechanism for wearable devices, where a hidden Markov model (HMM) is devised to predict privacy risks of patient data leakage. Besides, the security of \cite{9290438} is analyzed using Burrows–Abadi–Needham (BAN) logic.

In the metaverse, Bluetooth may play an important role in short-range communications for wearables.
Aksu \emph{et al}. \cite{8299447} study the wearable device identification issue using the Bluetooth protocol. In their work, a smart wearable fingerprinting method tailored to Bluetooth is devised using a series of AI algorithms, and real tests on wearables validate its functionality and feasibility.
By using two representatives (i.e., Google Nest Learning Thermostat and Nike+ Fuelband Fitness Tracker) as test devices, Arias \emph{et al}. \cite{7321811} present a real attack using a hardware with particular attack vectors to bypass software authentications and compromise the two devices. Lessons show that it is necessary to secure all update channels and disable the microcontroller's external reprogrammability and any debug interface for wearable devices.

\subsubsection{Cross-Domain Identity Authentication}
The metaverse typically contains various administrative security domains created by distinct operators/standards. Identity authentication across distinct {administrative domains (e.g., VR/AR services run by distinct VSPs)} in the metaverse is critical to deliver seamless metaverse services for users/avatars.
Traditional cross-domain authentication mechanisms mainly rely on a trusted intermediary and bring heavy overhead in key management.
To address this issue, Shen \emph{et al}. \cite{9036971} employ blockchain technology to design a decentralized and transparent cross-domain authentication scheme for industrial IoT devices in different domains (e.g., factories). In their work, a consortium blockchain is employed to establish trust among distinct domains, and identity-based encryption (IBE) is used for device authentication. Besides, an anonymous authentication protocol with identity revocation capability is proposed to remedy the drawback of IBE in terms of identity revocation. In addition, real domain-specific information are moved to off-chain storage to reduce storage burdens in the blockchain system.

In the PKI system, it only identifies certificates in its domain. In accessing services in other domains such as Kerberos, users' identities usually could not be recognized or it involves extremely complex operations for cross-domain authentication. By leveraging the distributed consensus of the blockchain, Chen \emph{et al}. \cite{9465710} propose an efficient cross-domain authentication scheme named XAuth. In their work, to improve the response speed arising from the low throughput of blockchains as well as protect user privacy, the authors design an optimized blockchain approach and privacy preservation functions in cross-domain authentication. 
An anonymous authentication protocol based on zero-knowledge proof is also devised to ensure privacy protection. An implemented proof-of-concept (PoC) prototype proves its functionality and feasibility.

\begin{table*}[!t]
\centering \setlength{\abovecaptionskip}{0cm}
    \caption{Summary of Existing/Potential Security Countermeasures to Identity Authentication And Access Control In Metaverse}\label{Summary1}
\begin{tabular}{cclc}\hline
\textbf{Ref.} & \textbf{\begin{tabular}[c]{@{}c@{}}Security\\ Threat\end{tabular}} &  \multicolumn{1}{c}{\textbf{\begin{tabular}[c]{@{}l@{}}$\star$ Purpose\\$\bullet$ Advantages\\$\circ$ Limitations\end{tabular}}} & \textbf{\begin{tabular}[c]{@{}c@{}}Utilized\\ Technology\end{tabular}} \\\hline

{\cite{8089389}}& {{\begin{tabular}[l]{@{}c@{}}Eavesdropping, RSS\\trajectory prediction\end{tabular}}} &{{\begin{tabular}[l]{@{}l@{}}$\star$RSS trajectoriy based secret key establishment for wearables\\$\bullet$Defense of eavesdropping and high efficiency in indoor/outdoor scene\\$\circ$Only work for wearables with short-range communications\end{tabular}}} &{RSS trajectory}\\\hline

\cite{sun2020accelerometer} & {\begin{tabular}[l]{@{}c@{}}Robust key sequence\\generation\end{tabular}}  & {\begin{tabular}[l]{@{}l@{}}$\star$Gait-based biometric group key management for wearable devices\\$\bullet$Pass both Dieharder and NIST tests with high efficiency\\$\circ$Lack real-world thorough test\end{tabular}} & Fuzzy vault \\\hline

\cite{chen2018lirek} & Gait predictability  & {\begin{tabular}[l]{@{}l@{}}$\star$Real-time and lightweight key establishment for wearable devices\\$\bullet$High matching rate of shake-to-generate secret keys\\$\circ$Lack complete and thorough evaluation (e.g., NIST tests)\end{tabular}} & HCI \\\hline

\cite{zheng2018critical} & Hijack of WIMDs  & {\begin{tabular}[l]{@{}l@{}}$\star$Efficient ECG-based key distribution for WIMDs\\$\bullet$High false acceptance rate\\$\circ$Relatively low precision in ECG signal processing\end{tabular}} & {\begin{tabular}[c]{@{}c@{}}Fuzzy commitment,\\fuzzy vault\end{tabular}}\\\hline

\cite{srinivas2018cloud} & Dolev-Yao threat&  {\begin{tabular}[l]{@{}l@{}}$\star$Low-cost mutual authentication for wearable medical devices\\$\bullet$Efficient authentication with low communication cost\\$\circ$Without consideration of the immersiveness of users\end{tabular}} & {\begin{tabular}[c]{@{}c@{}}Real-or-Random\\model\end{tabular}} \\\hline

\cite{zhao2020trueheart} & {\begin{tabular}[l]{@{}c@{}}Random attack,\\synthesis attack\end{tabular}} & {\begin{tabular}[l]{@{}l@{}}$\star$Low-cost PPG-based continuous authentication for wearables\\$\bullet$Low communication overhead and computation cost\\$\circ$Unscalable to large-scale networks\end{tabular}} & {\begin{tabular}[c]{@{}c@{}}Motion artifacts,\\gradient boosting tree\end{tabular}} \\\hline


\cite{9036971} & {\begin{tabular}[l]{@{}c@{}}Eavesdropping,\\impersonation,\\man-in-the-middle\end{tabular}}& {\begin{tabular}[l]{@{}l@{}}$\star$Decentralized cross-domain authentication in industrial IoT\\$\bullet$Anonymous identity authentication and low overhead \\$\circ$Low response speed due to the low throughput of blockchains \end{tabular}} & Blockchain  \\\hline

\cite{9465710} & {\begin{tabular}[l]{@{}c@{}}Impersonation\end{tabular}}& {\begin{tabular}[l]{@{}l@{}}$\star$Efficient cross-domain authentication in optimized blockchain\\$\bullet$Fast response, anonymous authentication, and low overhead\\$\circ$Lack large-scale real-world test  \end{tabular}} & {\begin{tabular}[l]{@{}c@{}}Blockchain,\\multiple Merkle tree\end{tabular}}  \\\hline


\cite{7422115}   &{\begin{tabular}[l]{@{}c@{}}Unauthorized UGVC \\access\end{tabular}} &{\begin{tabular}[l]{@{}l@{}}$\star$Time-domain access control with provable security for UGVC sharing\\$\bullet$Support time-domain UGVC access control\\$\circ$Lack consideration of illegal UGC redistribution\end{tabular}}     & CP-ABE   \\\hline

\cite{8432128}   &{\begin{tabular}[l]{@{}c@{}}Illegal UGC \\redistribution\end{tabular}} &{\begin{tabular}[l]{@{}l@{}}$\star$Secure encrypted UGMC sharing scheme with fair traitor tracing\\$\bullet$High traitor tracing accuracy and perceptual quality\\$\circ$Ignore UGMC usage control\end{tabular}}                                                       & {\begin{tabular}[l]{@{}c@{}}Proxy re-encryption, \\fair watermarking\end{tabular}}   \\\hline

\cite{9268472}   &{\begin{tabular}[l]{@{}c@{}}Unintended UGC \\usage\end{tabular}} &{\begin{tabular}[l]{@{}l@{}}$\star$Fine-grained and transparent UGC usage/processing audit\\$\bullet$Low computational overheads in UGC usage/processing audit\\$\circ$Lack large-scale and real-world performance test\end{tabular}}   & {\begin{tabular}[l]{@{}c@{}}Smart contract, \\trusted computing\end{tabular}}   \\\hline

\end{tabular} 
\end{table*}
\subsubsection{Fine-grained Access Control and Usage Audit for {Wearables and} UGCs}\label{subsubsec:Access}
The massive personally identifiable information (PII) handled by wearables can pose a huge risk of unauthorized exposure.
To address this issue, Ometov \emph{et al}. \cite{7518650} propose a novel delegation-of-use mode for wearable devices with privacy guarantees, where owners can lend their personal devices to others for temporary use. However, the associated attacks along with scalability and efficiency issues still need more investigations in real-world implementation.

The native content creation (e.g., UGCs) produced by avatars is essential to maintain the creativity and sustainability of the metaverse. As UGCs inevitably contain sensitive and private user information, efficient UGC access control and usage audit schemes should be designed. The following works \cite{8399560,7422115,8432128} discuss the UGC access control.
Different from conventional access control schemes which enforce a single access policy for a specific content, Ma \emph{et al}. \cite{8399560} design a scalable access control scheme to allow multiple levels of access privileges for sharing user-generated media contents (UGMCs) in the cloud. The detailed construction based on scalable CP-ABE mechanism is also presented with formal security proof.
However, the above scheme cannot support time-domain UGMC access control.
To address this issue, Yang \emph{et al}. \cite{7422115} propose a time-domain attribute-based access control mechanism with provable security for sharing user-generated video contents (UGVCs) in the cloud.
In their mechanism, the allowed time slots for access are embedded into both ciphertexts and keys in CP-ABE, thereby only authorized users in specific time slots can decrypt the UGVCs. Moreover, queries on UGVCs created in previous time slots along with efficient attribute updating and revoking are supported.
Nevertheless, the above works overlook that authorized entities may become traitors to illegally redistribute UGCs to the public, i.e., \emph{illegal UGC redistribution}.
To address this realistic threat, Zhang \emph{et al}. \cite{8432128} propose a novel secure encrypted UGMC sharing scheme with traitor tracing in the cloud via the proxy re-encryption mechanism (for secure UGMC sharing) and watermarking mechanism (for traitor tracing).

The above works mainly focus on the access control of UGCs, while the usage control (i.e., shared UGCs can be only used for intended purposes) is ignored.
To bridge this gap, Wang \emph{et al}. \cite{9268472} propose a novel data processing-as-a-service (DPaaS) mode to complement the current data sharing ecosystem and exploit blockchain technologies for fine-grained data usage policy making on the user's side, policy execution atop smart contracts, and policy audit on transparent ledgers.
Yu \emph{et al}. \cite{8249924} combine both sensitiveness of UGMC (to be shared) and trustworthiness of user (being granted) to train a tree classifier for fine-grained privacy setting configurations. In their scheme, a deep network is utilized to extract discriminative features and identify privacy-sensitive object classes/events, and users are clustered into social groups for trustworthiness characterization.

\vspace{-1mm}
\subsection{Summary and Lessons Learned}\label{subsec:summary1}
{The metaverse requires users to autonomously control their identity and behavioral data, where users can independently manage the UGCs, assets and behavior data generated in different sub-metaverses, avoiding the risk of private data being abused. Moreover, under the premise of autonomous authorization, users can provide data to other subjects to share the benefits generated by these data.}
For identity authentication and access control in the metaverse, we have learned that apart from traditional cryptography system design, the fusion of sensory signals (e.g., ECG and PPG) of wearable devices and biometrics (e.g., face and gait) of users can be beneficial for efficient key generation and identity authentication in the metaverse. Besides, blockchain can build trust-free digital identities for metaverse users. Moreover, continuous-time dynamic authentication, as well as cross-chain and cross-domain authentication need further investigation under the metaverse environment.
A comparison of existing/potential security countermeasures to identity authentication and access control in the metaverse is presented in Table~\ref{Summary1}.

\section{Threats and Countermeasures to Data Management in Metaverse}\label{sec:Threat2}
\subsection{Threats to Data Management in Metaverse}\label{subsec:Datathreat}
The data collected or generated by {wearable devices and users/avatars} may suffer from threats in terms of {data tampering, false data injection, low-quality UGC, ownership/provenance tracing, and intellectual property violation} in the metaverse.

\emph{1) Data Tampering Attack}. Integrity features ensure effective checking and detection of any modification during data communication among the ternary worlds and various sub-metaverses. Adversaries may modify, forge, replace, and remove the raw data {throughout the life-cycle of metaverse data services} to interfere with the normal activities of users, avatars, or physical entities \cite{9035635}. Besides, adversaries may remain undetected by falsifying corresponding log files or message-digest results to hide their criminal traces in the virtual space.

\emph{2) False Data Injection Attack}. Attackers can inject falsified information such as false messages and wrong instructions to mislead metaverse systems \cite{7752958}. For example, AI-aided content creation can help improve user immersiveness in the early stage of the metaverse, and adversaries can inject adversary training samples or poisoned gradients during centralized or distributed AI training, respectively, to generate biased AI models. {The returned wrong feedbacks or instructions may also threaten the safety of physical equipment and even personal safety. For example, falsified feedbacks such as excessive voltage can cause damage and malfunction of wearable XR devices. Another example is that the tampered hundredfold magnifications of bodily pain in being shot in Fortnite (a metaverse game) may cause the death of human user.}

{\emph{3) Issues in Managing New Types of Metaverse Data}. Compared with the current Internet, the metaverse requires new hardware and devices to gather various new types of data (e.g., eye movement, facial expression, and head movement), which is previously uncollected, to make fully immersive user experiences \cite{4623222}. Besides, end-devices in the metaverse (e.g., VR glasses and haptic gloves) can be capable of capturing iris biometrics, fingerprints, or other user-sensitive biometric information.
Consequently, it raises new challenges in collecting, managing, and storing these enormous user-sensitive metaverse data, as well as the cyber/physical security of metaverse devices.}

{For each virtual world (e.g., Horizon and Fortnite), the corporations (e.g., Meta and Epic Games) that create and manage it can monetize these private data to streamline and tailor their services or products towards users' expectations, thereby facilitating precision marketing for benefits. Other relevant issues to be addressed include who will be the subject of responsibility for collecting, handling, storing, securing, and destroying these data.}

\emph{4) Threats to Data Quality of UGC and Physical Input}. In metaverse, selfish users/avatars may contribute low-quality contents under the UGC mode to save their costs, thereby {undermining user experience such as unreal experience in the synthesized environment}. 
For example, they may share unaligned and severe non-IID data during the collaborative training process of the content recommendation model in the metaverse{, causing inaccurate content recommendation}. Another example is that uncalibrated wearable sensors can generate inaccurate and even erroneous sensory data to mislead the creation of digital twins in the metaverse{, causing poor user experience}.

\emph{5) Threats to UGC Ownership and Provenance}. Different from the asset registration procedure supervised by the government in the real world, the metaverse is an open and fully autonomous space and there exists no centralized authority. Due to the lack of authority, it is hard to trace the ownership and provenance of various UGCs produced by massive avatars under different virtual worlds in the metaverse, as well as turn UGCs into protected assets \cite{liang2017provchain}. {Besides, UGCs can be shared in real time within the virtual world or across various virtual worlds and unlimitedly replicated due to the digital attributes, making it harder for efficient provenance and ownership tracing.}

{\emph{6) Threats to Intellectual Property Protection}. Different from the actual world, the definition of intellectual property in the metaverse should be adapted to enforce licensing boundaries and usage rights for the owners with the evolvement and expanding scale of the metaverse \cite{4428344}. Moreover, severe challenges may arise in defining and protecting intellectual property (e.g., avatars, UGCs, and AIGCs) in the new metaverse ecology, as the geographic boundaries of countries are broken down in the metaverse. For example, there have already been disputes owing to the use of celebrity lookalikes in video games \cite{btlj2014Lawsuits}. Given the commercial value created by avatars, such kinds of disputes may spike exponentially in the future metaverse.}

\subsection{Security Countermeasures to Metaverse Data Management}\label{subsec:defense2}
The metaverse is a digital world built on digital copies of the physical environment and avatars' digital creations. Analogy to the value created by human activities in the real world, digital twins and UGCs as well as avatars' behaviors (e.g., chat records and browsing records) will produce certain value in the metaverse \cite{Metaverse2021Duan}. Information security is an important prerequisite for the development and prosperity of the metaverse.
In the following, we discuss the data security in metaverse in terms of data reliability, data quality, and provenance.
\subsubsection{Data Reliability of AIGC, Digital Twin, and Physical Input}

\begin{figure}[!t]\setlength{\abovecaptionskip}{-0.0cm}
\centering
  \includegraphics[width=10.cm]{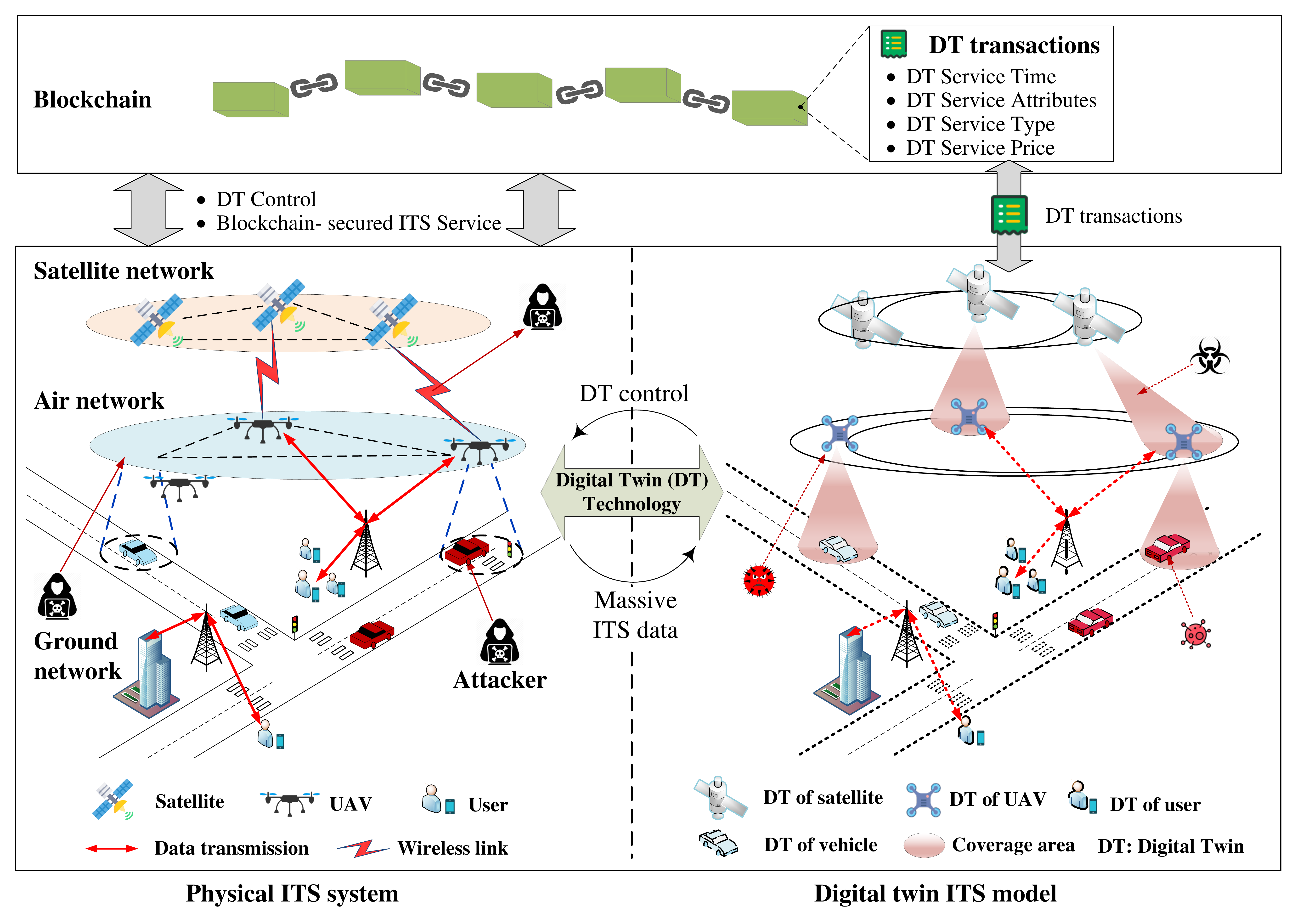}
  \caption{{Illustration of blockchain-enabled digital twin (DT)-as-a-service (DTaaS) in intelligent transportation systems (ITS) \cite{9660773}.}}\label{fig:BCdigitaltwin}\vspace{-2mm}
\end{figure}

In the metaverse, AI such as generative adversarial network (GAN) can help generate high-quality dynamic game scenarios and context images, but also poses security threats such as adversarial and poisoned samples which is hard to detect for humans. In the literature, by taking adversarial samples as part of training data, various efforts have been done to resist adversarial samples via virtual adversarial learning \cite{8417973}, adversarial representation learning \cite{2019Modality}, adversarial reinforcement learning \cite{2020Stealthy}, adversarial transfer learning \cite{2020Efficient}, and so on, which can be beneficial to resist adversarial threats in the construction of the metaverse.

The works \cite{8822494,9660773} discuss the data reliability of digital twins in the metaverse.
Gehrmann \emph{et al}. \cite{8822494} propose a reliable state replication method for digital twin synchronization {in industrial applications} and identify seven key requirements in security architecture design. Besides, the authors formally define the \emph{synchronization consistency} as a metric of the robustness of digital twin synchronization. A PoC implementation using programmable logic controllers (PLCs) validates its effectiveness.
However, the trustworthiness of data collected from disparate data silos is not studied in \cite{8822494}.
To address this issue in the metaverse, as shown in Fig.~\ref{fig:BCdigitaltwin}, Liao \emph{et al}. \cite{9660773} leverage permissioned blockchain technology for trusted digital twin (DT) service transactions between VSPs and service requesters in intelligent transportation systems (ITS). A DT-DPoS (delegated proof of stake) consensus protocol is devised to improve consensus efficiency by using distributed DT servers to form the validator committee. Besides, to facilitate users' customized DT services, an on-demand DT-as-a-service (DTaaS) architecture is presented for fast response to meet diverse DT requirements in ITS.

\begin{figure}[!t]\setlength{\abovecaptionskip}{-0.0cm}
\centering
  \includegraphics[width=4.5cm,height=3.1cm]{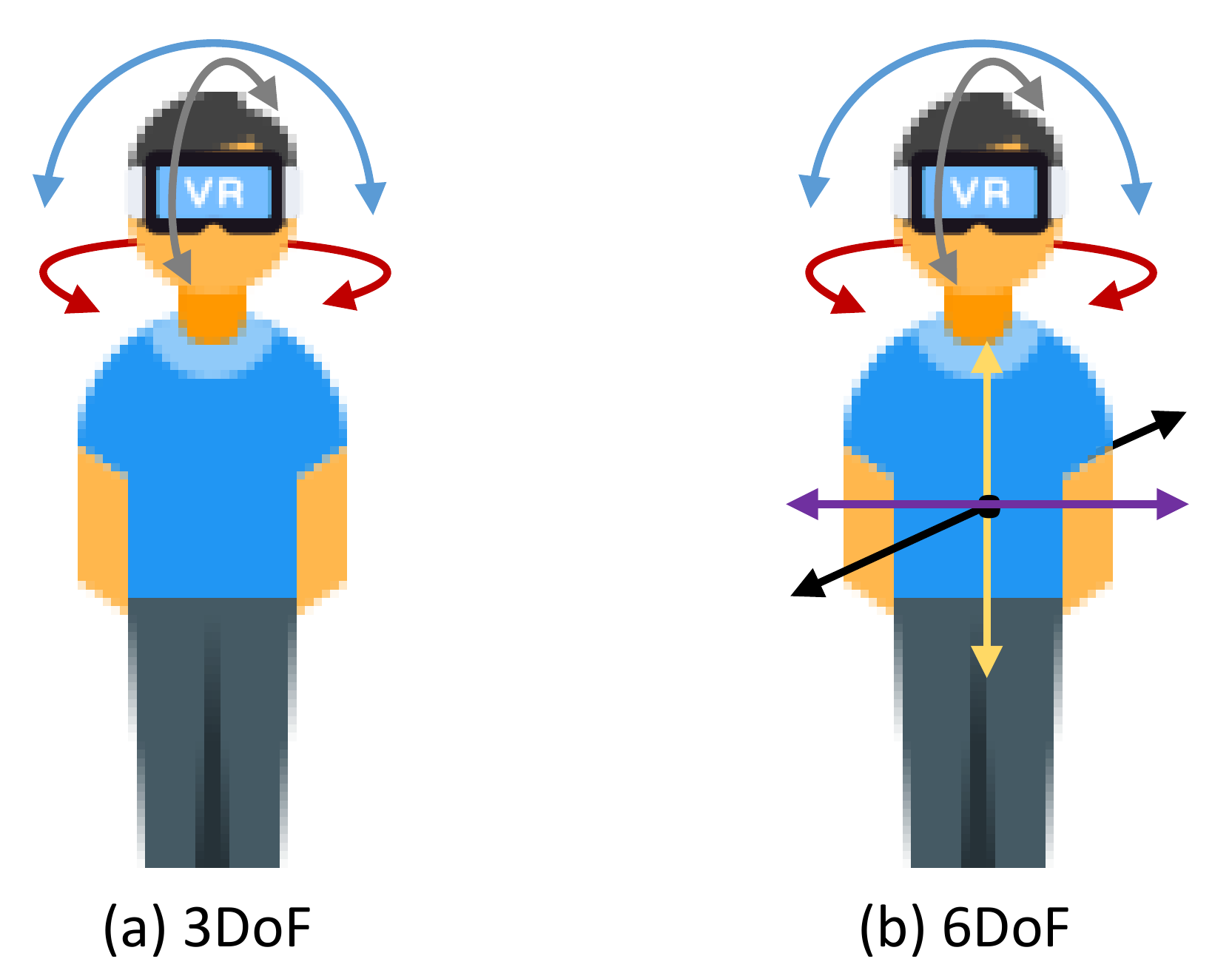}
  \caption{Illustration of (a) 3DoF and (b) 6DoF. 3DoF means an object can rotationally move around the 3D space (i.e., x, y, and z axes), while 6DoF has additional translational movement along those axes (i.e., moving forward/backward, up/down, and left/right).}\label{fig:6DoF}\vspace{-2mm}
\end{figure}

The works \cite{2008Spatialized,Jean2019Rendering} investigate parametric audio rendering to match and improve the visual experience in 3D virtual worlds.
Zimmermann \emph{et al}. \cite{2008Spatialized} present an interactive audio streaming mechanism with high scalability based on peer-to-peer (P2P) topology for immersive interaction in NVEs. Their mechanism combines two concepts: \emph{area of interest (AoI)} and \emph{aural soundscape} to make proximal and spatialized audio interactions. Specifically, AoI limits the distribution area of audio streams as avatars are more likely to interact with others in proximity (the distance is measured by virtual coordinates), and aural soundscape allows distributively audio rendering from different sources to match the visual landscape.
Jot \emph{et al}. \cite{Jean2019Rendering} design an interactive audio engine based on 6-degree-of-freedom (6DoF) object for parametric audio scene programming (i.e., controllable acoustic orientation, size, orientation, and other properties) in audiovisual metaverse experiences. Fig.~\ref{fig:6DoF} illustrates the difference of 6DoF with conventional 3DoF in using VR devices.
Simulation results in \cite{2008Spatialized,Jean2019Rendering} show the feasibility of their design.

\subsubsection{Data Quality of UGC and Physical Input}
Low-quality data input from physical sensors and the UGCs produced by avatars can deteriorate the quality-of-service (QoS) of metaverse services and the QoE of users. 
Effective quality control mechanisms are important to offer efficient metaverse services and maintain sustainability of the creator economy.
Dickinson \emph{et al}. \cite{Patrick2021Experiencing} give a user study on 68 participants in a VR environment and show that user perception of character believability is influenced positively by behavioral features while negatively by visual elements.

In the literature, game theory and AI methods have been widely utilized to motivate users' high-quality data contribution or service offering, which can offer some lessons in the metaverse design. For example, Xu \emph{et al}. \cite{9036917} propose a dynamic Stackelberg game to model the interactions between the content provider and edge caching devices (ECDs), where content provider is the game-leader which makes its payment strategy of caching service while each ECD serves as the game-follower to decide its strategy on quality of caching service. A two-tier Q-learning based mechanism is devised in \cite{9036917} to dynamically derive the optimal strategies for each side. In \cite{9478223}, Su \emph{et al}. propose a deep RL (DRL)-based incentive mechanism to encourage users' high-quality model contribution in distributed AI paradigms with consideration of both non-IID effects and collaboration between edge/cloud servers.

The works \cite{Han2022ADynamic,du2021optimal} study the data availability in metaverse in terms of data synchronization and QoS, respectively. 
For accurate DT synchronization with its physical counterpart, Han \emph{et al}. \cite{Han2022ADynamic} propose a hierarchical game for dynamic DT synchronization in the metaverse, where end devices collectively gather the status information of physical objects and VSPs decide proper synchronization intensities. In their work, every user selects the optimal VSP in the lower-level evolutionary game, and every VSP makes the optimal synchronization strategy in the upper-level differential game based on users' strategies and value of DT.
Simulation results demonstrate that the proposed mechanism attains a higher accumulated revenue for VSP.
By leveraging covert communication methods, Du \emph{et al}. \cite{du2021optimal} propose an optimal targeted advertising strategy for the VSP to maximize its payoff in offering high-quality access services for end-users while attaining close-to-one detection error for attackers. In their work, the Vidale-Wolfe advertising model is exploited, and a novel metric \emph{meta-immersion} is introduced to measure users' feelings in metaverse experience. Simulation results show that the VSP can boost its payoff in comparison with that without advertising.
For dynamic metaverse applications, the information freshness (e.g., age of information) can be further considered in data/service offering.

\subsubsection{Secure Data Sharing in XR Environment}
Metaverse applications are usually multi-user such as multi-player gaming and remote collaboration.
Aimed for secure content sharing under multi-user AR applications, Ruth \emph{et al}. \cite{Kimberly2019Secure} study an AR content sharing control mechanism and implement a prototype on HoloLens to allow AR content sharing among remote or co-located users with inbound and outbound control. By rigorously exploring user's design space on various AR apps, the authors also define various mapping manners of AR contents into the real world.
In WebVR (a VR-based 3D virtual world on HTML canvases), Lee \emph{et al}. \cite{Hyunjoo2021AdCube} identify three new ad fraud threats (i.e., blind spot tracking, gaze and controller cursor-jacking, and abuse of an auxiliary display) in content sharing. User studies on 82 participants show the success rates range from 88.23\% to 100\%. Besides, a defense mechanism named AdCube is presented in \cite{Hyunjoo2021AdCube} via visual confinement of 3D ad entities and sandboxing technique. Experimental results show the defense effectiveness of AdCube with a small system cost for 9 WebVR demo sites.
\begin{table*}[!t]
\centering \setlength{\abovecaptionskip}{0cm}
    \caption{Summary of Existing/Potential Security Countermeasures To Data Management In Metaverse}\label{Summary2}
\begin{tabular}{cclc}\hline
\textbf{Ref.} & \textbf{\begin{tabular}[c]{@{}c@{}}Security\\ Threat\end{tabular}} & \multicolumn{1}{c}{\textbf{\begin{tabular}[c]{@{}l@{}}$\star$ Purpose\\$\bullet$ Advantages\\$\circ$ Limitations\end{tabular}}} & \textbf{\begin{tabular}[c]{@{}c@{}}Utilized\\ Technology\end{tabular}} \\\hline

\cite{8822494}   &{\begin{tabular}[l]{@{}c@{}}Threats to\\digital twin\end{tabular}} &{\begin{tabular}[l]{@{}l@{}}$\star$Reliable state replication method for digital twin synchronization\\$\bullet$Low computational cost and synchronization latency\\$\circ$Lack trustworthiness guarantee of data gathered from disparate data silos\end{tabular}}                                                       & {\begin{tabular}[l]{@{}c@{}}Cloud computing,\\digital twin\end{tabular}}   \\\hline

\cite{9660773}   &{\begin{tabular}[l]{@{}c@{}}Trustworthiness \\of digital twin\end{tabular}} &{\begin{tabular}[l]{@{}l@{}}$\star$Trustworthy data dissemination for digital twins on customized DTaaS\\$\bullet$High reliability of data sources in digital twin creation\\$\circ$Lack accurate representation of digital footprints\end{tabular}}    & Blockchain  \\\hline

{\cite{Han2022ADynamic}} &{{\begin{tabular}[l]{@{}c@{}}Synchronization \\of digital twin\end{tabular}}} &{{\begin{tabular}[l]{@{}l@{}}$\star$Dynamic and optimized DT synchronization strategies of VSPs\\$\bullet$Higher accumulated revenue for VSP\\$\circ$Interoperability issues among VSPs\end{tabular}}}    & {Hierarchical game} \\\hline

{\cite{Kimberly2019Secure}} &{{\begin{tabular}[l]{@{}c@{}}Insecure AR\\content sharing\end{tabular}}} &{{\begin{tabular}[l]{@{}l@{}}$\star$Content sharing control module in multi-user AR apps\\$\bullet$Feasibility via prototype validation on Microsoft HoloLens\\$\circ$Lack location privacy protection in AR applications\end{tabular}}}    & {Multi-user AR} \\\hline

{\cite{Hyunjoo2021AdCube}} &{{\begin{tabular}[l]{@{}c@{}}Cursor-jacking attack,\\blind spot attack\end{tabular}}} &{{\begin{tabular}[l]{@{}l@{}}$\star$Allow behavior specification and enforcement of TTP's ad code\\$\bullet$High defense success rate with low page loading time and frame-per-second drop\\$\circ$Lack visibility reporting\end{tabular}}}    & {{\begin{tabular}[l]{@{}c@{}}WebVR,\\Sandbox\end{tabular}}} \\\hline

\cite{8070948}   &{\begin{tabular}[l]{@{}c@{}}Low data quality\end{tabular}} &{\begin{tabular}[l]{@{}l@{}}$\star$Quality-aware vehicular service access with mobility support\\$\bullet$High average service quality and network success rate\\$\circ$Lack impact analysis on trust management and security issues\end{tabular}}   & {\begin{tabular}[l]{@{}c@{}}Generation tree, \\bi-direction buffering\end{tabular}}   \\\hline

\end{tabular} 
\end{table*}
\subsubsection{Provenance of UGC}
Data provenance can realize the traceability of historical archives of a piece of UGC, which is essential to evaluate data quality, trace data source, reproduce data generation process, and conduct audit trail to quickly identify data responsible subjects.
In the metaverse, UGC provenance information such as the source, circulation, and intermediate processing information is often stored in disparate data silos (e.g., distinct blockchains), making it difficult to monitor and track in real time. 
Existing works on IoT data provenance can offer some lessons for UGC provenance design in the metaverse.

Satchidanandan \emph{et al}. \cite{7738534} design a dynamic watermarking techniques which exploits indelible patterns imprinted in the medium to detect misbehaviors (e.g., signal tampering) of malicious sensors or actuators. Besides, advanced watermarking technique can be utilized for intellectual property protection and ownership authentication in the metaverse.
Liang \emph{et al}. \cite{liang2017provchain} present a blockchain-based cloud file provenance architecture named ProvChain with three stages, i.e., collection, storage, and verification of provenance information. ProvChain ensures source tamper resistance, user privacy, and reliability of cloud storage.
For multi-hop IoT, Mohsin \emph{et al}. \cite{kamal2018light} design a lightweight protocol to enable data provenance in wireless communications, where the RSS indicator of the communicating IoT node is exploited to produce the unique link fingerprint.

In the metaverse, the life-cycle of UGCs involves the ternary worlds and multiple sub-metaverses, which can be more complex than that in traditional IoT. Moreover, smart contracts are anticipated to play an important role in enforcing UGC provenance across various metaverse platforms, and more research efforts on its functionality, efficiency, and security are required.
Besides, the scalability, trust, and efficiency (e.g., response delay) are still challenging issues in the provenance of massive UGCs in the large-scale metaverse.

\vspace{-1.mm}
\subsection{{Summary and Lessons Learned}}\label{subsec:summary2}
For data management in the metaverse, we have learned that the integration of various cutting-edge technologies in the metaverse results in more attack surfaces on UGC, physical inputs, and metaverse outputs. Besides, blockchain offers a potential solution to ensure data reliability in digital twin creation and mitigation. With the flourishing and expanding scale of future metaverse systems, brand new threats emerged specifically under a metaverse setting can breed, where new defenses for them need to be designed. Essentially, as various emerging technologies are incorporated by the metaverse as its foundation, their intrinsic flaws and vulnerabilities may also be inherited by the metaverse. In addition, the effects of existing threats can be amplified and become more severe in the metaverse, driven by the interweaving of various technologies.
A comparison of existing/potential security countermeasures to metaverse data management is presented in Table~\ref{Summary2}.

\section{Privacy Threats and Countermeasures in Metaverse}\label{sec:Threat3}
\subsection{Privacy Threats in Metaverse}\label{subsec:Datathreat}
When enjoying digital lives in the metaverse, user privacy including location privacy, habit, living styles, and so on may be offended during the life-cycle of data services including data perception, transmission, processing, governance, and storage.

\emph{1) Pervasive Data Collection}. {To immersively interact with an avatar, it} requires pervasive user profiling activities {at an unreasonably granular level} \cite{Falchuk8371577} including facial expressions, eye/hand movements, speech and biometric features, and even brain wave patterns. {Besides, via advanced XR and HCI technologies, it can facilitate the analysis of physical movements and user attributes and even enable user tracking \cite{9134928}.} For example, the motion sensors and four built-in cameras in the Oculus helmet help track the head direction and movement, draw our rooms, as well as monitor our positions and environment in real time with submillimeter accuracy{, when we browse the Roblox and interact with other avatars}. If this device is hacked by attackers, severe crimes can be committed on the basis of these large-volume of sensitive data.

{Another example is the attractive virtual office (e.g., Horizon Workroom and Microsoft Mesh), which may arise significant security and privacy risks to employees. On one hand, employee conversations, the emails they send, the URLs they visit, their behaviors, and even the tones of their voices may be monitored by the managers. On the other hand, the immersive workplace may be prone to other security and privacy issues such as intrusions, snooping, and impostors.}

\emph{2) Privacy Leakage in Data Transmission}. In metaverse systems, {abundant personally identifiable information collected from wearables (e.g., HMDs)} are transferred via wired and wireless communications, the confidentiality of which should be prohibited from unauthorized individuals/services \cite{7518650}.
Although communications are encrypted and information is confidentially transmitted, adversaries may still access the raw data by eavesdropping on the specific channel and even track users' locations via differential attacks \cite{9205204} and advanced inference attacks \cite{4492778}.

\emph{3) Privacy Leakage in Data Processing}. In metaverse services, the aggregation and processing of massive data collected from human bodies and their surrounding environments are essential for the creation and rendering of avatars and virtual environments, in which users' sensitive information may be leaked \cite{9531392}. For example, the aggregation of private data (belonging to different users) to a central storage for training personalized avatar appearance models may offend user privacy and violate existing real-world regulations such as General Data Protection Regulation (GDPR) \cite{GDPR}. Besides, adversaries may infer users' privacy (e.g., preferences) by analyzing and linking the published processing results (e.g., synthetic avatars) in various virtual spaces such as Roblox and Fortnite.

\emph{4) Privacy Leakage in Cloud/Edge Storage}. The storage of the private and sensitive information (e.g., user profiling) of massive users in cloud servers or edge devices may also raise privacy disclosure issues. For example, hackers may deduce users' privacy information by frequent queries via differential attacks \cite{9205204} and even compromise the cloud/edge storage via distributed denial-of-service (DDoS) attacks \cite{7842850}. In 2006, a customer database of the Second Life (a metaverse game) was hacked and the user data was breached including unencrypted usernames and addresses, as well as encrypted payment details and passwords \cite{SL2006breached}.

\emph{5) Rogue or Compromised End Devices}. In the metaverse, more wearable sensors will be placed on human bodies and their surroundings to allow avatars to make natural eye contact, capture hand gesture, reflect facial expression, and so on in real time. A significant risk is that these wearable devices can have a completely authentic sense of who you are, how you talk, behave, feel, and express yourself.
The use of rogue or compromised wearable end devices (e.g., VR glasses) in the metaverse is becoming an entryway for data breaches and malware invasions, and the problem may be more severe with the popularity of wearable devices for entering the metaverse \cite{9134928}. Under the manipulation of rogue or compromised end devices, the avatars in the metaverse may turn into a source of data collection, thereby infringing user privacy.
For example, as advanced wearable devices such as Oculus helmets and haptic gloves can track eye movements and hand gestures, hackers can recreate user actions and even sensitive passwords for personal accounts by following the eye and finger movements in tapping in codes on a virtual keypad.

\emph{6) Threats to Digital Footprints}. As the behavior pattern, preferences, habits, and activities of avatars in the metaverse can reflect the real statuses of their physical counterparts, attackers can collect the digital footprints of avatars and exploit the similarity linked to real users to facilitate accurate user profiling and even illegal activities \cite{ning2021survey}. Besides, metaverse usually offers the third person view with a wider viewing angle of their avatar's surroundings than that in the real world \cite{2008Privacy}, which may infringe on other players' behavior privacy without awareness. For example, an avatar may conduct the virtual stalking/spying attack in Roblox by following your avatar and recording all your digital footprints, e.g., purchasing behaviors, to facilitate social engineering attacks.

\emph{7) Identity Linkability in Ternary Worlds}. As the metaverse assimilates the reality into itself, the human, physical, and virtual worlds are seamlessly integrated into the metaverse, causing identity linkability concerns across the ternary worlds \cite{4428344}. For example, a malicious player \emph{A} in Roblox can track another player \emph{B} by the name appeared above the corresponding avatar of player \emph{B} and infer his/her position in the real world. Another example is that hackers may track the position of users via compromised VR headsets or glasses \cite{9134928}.

\emph{8) Threats to Accountability}. XR and HCI devices intrinsically gather more sensitive data such as locations, behavior patterns, and surroundings of users than traditional smart devices. For example, in Pok{\'e}mon Go (a location-based AR game), players can discover, capture, and battle Pok{\'e}mon using mobile devices with GPS. The accountability in the metaverse is important to ensure users' sensitive data are handled with privacy compliance. For metaverse service providers, the audit process of the compliance of privacy regulations (e.g., GDPR) for accountability can be clumpy and time-consuming under the centralized service offering architecture. Besides, it is hard for VSPs to ensure the transparency of regulation compliance during the life-cycle of data management \cite{9268472}, especially in the new digital ecology of metaverse.

\emph{9) Threats to Customized Privacy}. Similar to existing Internet service platforms, distinct users generally exhibit customized privacy preferences for different services or interaction objects \cite{8933081} under distinct sub-metaverses. For example, a user in Roblox may be more sensitive to monetary trading activities than social activities. Besides, users/avatars may be more sensitive in interacting with strangers than acquaintances, friends, or relatives. However, challenges exist in developing customized privacy preservation policies for personal data management while considering avatars in the metaverse as individual information subjects \cite{7486070}, as well as the characteristics of users and sub-metaverses.

\subsection{Privacy Countermeasures in Metaverse}\label{subsec:defense3}
\subsubsection{Privacy in Metaverse Games}
AR/VR games are the current most popular metaverse application for users.
AR/VR games usually contain three steps: the game platform (i) collects sensory data from users and their surroundings, (ii) identifies objects according to these contexts, and lastly (iii) performs rendering on game senses for immersiveness.

Existing works have demonstrated the security and safety concerns related to metaverse games using case studies \cite{5054904} and qualitative studies \cite{8418615,9134928}.
Bono \emph{et al}. \cite{5054904} offer two case studies (i.e., \emph{Second Life} and \emph{Anarchy Online}) and show that a hacker can exploit the features and vulnerabilities of MMO metaverse games to fully compromise and take over players' devices (e.g., laptops).
Lebeck \emph{et al}. \cite{8418615} carry out a qualitative lab study using Microsoft HoloLen (an AR headset), whose result shows that players can easily be immersed in AR experiences and treat virtual objects as real, as well as various security, privacy, and safety issues are uncovered.
Shang \emph{et al}. \cite{9134928} identify a novel user location tracking attack in location-based AR games (e.g., Pok{\'e}mon Go) by solely exploiting the network traffic of the player, and real-world experiments on 12 volunteers validate that the proposed attack model attains fine-grained geolocation of any player with high accuracy.
Besides, three possible mitigation approaches are presented in \cite{9134928} to alleviate attack effects.

To prevent potential privacy issues in metaverse games, Laakkonen \emph{et al}. \cite{7535117} introduce privacy-by-design principles in digital games from both qualitative and quantitative perspectives, where nineteen privacy attributes divided into three levels are accounted for privacy evaluation.
In \cite{8516484}, Corcoran \emph{et al}. distinguish the \emph{individual privacy} and \emph{group privacy} in privacy-preserving interactive metaverse game design. The former refers to the purchasing patterns, behavioral traits, communication, image/video data, and location/space related to an individual, while the latter refers to the privacy associated with a group of individuals (e.g., a social group, an organization, and a nation).

\subsubsection{Privacy-Preserving UGC Sharing and Processing}\label{subsubsec:Sharing}
Existing privacy-preserving schemes for data sharing and processing mainly focus on four fields: differential privacy (DP), federated learning (FL), cryptographic approaches (e.g., secure multi-party computation (SMC), homomorphic encryption (HE), and zero-knowledge proof (ZKP)), and trusted computing.
The following works \cite{9205204,9488776,9200775,7120947} discuss privacy-preserving UGC sharing in the metaverse.
To offer privacy-preserving trending topic recommendation services in the metaverse, Wei \emph{et al}. \cite{9205204} propose a graph-based local DP mechanism, where a compressive sensing indistinguishability method is devised to produce noisy social topics to prevent user-linkage association and protect keyword correlation privacy with high efficiency.
To enable smart health sensing without violating users' private data in the metaverse, Zhang \emph{et al}. \cite{9488776} present a FL-based secure data collaboration framework where wearable sensors periodically send local model updates trained on their private sensory data to the server which synthesizes a global abnormal health detection model. To resolve class imbalance concerns of participants under FL, the authors in \cite{9488776} further design a novel local update method based on reinforcement learning (RL) and an adaptive global update method via online regret minimization.
To enhance privacy protection in the blockchain-based metaverse, Guan \emph{et al}. \cite{9200775} utilize ZKP to empower current account-model blockchains (e.g., Ethereum) with privacy preservation functions in terms of hiding sender-recipient linkage, account balances, and transaction amounts.
Xu \emph{et al}. \cite{7120947} identify the \emph{co-photo privacy} threat in social metaverse that a shared photo may contain not only the individual privacy but also the privacy of others in photos. Besides, by utilizing SMC and SVM techniques, the authors design a personalized facial recognition method to differentiate photo co-owners without disclosing their privacy in users' private photos.

Privacy-preserving UGC processing in the metaverse has also attracted various attention. Based on Okamoto-Uchiyama HE, Li \emph{et al}. \cite{9531392} present a verifiable privacy-preserving method for data processing result prediction in edge-enabled CPSSs. Besides, batch verification is supported for multiple prediction results at one time to reduce communication burdens.
Wang \emph{et al}. \cite{9268472} leverage the trusted computing technique to design a privacy-preserving off-chain data processing mechanism, where private UGC datasets are processed in an off-chain trusted enclave and the exchange of processed results and payment are securely executed via the designed fair exchange smart contract.

\subsubsection{Confidentiality Protection of UGC and Physical Input}
The confidentiality of UGCs (inside the metaverse) along with physical inputs (to the metaverse) should be ensured to prevent private data leakage and sensitive data exposure. The authentication \&access control (in Sect.~\ref{subsec:defense1}) and privacy computing technologies (in Sect.~\ref{subsubsec:Sharing}) are enablers to maintain UGC confidentiality in the metaverse.
For confidentiality of physical inputs, Raguram \emph{et al}. \cite{6509878} propose a novel threat named \emph{compromising reflections}, which can automatically reconstruct user typing on virtual keyboards, thereby compromising data confidentiality and user privacy. Experiment results show that compromising reflections of a device's screen (e.g., sunglass reflections) are sufficient for automatic and accurate reconstruction with no limitation on the motion of handheld cameras even in challenging scenarios such as a bus and even at long distances (e.g., 12m for sunglass reflections).

\subsubsection{Digital Footprints Protection}
In the metaverse, privacy inside avatars' digital footprints can be classified into three types \cite{Falchuk8371577}: (i) personal information (e.g., avatar profiling), (ii) virtual behaviors, and (iii) interactions or communications between avatars or between avatar and NPC.
Avatars' digital footprints can be tracked via virtual stalking/spying attacks in the metaverse to disclose user's real identity and other private information, e.g., shopping preferences, location, and even banking details. A potential solution is \emph{avatar clone} \cite{ning2021survey}, which creates multiple virtual clones of the avatar which appear
identical to confuse the attackers. Nevertheless, it brings other challenging issues such as managing multiple representations of each user and managing millions of clones roaming around the metaverse.

Another potential solution is to \emph{disguise} by periodically changing avatar's appearance to confuse attackers, or \emph{mannequin} by replacing the avatar with a single clone (e.g., bot) which imitates user's behavior and \emph{teleport} user's true avatar to another location when being tracked.
Other privacy preservation mechanisms \cite{Falchuk8371577} include invisibility, private enclave, lockout.
\emph{Invisibility} indicates the avatar is made to be temporarily invisible in case of suspected stalking.
\emph{Private enclaves} allow certain locations inside the metaverse to be occupied by individuals, which are unobserved by others. In private enclaves, owners have control over who can enter into the enclave by teleporting, thereby offering a maximum level of privacy.
\emph{Lockout} means certain areas inside the metaverse are temporarily locked out for private use. After the lock expires, the restriction is lifted and other users are allowed to enter the area.

\subsubsection{{Personalized Privacy-Preserving Metaverse}}
{As users/avatars are featured with personalized privacy demands and service preferences, existing privacy computing technologies (in Sect.~\ref{subsubsec:Sharing}) should also take their customized privacy/service profiles into account in designing privacy-enhanced metaverse. Existing works on personalized privacy computing are mainly based on similarity \cite{7486070}, randomized response \cite{8933081}, personalized FL \cite{9492755}, and so on. With the growth of metaverse, more research on new personalized privacy preservation methods is required to serve new applications and the new ecology in the metaverse.}

\subsubsection{{Privacy-Enhancing Advances in Industry}}
\begin{figure}[!t]\setlength{\abovecaptionskip}{-0.0cm}
\centering
  \includegraphics[width=8.3cm]{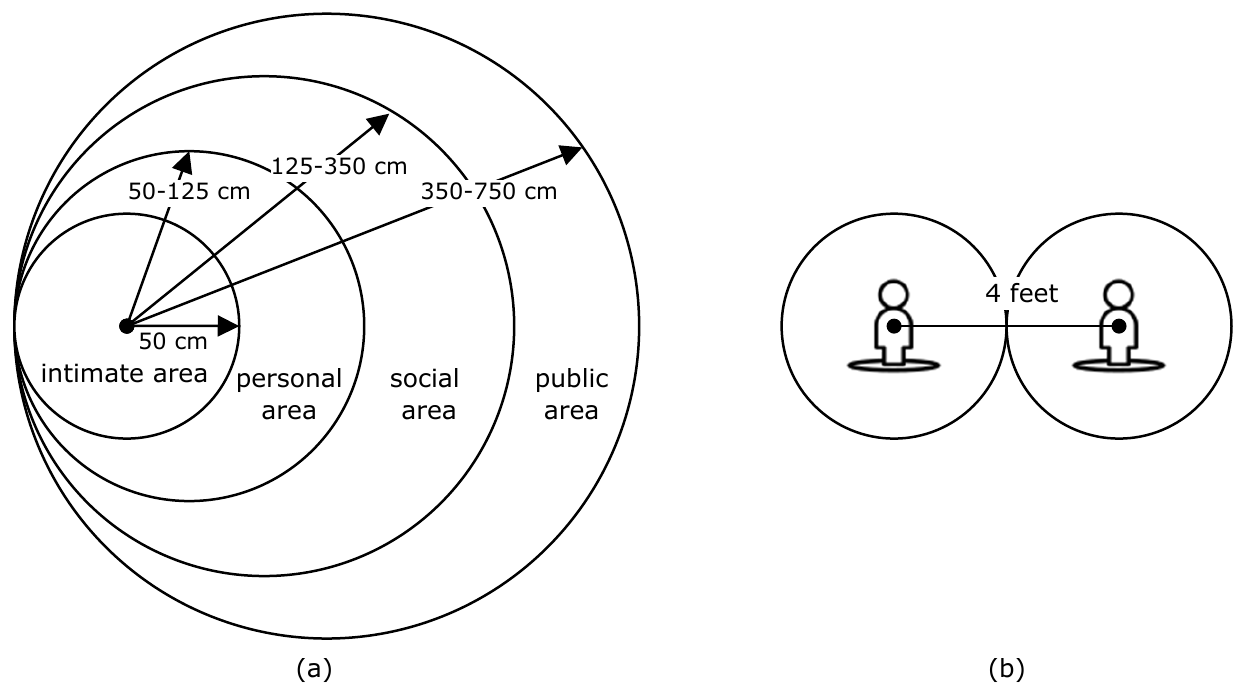}
  \caption{{Illustration of personal space in real and virtual worlds. (a) Four types of personal spaces: public area (350-750 cm), social area (125-350 cm), personal area (50-125 cm), and intimate area (within 50 cm). (b) Meta's personal boundary function for avatars with default private border of 2-foot.}}\label{fig:distance}\vspace{-2mm}
\end{figure}

{In the metaverse, there have been incidents such as VR groping and VR sexual harassments in Horizon Worlds \cite{groping2021MIT}.
In the real world, people potentially keep an appropriate distance from others to maintain personal spaces when socializing.
According to the interpersonal intimacy, psychologist Stanley Hall quantified and divided four types of personal spaces: public area (350-750 cm), social area (125-350 cm), personal area (50-125 cm), and intimate area (within 50 cm), as shown in Fig.~\ref{fig:distance} (a).
It means that for less familiar people, the more personal space we require. Similarly, each avatar also requires personal space even in the virtual world.}
{Recently, Meta announced the \emph{private boundary} function in its metaverse platforms Horizon Venues and Horizon Worlds to avoid groping and harassments, where the default personal border for every avatar is a 2-foot circle \cite{distance2022Meta}. As shown in Fig.~\ref{fig:distance} (b), avatars need to keep at least 4 feet (about 1.2 m) away from others to maintain private space.}

{Google has built a \emph{Privacy Sandbox} on Android apps in 2022 to promote private advertising solutions by curbing the sharing of private information with third parties and removing cross-app identifiers (including advertising ID) \cite{privacysandbox2022}. Besides, Google debuts its open-source DP tool named PipelineDP with Python library in 2022 by creating pipelines which aggregate personal data to derive valuable insights in a differentially private manner.
Apple also utilizes local DP to gather individual data from end devices running on macOS or iOS for privacy-preserving services \cite{DPpaperApple} such as lookup hints, Emoji suggestions, QuickType suggestions, and Safari autoplay intent detection.}

\subsection{{Summary and Lessons Learned}}\label{subsec:summary3}
{In traditional Internet services, the platform operators (e.g., enterprises) control the user data for commercial purposes. Such a centralized management pattern has intrinsic principal-agent risks, and is more prone to privacy leakage and data abuse. In the metaverse, users (as privacy subjects) need to take back control of their private data.}
Notably, the PII (including user profiling and biometric data) collected and processed in the metaverse can be more granular and unprecedentedly pervasive to make fully immersive experiences, where the device for acquisition massive user-sensitive data, as well as the transmission, storage, processing, access control, and destruction process should be well-protected in the life cycle of private data. For privacy in the metaverse, we have learned that existing privacy threats can be amplified, and new threats related to digital footprints can emerge. Therefore, users may suffer more privacy exposure and higher leakage impact, and require stricter privacy protection in the metaverse.
A comparison of existing/potential security countermeasures to metaverse privacy issues is presented in Table~\ref{Summary3}.

\begin{table*}[]
\centering \setlength{\abovecaptionskip}{0cm}
    \caption{Summary of Existing/Potential Privacy Countermeasures In Metaverse}\label{Summary3}
\begin{tabular}{cclc}\hline
\textbf{Ref.} & \textbf{\begin{tabular}[c]{@{}c@{}}Security\\ Threat\end{tabular}} & \multicolumn{1}{c}{\textbf{\begin{tabular}[c]{@{}l@{}}$\star$ Purpose\\$\bullet$ Advantages\\$\circ$ Limitations\end{tabular}}} & \textbf{\begin{tabular}[c]{@{}c@{}}Utilized\\ Technology\end{tabular}} \\\hline

\cite{9134928}   &{\begin{tabular}[l]{@{}c@{}}Location tracking\\in AR games\end{tabular}} &{\begin{tabular}[l]{@{}l@{}}$\star$Attack model construction and possible mitigation design\\$\bullet$Fine-grained and high-accuracy location tracking attack modeling\\$\circ$Lack complete defense analysis under real-world test\end{tabular}}                                                       & {\begin{tabular}[l]{@{}c@{}}Cloud, AR, \\access control\end{tabular}}   \\\hline

\cite{9205204}   &{\begin{tabular}[l]{@{}c@{}}Privacy exposure \\in UGC sharing\end{tabular}} &{\begin{tabular}[l]{@{}l@{}}$\star$Graph-based local DP for privacy-preserving topic recommendation\\$\bullet$High-level privacy and high efficiency in user-linkage unassociation\\$\circ$Lack image indistinguishability mechanism in practical use\end{tabular}}   & Local DP   \\\hline

\cite{9488776}   &{\begin{tabular}[l]{@{}c@{}}Privacy exposure \\in UGC sharing\end{tabular}} &{\begin{tabular}[l]{@{}l@{}}$\star$Secure data collaboration with class imbalance scenarios\\$\bullet$High accuracy in abnormal health detection\\$\circ$Lack Byzantine robustness in FL\end{tabular}}   & FL   \\\hline

\cite{7120947}   &Co-photo privacy   &{\begin{tabular}[l]{@{}l@{}}$\star$Personalized facial recognition with privacy protection in photo sharing\\$\bullet$High recognition ratio and efficiency in OSNs\\$\circ$Lack implementation and test on personal clouds (e.g., Dropbox)\end{tabular}}   & Facial recognition  \\\hline

\cite{6509878}   &{\begin{tabular}[l]{@{}c@{}}Compromising \\reflections\end{tabular}} &{\begin{tabular}[l]{@{}l@{}}$\star$Automatically reconstruct user typing on virtual keyboards\\$\bullet$Effective attack execution with high robustness and accuracy\\$\circ$Lack effective defense design\end{tabular}}    & {\begin{tabular}[l]{@{}c@{}}Feature extraction \\and matching\end{tabular}}   \\\hline

\cite{Falchuk8371577}   &{\begin{tabular}[l]{@{}c@{}}Threats to \\digital footprints\end{tabular}}    &{\begin{tabular}[l]{@{}l@{}}$\star$Privacy preservation tools for digital footprints in social metaverse\\$\bullet$Offer complete confusion and private copy tools for avatars\\$\circ$Lack user experience analysis and practical deployment of such tools\end{tabular}}   & {\begin{tabular}[l]{@{}c@{}}Avatar confusion,\\ private copy\end{tabular}} \\\hline

\end{tabular} 
\end{table*}

\section{Network-related Threats and Countermeasures in Metaverse}\label{sec:Threat4}
\begin{figure}[!t]\setlength{\abovecaptionskip}{-0.0cm}
\centering
  \includegraphics[width=7.8cm,height=9cm]{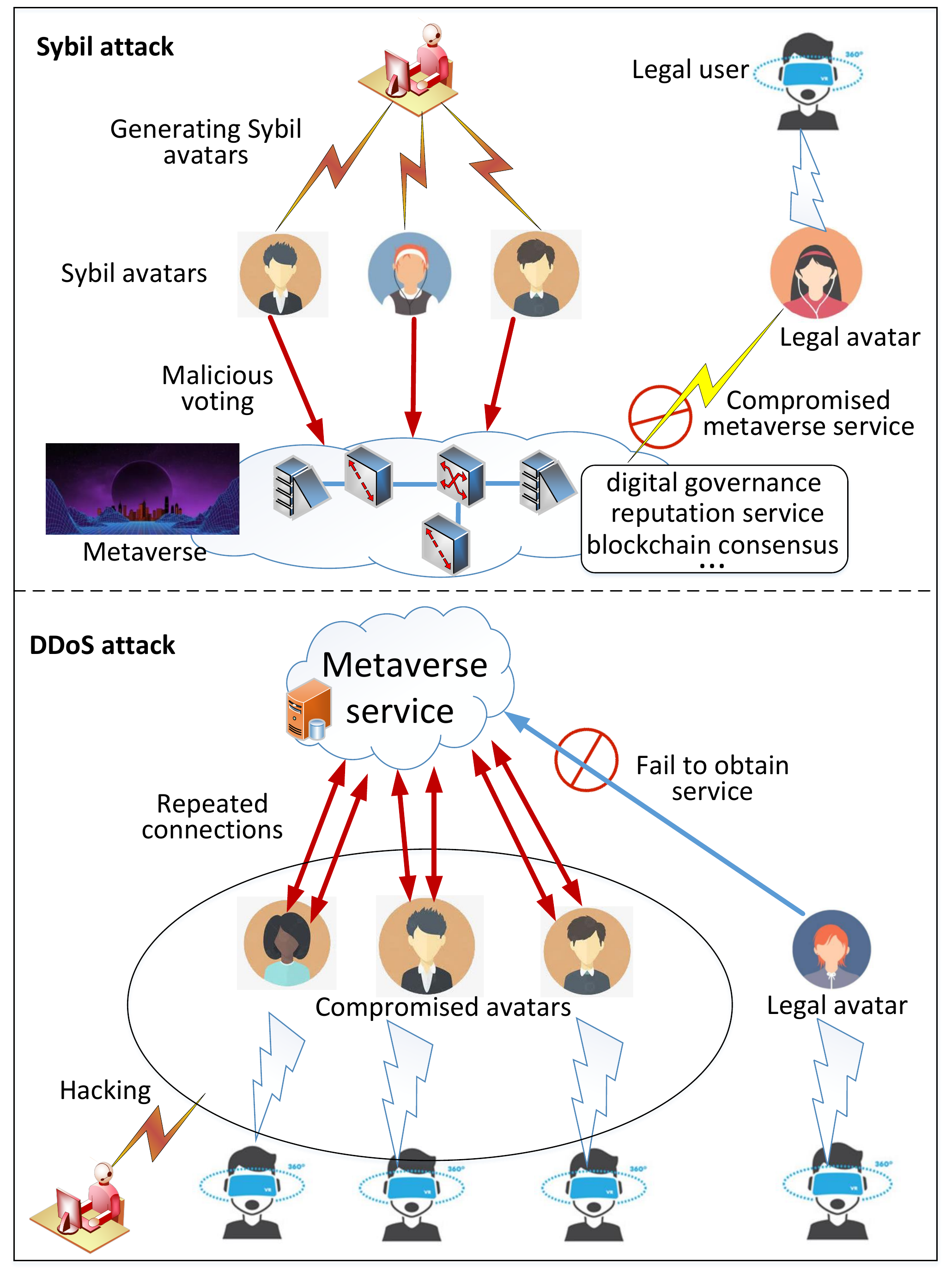}
  \caption{{An illustrative example of Sybil attack and DDoS attack in metaverse.}}\label{fig:SybilDDos}\vspace{-3mm}
\end{figure}
\subsection{Threats to Metaverse Network}\label{subsec:Networkthreat}
In the metaverse, traditional threats (e.g., physical-layer security) to the communication networks can also be effective, as the metaverse evolves from the current Internet and incorporates existing wireless communication technologies. Here, we list some typical threats as below.

\emph{1) SPoF}. In the construction of metaverse systems, the centralized architecture (e.g., cloud-based system) brings convenience for user/avatar management and cost saving in operations. Nevertheless, it can be prone to the SPoF caused by the damage of physical root servers and DDoS attacks \cite{9631953}. Besides, it raises {trust and transparency} challenges {in trust-free exchange of virtual goods, virtual currencies, and digital assets} across various virtual worlds {in the metaverse}.

\emph{2) DDoS}. {As the metaverse includes massive tiny wearable devices, adversaries may compromise these metaverse end-devices and make them part of a botnet \cite{7842850} (e.g., Mirai)} to conduct DDoS attacks to make network outage and service unavailability by overwhelming the centralized server with giant traffic within short time periods, as depicted in the upper part of Fig.~\ref{fig:SybilDDos}. {Besides, owing to the constrained communication pressure and storage space on the blockchain, part of NFT functions may be performed on off-chain systems in practical applications \cite{NFT2022hashkey}, where adversaries may launch DDoS attacks to cause service unavailability of the NFT system.}

\emph{3) Sybil Attacks}. Sybil adversaries may manipulate multiple faked/stolen identities to gain disproportionately large influence on metaverse services (e.g., reputation service{, blockchain consensus, and voting-based service in digital governance) and even take over the metaverse network}, thereby compromising system effectiveness, as shown in the lower part of Fig.~\ref{fig:SybilDDos}. {For example, adversaries may be able to out-vote genuine nodes by producing sufficient Sybil identities to refuse to deliver or receive some blocks, thereby effectively blocking other nodes from a blockchain network in the metaverse.} 

\subsection{Situational Awareness in Metaverse}\label{subsec:defense4}
Situational awareness is an effective tool for security monitoring and threat early-warning in large-scale complex systems such as the metaverse \cite{9676467}.
In the metaverse, local situational awareness is essential for monitoring a single security domain and global situational awareness can assist early-warning of large-scale distributed threats targeted at multiple sub-metaverses.
\subsubsection{Local Situational Awareness}
Situational awareness for devices and systems built on XR technology has received increasing attention in the metaverse \cite{9676467,9261134,9130693}.
Woodward \emph{et al}. \cite{9676467} review the presentation of information in AR headsets, and discuss the potential in applying AR technologies to enhance users' situational awareness in perception and understanding the surroundings. Apart from AR technology, VR technology can enhance situational awareness capacities in various applications.
Ju \emph{et al}. \cite{9261134} carry out realistic and immersive driving simulations, whose findings validate that acoustic cues can help VR drivers remain alert in emergencies (e.g., accidents) under VR car-driving scenarios.
Lv \emph{et al}. \cite{9130693} present a smart intrusion detection model to detect attack behaviors {on 3D VR-based industrial control systems} based on support vector machine (SVM). Experimental results on a simulated VR industrial scenario show that its average accuracy can keep above 90\%.
However, the proposed model cannot resist unknown/new attack types.

To effectively detect unknown/new threats, Vu \emph{et al}. \cite{9199886} design a representation learning approach for better prediction of unknown attacks, where three regularized autoencoders (AEs) are deployed to learn the latent representation. The effectiveness of their work is evaluated on nine recent IoT datasets.
To be further adaptive to wearable devices with extreme size and energy constraints, Heartfield \emph{et al}. \cite{6507636} propose a multi-layered lightweight anomaly detection method by exploiting radio-frequency wireless communications to/from them to identify potentially malicious transactions. In \cite{9277640}, RL methods are employed for intrusion detection in small-scale applications such as smart homes. In practical applications, it is usually hard and costly to label massive attack samples. To deal with the challenges of few labeled data and the corresponding over-fitting issues, Zhou \emph{et al}. \cite{9311786} combine few-shot learning and Siamese neural network to mitigate over-fitting and intelligently detect diverse attack types in industrial systems.

To summarize, existing security measures can be categorized into two groups: \emph{reactive} approaches (aim to counter past known attacks) and \emph{proactive} approaches (aim to mitigate future unknown attacks). In general cases, reactive defenses built on timely attack trapping, frequent retraining, and decision verification can be more convenient and effective than pure proactive defenses. Besides, proactive defenses can be classified into two paradigms \cite{2017Wild}: \emph{security by design} defenses (against white-box attacks) and \emph{security by obscurity} defenses (against black-box attacks).
The above defense approaches can provide some lessons to resist unknown/new threats in the metaverse.

\begin{figure}[!t]\setlength{\abovecaptionskip}{-0.0cm}
\centering
  \includegraphics[width=9.8cm]{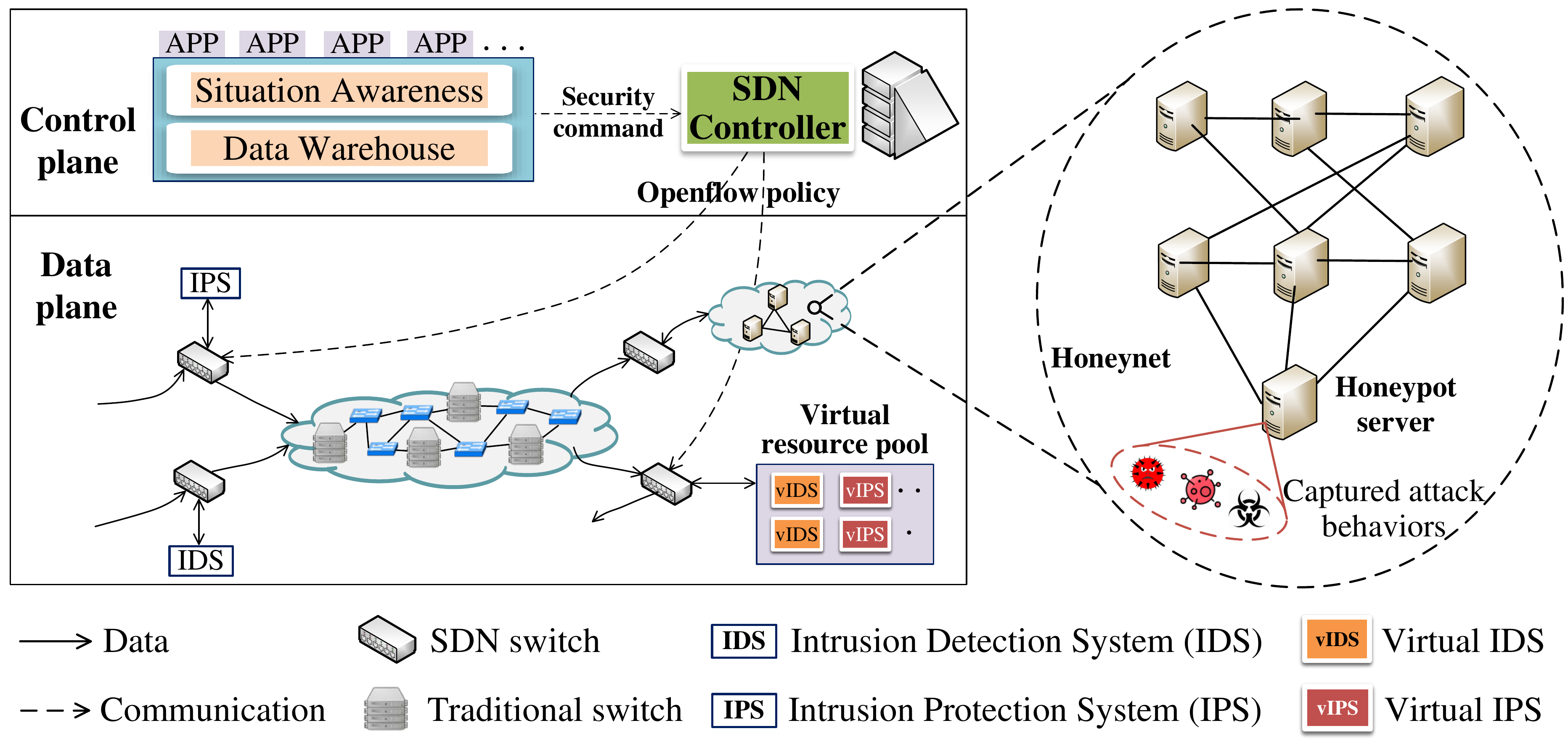}
  \caption{{Illustration of SDN-enabled virtual honeynet services for collaborative situational awareness \cite{9060972}.}}\label{fig:SDNhoneynet}\vspace{-3mm}
\end{figure}

\subsubsection{Global Situational Awareness}
\begin{table*}[!t]
\centering \setlength{\abovecaptionskip}{0cm}
    \caption{Summary of Existing/Potential Security Countermeasures To Network-Related Threats In Metaverse}\label{Summary4}
\begin{tabular}{cclc}\hline
\textbf{Ref.} & \textbf{\begin{tabular}[c]{@{}c@{}}Security\\ Threat\end{tabular}} & \multicolumn{1}{c}{\textbf{\begin{tabular}[c]{@{}l@{}}$\star$ Purpose\\$\bullet$ Advantages\\$\circ$ Limitations\end{tabular}}} & \textbf{\begin{tabular}[c]{@{}c@{}}Utilized\\ Technology\end{tabular}} \\\hline

\cite{9130693}   &{\begin{tabular}[l]{@{}c@{}}Intrusion of VR \\control system\end{tabular}}    &{\begin{tabular}[l]{@{}l@{}}$\star$Smart intrusion detection to detect attacks in 3D VR environments\\$\bullet$High classification and detection accuracy\\$\circ$Cannot resist unknown/new attack types\end{tabular}}   & SVM \\\hline

\cite{8640020}   &{\begin{tabular}[l]{@{}c@{}}Malicious events \\in distribution grid\end{tabular}}    &{\begin{tabular}[l]{@{}l@{}}$\star$Data-driven situational awareness in large-scale distributed power grids\\$\bullet$High accuracy in malicious event labeling \\$\circ$Rely on additional expert knowledge for costly event labeling\end{tabular}}   & Multi-class SVM \\\hline

\cite{9540743}   &{\begin{tabular}[l]{@{}c@{}}Intrusion of industrial \\control system\end{tabular}}  &{\begin{tabular}[l]{@{}l@{}}$\star$Monitoring and profiling of potential attack behaviors\\$\bullet$High detection/prediction accuracy and low response time\\$\circ$Lack merging other cutting-edge technologies into this framework\end{tabular}}   & {\begin{tabular}[l]{@{}c@{}}SDN, \\digital twin\end{tabular}} \\\hline

\cite{8915712}   &{\begin{tabular}[l]{@{}c@{}}Large-scale \\network intrusion\end{tabular}} &{\begin{tabular}[l]{@{}l@{}}$\star$Honeynet-based situational awareness to deceive attackers\\$\bullet$Rapid honeynet deployment with adaptability to unknown threats\\$\circ$Low scalability and programmability in large-scale deployment\end{tabular}}   & Honeynet \\\hline

\cite{9060972}   &{\begin{tabular}[l]{@{}c@{}}Large-scale \\network intrusion\end{tabular}} &{\begin{tabular}[l]{@{}l@{}}$\star$SDN-enabled virtual honeynet with high scalability and flexibility\\$\bullet$Successful implementation and test in real-world EU project\\$\circ$Lack resilience of compromised domain operators\end{tabular}}   & {\begin{tabular}[l]{@{}c@{}}SDN, \\honeynet\end{tabular}} \\\hline

\end{tabular} 
\end{table*}

The above works mainly focus on situational awareness in a local security domain. Global situational awareness can facilitate understanding global security statuses in defending large-scale attacks in the metaverse.
Both works \cite{8640020,7587350} utilize data-driven approaches for global situational awareness in large-scale distributed power grids. In \cite{8640020}, Shahsavari \emph{et al}. propose a multi-class SVM classifier to extract malicious events from collected raw metering data. However, their approach relies on additional expert knowledge for costly event labeling. To resolve this issue, Wu \emph{et al}. \cite{7587350} further model legitimate users and attackers as an evolutionary game and devise a two-phase RL algorithm to solve the game.
Profiling of potential attack behaviors is another challenge in the metaverse.
Krishnan \emph{et al}. \cite{9540743} combine digital twin and SDN to build a behavioral monitoring and profiling system where security strategies are evaluated on digital twins before being deployed in the real network.

Honeynets consisting of collaborative honeypots offer an alternative solution for building a secure metaverse to defend against large-scale distributed attacks.
Zhang \emph{et al}. \cite{8915712} propose a honeynet-based situational awareness system where each honeypot built on the Docker environment traps attackers, monitors their attack behaviors, and exchanges these information with each other coordinated by the honeynet controller.
However, the work \cite{8915712} has a drawback in terms of scalability and programmability in large-scale deployment.
Zarca \emph{et al}. \cite{9060972} further propose SDN-enabled virtual honeynet services with higher degree of scalability and flexibility, and the efficiency of the proposed approach is validated using real implementations and tests. As shown in Fig.~\ref{fig:SDNhoneynet}, based on specific security policies, security virtual network functions (VNFs) (e.g., virtual honeynet, IDS, IPS, and firewall) can be configured and instanced on demand reactively or proactively, coordinated by the SDN controller. Thereby, appropriate defense mechanisms (including situation monitoring, attack trapping, and security resource allocation) can be provisioned quickly and feasibly to enable self-protection, self-repair, and self-healing.
However, the trust issues and resilience of compromised domain operators in aggregating local situational awareness into the global one require further investigation.

\subsection{{Summary and Lessons Learned}}\label{subsec:summary4}
For situational awareness in the metaverse, we have learned that AR, AI, honeypot, and SDN technologies can help build situational awareness systems in the metaverse. Besides, global situational awareness can assist monitoring and early-warning of large-scale distributed threats targeted at multiple sub-metaverses.
A comparison of existing/potential security countermeasures to network-related threats in the metaverse is presented in Table~\ref{Summary4}.

\section{Economy-related Threats and Countermeasures in Metaverse}\label{sec:Threat5}

\subsection{Threats to Metaverse Economy}\label{subsec:Economicthreat}
Various attacks may threaten the creator economy in the metaverse from the service trust, digital asset ownership, and economic fairness aspects.

\emph{1) Service Trust Issues in UGC \& Virtual Object Trading}. In the open metaverse marketplace, avatars may be distrustful entities without historical interactions. There exist inherent fraud risks (e.g., repudiation and refusal-to-pay) during UGC and virtual object trading among different stakeholders in the metaverse. Besides, in the construction of virtual objects via digital twin technologies, the metaverse has to guarantee that the produced and deployed digital copies are authentic and trustworthy \cite{9660773}. For example, malicious users/avatars may buy UGCs or virtual objects in Roblox and illegally sell the digital duplicates of them to others to earn profits. In addition, adversaries may exploit vulnerabilities in metaverse systems to commit fraud and undermine service trust. An example is that the metaverse project \emph{Paraluni} based on Binance Smart Chain (BSC) lost over \$1.7 million in 2022 due to the reentrancy flaw in smart contracts \cite{Paraluni2022Reen}.

\emph{2) Threats to Digital Asset Ownership}. Due to the lack of central authority and the complex circulation and ownership forms (e.g.,  collective ownership and shared ownership \cite{8360480}) in the distributed metaverse system, it poses huge challenges for the generation, pricing, trusted trading, and ownership traceability in the life-cycle of digital assets in the creator economy. Empowered by blockchain technology, the indivisible, tamper-proof, and irreplaceable NFT offers a promising solution for asset identification and ownership provenance in the metaverse \cite{wang2021NFT}. However, NFTs also face threats such as ransomware, scams, and phishing attacks. For example, adversaries may mint the same NFT on multiple blockchains at the same time. Besides, evil actors may cash out their shares after inflating the value of NFTs, or they may sell NFTs to gain benefits before minting anything, where these De-Fi scams cause \$129 million lost in 2020 \cite{NFT2021lost}.

\emph{3) Threats to Economic Fairness in Creator Economy}. Well-designed incentives \cite{9354053,6725569} are benign impetuses to promote user participation and open creativity in resource sharing and digital asset trading in the creator economy. The following three adversaries who threaten economic fairness are considered.
\begin{itemize}
  \item \emph{Strategic} users/avatars may manipulate the digital market in the metaverse to make enormous profits by breaking the supply and demand status \cite{9354053}. {For example, in metaverse auctions, strategic avatars may overclaim its bid, instead of its true valuation, to manipulate the auction market and win the auction.}
  \item \emph{Free-riding} users/avatars may unfairly gain revenues and enjoy metaverse services without contributing to the metaverse market \cite{4359459}, thereby compromising the sustainability of creator economy. {For example, a free-riding avatar may submit meaningless local updates in collectively training an intelligent 3D navigation model under distributed AI and unfairly enjoy the benefits from the trained metaverse model.}
  \item \emph{Collusive} users/avatars in the metaverse may collude with each other or with the VSP to perform market manipulation and gain economic benefits \cite{6725569}. {For example, collusive avatars may collude to manipulate the results of metaverse auctions and earn illegal revenues.}
\end{itemize}

\subsection{Open and Decentralized Creator Economy}\label{subsec:defense5}
Creator economy is an essential component of the metaverse to maintain its sustainability and promote avatars' open creativity. Besides, it should be built on a decentralized architecture to prevent centralization risks, e.g., SPoF, non-transparency, and control by a few entities. {Specifically, the metaverse economy should simultaneously achieve three goals: (1) make data/assets from different sources mutually identifiable, trustworthy, and verifiable (see Sects.~\ref{sec:Threat1} and \ref{sec:Threat2}); (2) design suitable incentive mechanisms for data/assets circulation to form a benign data sharing and coordination pattern; (3) allow data subjects, data controller, data processor, and the user have the right to negotiate the rules and mechanisms of data protection and applications.}

\subsubsection{Trusted UGC/Asset/Resource Trading}
As shown in Fig.~\ref{fig:blockchainMeta}, blockchain technologies (e.g., NFT and smart contract) provide a decentralized solution to construct the sustainable creator economy.
NFT is the irreplaceable and indivisible token in the blockchain \cite{wang2021NFT} and is regarded as the unique tradable digital asset associated with virtual objects (e.g., land parcel and digital drawing).
For example, in the game Cryptokitties, players can buy virtual pet cats with unique genetic attributes identified by NFT and breed them.
Besides, smart contracts enable the automatic transaction enforcement and financial settlement in trading virtual objects, items, and assets.
The works \cite{8892660,8572758,8502858} discuss the usage of blockchain technology for virtual economy design.

Rehman \emph{et al}. \cite{8892660} discuss several design principles in cryptocurrency ecosystems including centrality, privacy, price manipulation, insider trading, parallel and shadow economy, governance, usability, and security.
Considering the cooperation of heterogeneous smart devices, Biase \emph{et al}. \cite{8572758} propose a swarm economy model for digital resource sharing which incorporates their spontaneous collaboration and dynamic organization in large-scale networks. A blockchain-based transaction model is also developed in \cite{8572758} for transparent and immutable currency audit, thereby ensuring trading trust among distrustful devices.
However, the work \cite{8572758} has drawbacks in terms of non-automatic transaction settlement, high computational overhead, and non-supervisability.
To address these issues, Liu \emph{et al}. \cite{8502858} propose a blockchain-based automatic transaction settlement framework, in which a three-layer sharding blockchain architecture is devised for enhanced system scalability. Moreover, the authors in \cite{8502858} devise an encryption scheme with keyword search to uncover criminal transactions and achieve crime traceability, where the supervision right is equally allocated among all participants.
{Jiang \emph{et al}. \cite{9606227} introduce FL-enabled digital twin (DT) edge networks, where access points (APs) serve as edge nodes to help end-user devices build DT models. In \cite{9606227}, a directed acyclic graph (DAG) blockchain is employed to securely record both local model updates and global model updates in FL, as well as the resource transactions between APs and users.}

{Apart from the trust-free blockchain approaches,} trust or reputation management offer a quantifiable solution to evaluate the trustworthiness of participants and services {with less computation/energy/storage consumption}.
Das \emph{et al}. \cite{6081879} propose dynamic trust models and metrics based on user interactions including direct/indirect trust (derived from local/recommendation experience) and recent/historical trust (considering time decay effects). 
To achieve ``trust without identify'', Wang \emph{et al}. \cite{wang2013enabling} present an anonymous trust and reputation management system in mobile crowdsensing.
However, most of the current works on trust or reputation evaluation may rely on the specific rules to determine trust scores and cannot intelligently learn from historical interaction information.
To cope with this issue, Jayasinghe \emph{et al}. \cite{8364607} exploit AI techniques to design an intelligent trust model, which classifies various individual trust attributes (e.g., frequency, duration, and cooperativeness) and aggregates them to produce final trust values.

\subsubsection{Economic Fairness for Manipulation Prevention}
{Collaboration is essential to the creator economy. Nevertheless, it is hard to promote collaboration among all individual users/avatars without sufficient incentives. Besides,} the economic fairness in metaverse markets may be violated by strategic, free-riding, and collusive users/avatars. Strategy-proof incentive mechanisms, e.g., truthful auctions \cite{7559774} and truthful contracts \cite{9664267}, can prevent strategic users/avatars from market manipulating. However, truthful participation also violates user's privacy, e.g., the true bid in auctions may reveal user's true valuation on the items.
Existing strategy-proof and privacy-preserving auctions mainly depend on cryptographic mechanisms (e.g., ZKP \cite{9354536}, HE \cite{8493354}), DP \cite{9354053}), which may bring large system burdens for energy-limited wearable devices or large data utility decrease in practical metaverse applications. {A trade-off mechanism between privacy and utility is needed for users/avatars with diverse preferences in the metaverse.}

Existing schemes to prevent free-riders (who try to enjoy benefits of the good/service without contributing to it) mainly focus on node behavior modeling \cite{4359459}, cryptographic mechanism \cite{7894274,8827301}, and contribution certification \cite{1709951}. 
For example, Li \emph{et al}. \cite{4359459} {observe that BitTorrent systems (account for 35\% of the traffic on the Internet) may fail to overcome free-riders if a large number of seeds (who have all pieces of the file) exist.
To bridge this gap, the authors} design a fluid model for non-free-riders and free-riders in P2P file sharing systems {(e.g., BitTorrent) to capture and mitigate free-riding effects by designing optimal seed bandwidth allocation strategies. Theoretical analysis shows the existence of Nash equilibrium (NE) in their strategy, and simulation results show its effectiveness in free-riding penalization and cooperation promotion.}

As the economic fairness may conflict with other vital metrics (such as economic efficiency and QoE) to some extent, Shin \emph{et al}. \cite{7894274} introduce two principles in incentive design: (i) strict economic fairness to forbid free-riders; and (ii) adaptive but non-exploitable newcomer bootstrapping for economic efficiency.
Based on symmetric key cryptography and pay-it-forward strategy, the authors in \cite{7894274} design a lightweight and easy-to-implement fairness algorithm named T-Chain to prevent free-riders and enforce reciprocity under fully distributed cooperative scenarios such as BitTorrent-like file sharing. Experiments on BitTorrent validate the efficiency of T-Chain in free-riding prevention, fast newcomer bootstrapping, and low efficiency loss (e.g., only 1\% additional overhead on bandwidth and storage).
To mitigate free-riding attacks, Li \emph{et al}. \cite{8827301} utilize smart contracts and ZKP to generate proof-of-ad-receiving commitments in blockchain systems with anonymity and conditional linkability guarantees.

To avoid tragedy of the commons in P2P networks, Ma \emph{et al}. \cite{1709951} propose a service differentiation framework with free-rider forbidden capabilities, where differentiated services are offered to peers based on their prior contribution levels in service offering. In their work, peers' competing resource request/distribution interactions are formulated as a dynamic competition game. Theoretical analysis proves its efficiency in reaching NE, and numerical examples illustrate its functionality in service differentiation and free-rider prevention.
As users/avatars in the metaverse may also exhibit free-riding behaviors, the above works can provide lessons for free-rider modeling, detection, and prevention in metaverse services.

Multi-user/avatar collusion prevention is also important for fairness in the creator economy. Existing collusion-resistant mechanisms mainly focus on AI-based collusion behavior detection \cite{7295625}, cryptographic approaches \cite{5752848}, game theory \cite{6725569}, and optimization theory \cite{9328528}, which can be beneficial for collusion defense in metaverse services. Besides, future research efforts are required in designing fair mechanisms with the combination of strategy-proofness, collusion-resistance, free-rider prevention, along with privacy preservation {in the metaverse}.

In the literature, various works leverage game theory and learning-based methods to improve economic efficiency for metaverse services, including iterative double auction for resource pricing in DT construction \cite{9660773,9606227}, DRL-based double Dutch auction for VR service trading \cite{Xu2021Wireless}, two-tier Q-learning for secure edge caching services \cite{9036917}, optimization theory for resource allocation in virtual education \cite{Wei2021Unified}, and hierarchical game for coded distributed computing services in metaverse \cite{Yuna2021Reliable}.

\subsubsection{Ownership Traceability of Digital Assets}
In the metaverse, blockchain provides a promising solution to manage the complex asset provenance and ownership tracing in the life-cycle of digital assets by recording the evidence of content/asset originality and involved operations on the public ledgers. As the recorded historical activities on blockchain ledgers are maintained by the majority of entities in the metaverse, it is ensured to be democratic, immutable, transparent, auditable, and non-repudiable. Besides, smart contracts offer an intelligent traceability solution by coding the ownership management logic into scripts which are run atop the blockchain. Existing works have utilized blockchain technologies for food supply \cite{8674550}, cloud storage \cite{liang2017provchain}, charging pile sharing \cite{9632411}, and ride sharing \cite{8939473}. In addition to private ownership, there can exist multiple types of ownership forms in the metaverse such as collective ownership and shared ownership \cite{8360480}, which raise extra challenges in ownership management of virtual objects and metaverse assets. In current metaverse projects, there have been increasing interest in utilizing NFT for asset identification and ownership provenance \cite{wang2021NFT}. Nevertheless, NFTs also face vulnerabilities such as cross-chain fraud, inflation attack, phishing, and ransomware. An example is that bad actors may concurrently mint the same NFT on multiple blockchains.

\begin{table*}[!t]
\centering \setlength{\abovecaptionskip}{0cm}
    \caption{Summary of Existing/Potential Security Countermeasures To Metaverse Economy}\label{Summary5}
\begin{tabular}{cclc}\hline
\textbf{Ref.} & \textbf{\begin{tabular}[c]{@{}c@{}}Security\\ Threat\end{tabular}} & \multicolumn{1}{c}{\textbf{\begin{tabular}[c]{@{}l@{}}$\star$ Purpose\\$\bullet$ Advantages\\$\circ$ Limitations\end{tabular}}} & \textbf{\begin{tabular}[c]{@{}c@{}}Utilized\\ Technology\end{tabular}} \\\hline

\cite{8572758}   &{\begin{tabular}[l]{@{}c@{}}Low cooperation \\in creator economy\end{tabular}} &{\begin{tabular}[l]{@{}l@{}}$\star$Swarm economy model for cooperative and dynamic digital resource sharing\\$\bullet$Real-world implementation of blockchain in such economy model\\$\circ$Non-supervisability in transaction settlement and high computational overhead \end{tabular}}   & {\begin{tabular}[l]{@{}c@{}}Blockchain\end{tabular}} \\\hline

\cite{8502858}   &{\begin{tabular}[l]{@{}c@{}}Lack supervisability \\on criminal transaction\end{tabular}} &{\begin{tabular}[l]{@{}l@{}}$\star$Three-layer sharding blockchain for scalable and automatic transaction\\$\bullet$Enhanced system scalability and traceability of criminal transactions\\$\circ$Lack vulnerability analysis and large-scale real-world simulations\end{tabular}}   & {\begin{tabular}[l]{@{}c@{}}Blockchain \\sharding\end{tabular}} \\\hline

{\cite{9606227}} &{{\begin{tabular}[l]{@{}c@{}}Fraud in\\DT construction\end{tabular}}} &{{\begin{tabular}[l]{@{}l@{}}$\star$Trusted and on-demand DT services in DT edge networks\\$\bullet$Transparent DT model training and resource trading\\$\circ$Lack efficiency and scalability analysis of their DAG blockchain\end{tabular}}}   & {FL, DT, DAG} \\\hline

\cite{8364607}   &{\begin{tabular}[l]{@{}c@{}}Compromised \\nodes/services\end{tabular}} &{\begin{tabular}[l]{@{}l@{}}$\star$Intelligent trust model to quantitatively evaluate user/service trustworthiness\\$\bullet$Aggregate multi-dimensional trust attributes for high-accuracy trust computing\\$\circ$Lack complexity and scalability analysis, as well as cold start issues\end{tabular}}   & {\begin{tabular}[l]{@{}c@{}}Machine learning\end{tabular}} \\\hline

\cite{8493354}   &{\begin{tabular}[l]{@{}c@{}}Economic fairness, \\strategic users\end{tabular}} &{\begin{tabular}[l]{@{}l@{}}$\star$Strategy-proof and privacy-preserving auction for heterogeneous spectrum\\$\bullet$Privacy protection, strategy-proofness, and high social welfare\\$\circ$Vulnerable to collusive bidders in auction\end{tabular}}   & {\begin{tabular}[l]{@{}c@{}}HE, \\auction\end{tabular}} \\\hline

{\cite{4359459}} &{{\begin{tabular}[l]{@{}c@{}}Economic fairness, \\free-riding attack\end{tabular}}} &{{\begin{tabular}[l]{@{}l@{}}$\star$Mitigate free-riding effects in BitTorrent by optimizing seed bandwidth allocation\\$\bullet$Effective free-rider penalization and cooperation promotion\\$\circ$Lack real-world tests on robustness and lack analysis of heterogeneous peers\end{tabular}}}   & {Fluid model} \\\hline

\cite{8827301}   &{\begin{tabular}[l]{@{}c@{}}Economic fairness, \\free-riding attack\end{tabular}}   &{\begin{tabular}[l]{@{}l@{}}$\star$Blockchain-based fair ad delivery among connected vehicles\\$\bullet$Enable anonymity and conditional linkability\\$\circ$Not support batch verification of aggregated dissemination proofs\end{tabular}}   & {\begin{tabular}[l]{@{}c@{}}Smart contracts, \\ZKP\end{tabular}} \\\hline

\cite{6725569}   &{\begin{tabular}[l]{@{}c@{}}Economic fairness, \\collusion attack\end{tabular}} &{\begin{tabular}[l]{@{}l@{}}$\star$Collusion-resistant auction design in cooperative communications\\$\bullet$Truthfulness, collusion-resistance, and budget-balance\\$\circ$Only apply to wireless cooperative communications\end{tabular}}   & {\begin{tabular}[l]{@{}c@{}}Game theory\end{tabular}} \\\hline

\end{tabular} 
\end{table*}

\vspace{-2mm}
\subsection{{Summary and Lessons Learned}}\label{subsec:summary5}
For creator economy in the metaverse, we have learned that blockchain technology is the key to build the decentralized virtual economy ecosystem from virtual currency creation and trusted UGC/asset/resource trading to economic fairness and ownership traceability. Moreover, the interoperability, resilience, and efficiency issues are prime concerns to construct a sustainable creator economy.
A comparison of existing/potential security countermeasures to metaverse economy is presented in Table~\ref{Summary5}.

\section{Threats to Physical World and Human Society and Countermeasures in Metaverse}\label{sec:Threat6}
The threats occurring in the metaverse may also affect the physical world and threaten human society.
\subsection{Threats to Physical World and Human Society}\label{subsec:PhySocthreat}
The metaverse is an extended form of the cyber-physical-social system (CPSS) \cite{8931796}, in which physical systems, human society, and cyber systems are interconnected with complex interactions. The threats in virtual worlds also severely affect physical infrastructures, personal safety, and human society.

\emph{1) Threats to Personal Safety}. In the metaverse, hackers can attack wearable devices, XR helmets, and other indoor sensors (e.g., cameras) to obtain the life routine and track the real-time position of users to facilitate burglary, which may threaten their safety \cite{8675340}. A report released by the XR Security Initiative (XRSI) shows that an adversary can manipulate a VR device to reset the hardware's physical boundaries \cite{XRSI2022Meta}. Thereby, a user in metaverse can be potentially pushed toward a flight of stairs or misdirected into dangerous physical situations (e.g., a street).

Besides, the metaverse can open up new opportunities for misconducts and crimes. In the metaverse, risks of physical trauma may be limited, but users could be mentally scarred. For example, due to the immersive realism of metaverse, hackers can suddenly display harmful and scary content (e.g., ghost pictures) in the virtual environment in front of the avatar, which may lead to the death of fright of the corresponding user. Moreover, the lack of laws and regulations can further increase the possibility of criminal or abusive actions.

\emph{2) Threats to Infrastructure Safety}. By sniffing the software or system vulnerabilities in the highly integrated metaverse, hackers may exploit the compromised devices as entry points \cite{6979242} to invade critical national infrastructures (e.g., power grid systems and high-speed rail systems) via APT attacks \cite{7218444}.

\emph{3) Social Effects}. Although metaverse offers an exciting digital society, severe side effects can also raise in human society such as user addiction \cite{9580681}, rumor prevention \cite{9107084}, child pornography, biased outcomes, extortion, cyberbullying, cyberstalkers \cite{2008Privacy}, and even simulated terrorist camps \cite{meta2022terrorist}.
For example, the immersive metaverse can provide future potentials for extremists and terrorists by making it easier to recruit and meet up, offering new ways for training and coordination, and lowering costs for finding new targets \cite{meta2022terrorist}. Essentially, the immersive training in digital clones of actual buildings can assist terrorists to plan attacks and escape routes.
Another example is that the metaverse, in its ultimate form, is fully controlled by AI algorithms (as depicted in the film \emph{Matrix}), in which the code can be the law to rule everything and severe ethical issues such as race/gender bias may arise.

\vspace{-2mm}
\subsection{Physical Safety}\label{subsec:defense6-1}
In this subsection, we review existing potential solutions to the physical safety in the metaverse from the cyber insurance and cyber-physical interaction aspects.
\subsubsection{Cyber Insurance-based Solutions}
Cyber insurance offers a financial instrument for risk mitigation of critical infrastructures in cyberthreats.
To resolve the high premium stipulation in traditional insurance offered by insurance companies, Lau \emph{et al}. \cite{9428549} propose the coalitional insurance in power systems where the coalitional premium is computed by considering loss distributions, vulnerabilities, and budget compliance in an insurance coalition.
Feng \emph{et al}. \cite{8496785} integrate cyber insurance into blockchain services to prevent potential damages under attacks, where a sequential game theoretical framework is developed to model the interactions among users, blockchain platform, and cyber-insurer. The user's optimal demand of blockchain service, blockchain platform's optimal pricing strategy, and cyber-insurer's optimal investment strategy are analytically derived by solving the joint market equilibrium problem.
However, when applying to the metaverse, the scalable and dynamic insurance coalition formation along with fair premium design under diverse cyber threats (e.g., anti-forensics) require further investigation.

\subsubsection{CPSS-based Solutions}
Apart from the single cyber perspective, existing CPSS-based solutions afford lessons for cyberthreat defense and physical safety protection in the metaverse from the perspective of interactions between cyber and physical worlds.
Vellaithurai \emph{et al}. \cite{6979242} introduce cyber-physical security indices for security measurement of power grid infrastructures. The cyber probes (e.g., IDS) are deployed on host systems to profile system activities, where the generated logs along with the topology information are to build stochastic Bayesian models using belief propagation algorithms.
To resolve the issues (e.g., low-level abstraction) in task-based programming paradigm, Tariq \emph{et al}. \cite{8051056} propose a service-oriented paradigm with QoS-aware operation and resource-aware deployment for better support of disruption-free incremental system implementation and reconfiguration.
Different from CPSSs, metaverse is an immersive and hyper spatiotemporal virtual space with a sustainable economy ecosystem, which adds extra challenges in solution migration.

\begin{table*}[htb]
\centering \setlength{\abovecaptionskip}{0cm}
    \caption{Summary of Existing/Potential Security Countermeasures To Physical and Social Threats In Metaverse}\label{Summary6}
\begin{tabular}{cclc}\hline
\textbf{Ref.} & \textbf{\begin{tabular}[c]{@{}c@{}}Security\\ Threat\end{tabular}} & \multicolumn{1}{c}{\textbf{\begin{tabular}[c]{@{}l@{}}$\star$ Purpose\\$\bullet$ Advantages\\$\circ$ Limitations\end{tabular}}} & \textbf{\begin{tabular}[c]{@{}c@{}}Utilized\\ Technology\end{tabular}} \\\hline

{\cite{8496785}} &{{\begin{tabular}[l]{@{}c@{}}Threats to cyber \\insurance\end{tabular}}} &{{\begin{tabular}[l]{@{}l@{}}$\star$Game theoretical modeling among users, blockchain platform, and cyber-insurer\\$\bullet$Analytically derive the market equilibrium with all participants' optimal strategies\\$\circ$Lack scalable and dynamic insurance coalition formation and fair premium design
\end{tabular}}}  & {Sequential game} \\\hline

\cite{6979242}  &{\begin{tabular}[l]{@{}c@{}}Stochastic risk \\on power system\end{tabular}} &{\begin{tabular}[l]{@{}l@{}}$\star$Cyber-physical security indices for security measurement of power systems\\$\bullet$Efficient indices computing under actual attacks in real-world test-bed\\$\circ$Lack merging other cutting-edge technologies into this framework\end{tabular}}   & {\begin{tabular}[l]{@{}c@{}}Graph theory\end{tabular}} \\\hline

\cite{9428549}   &{\begin{tabular}[l]{@{}c@{}}High premium \\stipulation\end{tabular}} &{\begin{tabular}[l]{@{}l@{}}$\star$Coalitional insurance with budget compliance for risk control in power grids\\$\bullet$High defense level with long-term reduced premiums\\$\circ$Lack dynamic insurance design and dependence analysis of cyberthreats\end{tabular}}   & {\begin{tabular}[l]{@{}c@{}}Cyber-insurance\end{tabular}} \\\hline

\cite{9107084}   &{\begin{tabular}[l]{@{}c@{}}Butterfly effect in \\information spreading\end{tabular}} &{\begin{tabular}[l]{@{}l@{}}$\star$Minimize misinformation influence via dynamic node blocking in OSNs\\$\bullet$Low misinformation spreading value and misinformation interactions\\$\circ$Challenging to be applied to the dynamic and time-varying metaverse\end{tabular}}   & {\begin{tabular}[l]{@{}c@{}}Heuristic greedy\end{tabular}} \\\hline

\cite{8675340}   &{\begin{tabular}[l]{@{}c@{}}Human joystick \\attack\end{tabular}} &{\begin{tabular}[l]{@{}l@{}}$\star$Construct human joystick attack model in immersive VR systems\\$\bullet$Deceive and move immersed players to intended physical locations unconsciously\\$\circ$Lack effective defense design\end{tabular}}   & {\begin{tabular}[l]{@{}c@{}}HCI, VR\end{tabular}} \\\hline

\end{tabular} 
\end{table*}

\vspace{-2mm}
\subsection{Society Management}\label{subsec:defense6-1}
In this subsection, we review existing works on society management in the metaverse from the following two perspectives.
\subsubsection{Misinformation Spreading Mitigation}
The extremely rapid information spreading (e.g., gossip) in the metaverse makes the so-called ``butterfly effect'' more challenging in social governance and public safety in the real world.
As an attempt to address this issue, Zhu \emph{et al}. \cite{9107084} propose to minimize the misinformation influence in online social networks (OSNs) by dynamically selecting a series of nodes to be blocked from the OSN.
However, it only works in traditional static OSNs and it is challenging to be applied in the fully interactive metaverse with a huge and time-varying social graph structure.
\subsubsection{Human Safety and Cyber syndromes}
The full immersiveness in metaverse can also raise immersion concerns, e.g., occlusion and chaperone attack, as well as cybersickness \cite{9697993}.
Casey \emph{et al}. \cite{8675340} investigate a new attack named \emph{human joystick attack} in immersive VR systems such as Oculus Rift and HTC Vive. In their work, adversaries can modify VR environmental factors to deceive, disorient, and control immersed human players and move them to other physical locations without consciousness.
Valluripally \emph{et al}. \cite{9580681} present a novel cybersickness mitigation method and several design principles in social VR learning scenarios via threat quantification and attack-fault tree model construction. However, the ethical issues and adaptations to different attack-defense strategies are not considered in their work, which is an important factor for future metaverse construction.
Besides, more research efforts are required on the mitigation of other immersion risks to human body and human society.

\subsubsection{{Society Acceptance Advances in Industry}}
To enforce age-appropriate interactions within its platforms, Meta has enhanced its age certification mechanism with GDPR-compliance, where a tool named Transfer Your Information (TYI) is developed in 2021 \cite{FYI2021Facebook}. In TYI, users are allowed to retract their personal information from Meta whenever they intend.

\subsection{{Summary and Lessons Learned}}\label{subsec:summary6}
For physical safety and social effect in the metaverse, we have learned that existing cyber-insurance and CPSS based approaches can offer some insights for protecting physical devices. More related technological and sociological efforts in this field considering the characteristics of metaverse are required.
A comparison of existing/potential security countermeasures to physical safety and social effect in the metaverse is presented in Table~\ref{Summary6}.

\section{Governance-related Threats and Countermeasures in Metaverse}\label{sec:Threat7}
Driven by the above threats, it raises huge governance demands and poses huge regulation challenges to metaverse lawmakers and regulators.
\subsection{Threats to Metaverse Governance}\label{subsec:Goverthreat}
In analogy to the social norms and regulations in the real world, the interactions among avatars (e.g., content creation, social activities, and virtual economy) in the metaverse should align with the digital norms and regulations to ensure compliance \cite{9351745}. In the supervision and governance process of metaverse, the following threats may deteriorate system efficiency and security.

\emph{1) New Laws \& Regulations for Virtual Crimes}. Essentially, it is difficult to decide whether a virtual crime is the same as a real one. Thereby, it is hard to directly apply the laws and regulations in real life to enforce penalization for criminal actions \cite{4428344} such as abusive language, virtual harassment, virtual stalking/spying, and so on. For example, if an avatar is verbally abusive in the metaverse, it can be easily regarded as verbal abuse either in virtual or real worlds. However, if an avatar attempts to virtually stalk or harass another user's avatar in the metaverse, the definitions of these crimes may be adapted from the real ones, as well as the appropriate punishments, which should be reconsidered for metaverse lawmakers and regulators.

\emph{2) Misbehaving Regulators}. Regulators may misbehave and cause system paralysis, and their authorities also need supervision. Dynamic and effective punishment/reward mechanisms should be enforced for misbehaving/honest regulators, respectively. To ensure sustainability, punishment and reward rules should be maintained by the majority of avatars in a decentralized and democratic manner \cite{9462451}. Automatic regulations implemented by smart contracts without reliance on trusted intermediaries may be a promising solution. However, it also raises new issues such as information disclosure, mishandled exceptions, and susceptibility to short address attacks and reentrancy attacks \cite{8976179}.

\emph{3) Threats to Collaborative Governance}. To avoid the concentration of regulation rights, collaborative governance under hierarchical or flat mode is more suitable for large-scale metaverse maintenance \cite{8885317}. Nevertheless, collusive regulators may undermine the {metaverse} system even under collaborative governance scenarios. For example, they can collude to make a certain regulator partitioned from the network via wormhole attacks.

\emph{4) Threats to Digital Forensics}. Digital forensics in the metaverse means the virtual reconstruction of cybercrimes by identifying, extracting, fusing, and analyzing evidences obtained from both real and virtual worlds \cite{9555823}. Nevertheless, due to the high dynamics and interoperability issues of various virtual worlds, it is challenging for efficient forensics investigation including entity-behavior association, identification, and tracing among anonymous users/avatars with diverse behavior patterns in the metaverse.
In addition, due to the blurred boundary between real and virtual worlds, the metaverse can make us confused to distinguish between the true and false. For example, bad actors may produce fake news, faces, audios, and videos via AI algorithms to mislead the public, just like the recent Deepfake event.

\subsection{Digital Governance in Metaverse}\label{subsec:defense7}
{Apart from the laws or regulations (i.e., ``hard law"), the ``soft law" is also significant to adjust social relations and regulate user's behaviors in public metaverse governance. The soft law refers to legal norms including autonomy and self-discipline norms and advocacy rules created by various organizations.}
Almeida \emph{et al}. \cite{9351745} highlight three principles in the digital governance of content moderation ecosystems: (i) open, transparent, and consensus-driven, (ii) respect human rights, and (iii) publicly accountable. Here, we review existing potential solutions to metaverse governance from the following three fields.
\subsubsection{AI Governance}
With the pervasive fusion of perception, computing, and actuation, AI will play a leading role to allow digital self-governance of individuals and society in the metaverse in a fully automatic manner.
AI approaches can be employed for detecting misbehaving entities and abnormal or Sybil accounts in the metaverse.
He \emph{et al}. \cite{9384217} exploit a multi-factor attention-enhanced LSTM model to dynamically reveal suspicious signals of malicious accounts in online dating applications by mining the user-generated textual information and the interplay of accounts' temporal-spatial activities. Experiments performed on the real-world dataset demonstrate its effectiveness in detection accuracy.

However, {as the work \cite{9384217} mainly focuses on AI-based malicious account detection, the association of massive avatar-activity-cluster needs further investigations.
Besides,} the outcomes of AI governance algorithms can be biased and unfair (e.g., race bias), thereby arising ethical concerns.
Gasser \emph{et al}. \cite{8114684} propose a three-layer AI governance model from the sociological perspective, where the bottom technical layer allows the data governance and algorithm accountability; the middle ethical layer guides decision-making and data processing via ethical criteria and norms; and the top social and legal layer addresses the allocation of responsibilities in regulation.
Zambonelli \emph{et al}. \cite{8371566} investigate the potential risks including interpretability, trust, autocracy, and ethic issues in delegating the governance of human activities and society to the algorithmic engines in the metaverse.
{Nevertheless, the concrete governance protocols and algorithms with ethic-compliance (e.g., how to define a malicious behavior/avatar) require more research efforts.}
To summarize, both technological and sociological insights are required to build an AI-governed future metaverse.

\subsubsection{Decentralized Governance}
For governance in the large-scale metaverse maintenance, {centralized regulatory can face multiple technical and standard obstacles and difficulty in the compatibility of transnational regulations.}
Collaborative governance can avoid concentration of regulation rights and promote democracy for avatars.
Blockchain technologies offer potential decentralized solutions for collaborative governance in the metaverse, where smart contracts offer a straightforward approach for decentralized governance in an automatic manner.
Febrero \emph{et al}. \cite{9462451} present a blockchain-based decentralized framework in digital city governance to encourage users' active engagement and witness in all administrative processes. In their approach, a verifier group is dynamically selected from digital citizens for transaction verification in the hybrid blockchain. A private-prior peer prediction mechanism is devised for collusion prevention among verifiers, and a Stackelberg game theoretical approach is designed to motivate citizens' participation.

Based on SDN, Bai \emph{et al}. \cite{8885317} design a decentralized data lifecycle governance architecture, where UGC owners can implement customized governance rules for data usage to VSPs, aiming to promote an open environment to satisfy users' diverse requirements.
To further defend against opportunistic attackers in market manipulation, Li \emph{et al}. \cite{7484666} study a Dirichlet-based probabilistic detection model to detect compromised local agents in decentralized power grid control systems by evaluating their reputation levels using historical operating observations.
The implementation of AI governance under decentralized architectures is a future trend for metaverse governance.
{Besides, tailored blockchain solutions to metaverse governance are required including metaverse-specific consensus protocols, new on/off-chain data storage mechanisms, law-compliant regulated blockchain, etc.}

\subsubsection{Trusted Digital Forensics}
Digital forensics is an enabler for accountability in the metaverse under disputes, which has been widely investigated in images and videos.
For example, Swaminathan \emph{et al}. \cite{4451097} develop a general forensic mechanism for digital camera images, according to the observation that in-camera and post-camera image processing leaves a series of distinct fingerprint traces on the digital camera image. The estimated post-camera fingerprints can be employed to validate image authenticity (i.e., whether a specific digital image is from a specific scanner, camera, or computer graphics program).
However, the use of anti-forensics makes trusted digital forensics challenging. To address this issue, Stamm \emph{et al}. \cite{6222325} propose an automatic video frame addition or deletion forensics method with anti-forensics detection, according to the observation that a modified video's motion vectors (i.e., fingerprint) can be imposed in the anti-forensics process.

An obstacle of digital forensics in the metaverse lies in trustworthiness {and labor cost especially for cross-platform operations}. Blockchain can offer a decentralized solution to establish trust {and enhance automation in multi-party cross-platform} digital forensics. For example, Li \emph{et al}. \cite{9555823} utilize blockchain to design a decentralized forensics method, where customized cryptography enables fine-grained forensics data access control and smart contracts enforce auditable forensics execution. {In the metaverse, smart contracts can enforce automated forensics procedure among multiple entities and platforms with improved convenience and mitigated cost, which still require more research efforts.}

Digital forensics can also be utilized for accountability of privacy violations. Zou \emph{et al}. \cite{8361414} propose a privacy leakage forensics scheme with taint analysis and RAM mirroring to obtain digital evidences without touching user's privacy data in a simulated virtual environment. More research efforts are required in terms of resilience, collaboration, QoS enhancement, and privacy preservation in the implementation of digital forensics for metaverse applications.

\begin{table*}[htb]
\centering \setlength{\abovecaptionskip}{0cm}
    \caption{Summary of Existing/Potential Security Countermeasures To Metaverse Governance}\label{Summary7}
\begin{tabular}{cclc}\hline
\textbf{Ref.} & \textbf{\begin{tabular}[c]{@{}c@{}}Security\\ Threat\end{tabular}} & \multicolumn{1}{c}{\textbf{\begin{tabular}[c]{@{}l@{}}$\star$ Purpose\\$\bullet$ Advantages\\$\circ$ Limitations\end{tabular}}} & \textbf{\begin{tabular}[c]{@{}c@{}}Utilized\\ Technology\end{tabular}} \\\hline

\cite{9384217}   &{\begin{tabular}[l]{@{}c@{}}Abnormal social \\accounts\end{tabular}} &{\begin{tabular}[l]{@{}l@{}}$\star$Dynamically reveal suspicious signals of malicious accounts in online dating\\$\bullet$High F1-score and AUC on a real-world dataset gathered from Momo\\$\circ$Challenging to be applied to dating services atop the blockchain\end{tabular}}                                                       & {\begin{tabular}[l]{@{}c@{}}Attention-based \\LSTM \end{tabular}}  \\\hline

\cite{9462451}  &{\begin{tabular}[l]{@{}c@{}}Centralized governance\\risks\end{tabular}}   &{\begin{tabular}[l]{@{}l@{}}$\star$Decentralized digital city governance with incentives for user engagement/witness\\$\bullet$High user utility and time efficiency in decentralized governance\\$\circ$Scalability and security issues in practical system deployment\end{tabular}}                                                       & {\begin{tabular}[l]{@{}c@{}}Blockchain,\\Stackelberg game \end{tabular}}  \\\hline

\cite{7484666}  &{\begin{tabular}[l]{@{}c@{}}Opportunistic attacks \\for price manipulation\end{tabular}}  &{\begin{tabular}[l]{@{}l@{}}$\star$Detect compromised local agents in decentralized power systems using reputation\\$\bullet$Fast aggressive attacker detection using the PowerWorld simulator\\$\circ$Lack credibility analysis for historical operations in reputation evaluation\end{tabular}}                                                       & {\begin{tabular}[l]{@{}c@{}}Dirichlet-based \\probabilistic model \end{tabular}}  \\\hline

\cite{4451097}     &Image authenticity    &{\begin{tabular}[l]{@{}l@{}}$\star$General camera image forensic via post-camera fingerprints\\$\bullet$High efficiency in non-intrusive digital image forensics\\$\circ$Absense of anti-forensics defense\end{tabular}}                                                       &  Image fingerprints         \\\hline

\cite{6222325}     &{\begin{tabular}[l]{@{}c@{}}Anti-forensics\\attack \end{tabular}}   &{\begin{tabular}[l]{@{}l@{}}$\star$Automatic video frame addition or deletion forensics with anti-forensics detection\\$\bullet$Able to automatically detect video tampering/forgeries with high accuracy\\$\circ$Lack trusted whole-process video forensics\end{tabular}}  &  {\begin{tabular}[l]{@{}c@{}}Anti-forensic, \\game theory \end{tabular}}  \\\hline

\cite{8361414}     &{\begin{tabular}[l]{@{}c@{}}Privacy violation \end{tabular}}   &{\begin{tabular}[l]{@{}l@{}}$\star$Privacy leakage forensics to ensure accountability of privacy violations\\$\bullet$High detection efficiency of privacy leakage paths on real malware samples\\$\circ$Only consider limited detection attributes and privacy leakage paths\end{tabular}}  &  {\begin{tabular}[l]{@{}c@{}}Cloud forensics \end{tabular}}  \\\hline

\end{tabular} 
\end{table*}

\vspace{-2.5mm}
\subsection{{Summary and Lessons Learned}}\label{subsec:summary7}
For digital governance in the metaverse, we have learned that AI-enabled governance and decentralized governance are two trends for future metaverse regulation. Moreover, trusted digital forensics offers a promising tool to regulate the metaverse. {Besides, it is important to leverage AI and blockchain technologies to promote the self-governance capabilities of metaverse communities, where each community forms an autonomous code of conduct and users can report the violation behavior according to the terms.} More research efforts are required from both technological and sociological perspectives. Due to the intrinsic characteristics (e.g., interoperability, decentralization, scalability, and heterogeneity) of the metaverse, a series of critical challenges may arise in directly applying existing security countermeasures into the metaverse. Advanced security solutions tailored to the metaverse setting are needed.
A comparison of existing/potential security countermeasures to metaverse governance is presented in Table~\ref{Summary7}.

\vspace{-0.03cm}
\section{Future Research Directions}\label{sec:FUTUREWORK}
In this section, we discuss several future research directions in the metaverse from the following aspects.

\vspace{-0.1cm}
\subsection{Endogenous Security Empowered Metaverse}\label{subsec:Endogenous}
Existing commercial metaverse systems mainly depend on the \emph{brought-in security} such as frequent security patch upgrades after the system deployment. Although security upgrades can enhance system security to an extent, the passive defense mechanisms built on security patching strategies inevitably result in the curse of being continuously broken.
With the continuity of ubiquitous cyber-physical attack surfaces in the metaverse, current bring-in security defenses can be fragile and costly in practical use, like the sword of Damocles hanging overhead.
Endogenous security theory offers a promising solution for provisioning \emph{built-in security} or called \emph{secure by design} mechanisms with self-protection, self-evolution, and autoimmunity capabilities \cite{9141216}, which takes security and privacy factors into account before the system design. Thereby, the future metaverse can resist the ever-increasing known/unknown security vulnerabilities and privacy threats.
An example of endogenous security is the quantum key distribution (QKD) \cite{9684555}, which utilizes channel-based secret keys to resolve information disclosure in wireless transmissions via quantum entanglement properties. Besides, quantum-resistant cryptography (QRC) for quantum secure metaverse applications is another promising research direction.

\begin{figure}[!t]\setlength{\abovecaptionskip}{-0.0cm}
\centering
  \includegraphics[width=9cm,height=6.4cm]{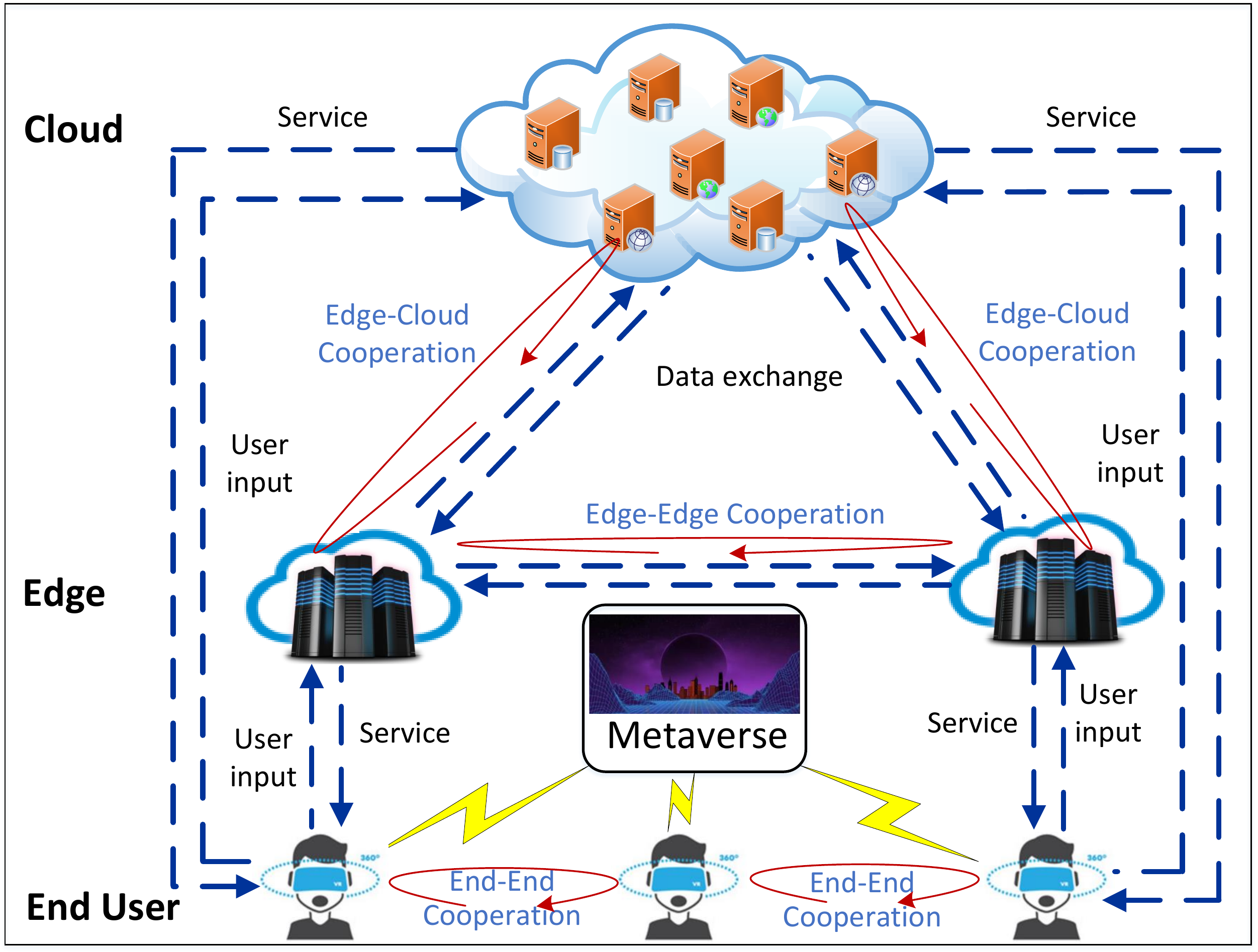}
  \caption{{Illustration of cloud-edge-end computing in metaverse service offering.}}\label{fig:EdgeCloud}\vspace{-2mm}
\end{figure}

\vspace{-0.1cm}
\subsection{{Cloud-Edge-End Orchestrated Secure Metaverse}}\label{subsec:CloudEdgeEnd}
Unlike the conventional 2D Internet, the metaverse gathers massive multi-sensory multimodal information from the real world to provide users with fully immersive 3D experiences. In the metaverse, different users/services have distinct QoE/QoS requirements, which incurs huge difficulty for the metaverse network to simultaneously offer these holographic services for massive users/avatars. For instance, VR usually requires downlink transmission and caching capabilities, AR mainly focuses on uplink transmission and computing capabilities, while the tactile Internet generally requires ultra-reliable low-latency communications \cite{Xu2022EdgeMeta}.
The orchestration of cloud-edge-end computing offers a potential solution by collaboratively and dynamically sharing computation, communication, and storage resources among various entities \cite{9171865}, thereby enhancing the QoE for users/avatars and QoS for metaverse services, as shown in Fig.~\ref{fig:EdgeCloud}. Besides, cloud-edge-end computing can assist edge intelligence and user privacy protection by aggregating and processing users' private data at edge devices (e.g., home gateways) via federated edge learning \cite{9478223}.
In addition, by analyzing the metaverse system as a whole, the cooperation among various sub-metaverses is essential to facilitate seamless security provision and privacy protection and requires further investigation. An attractive case is to dynamically allocate spatiotemporal security resources (e.g., firewall, IDS, and IPS) among heterogeneous sub-metaverses (consisting of various edge/cloud servers) in an on-demand manner. Future works to be investigated include the design of specific edge-edge, edge-cloud, and edge-end collaboration mechanisms in the metaverse.

\vspace{-0.25cm}
\subsection{Cross-Chain Interoperable and Regulatory Metaverse}\label{subsec:Cross-Chain}
\begin{figure}[!t]\setlength{\abovecaptionskip}{-0.1cm}\vspace{-0.1cm}
\centering
  \includegraphics[width=9.3cm,height=5.05cm]{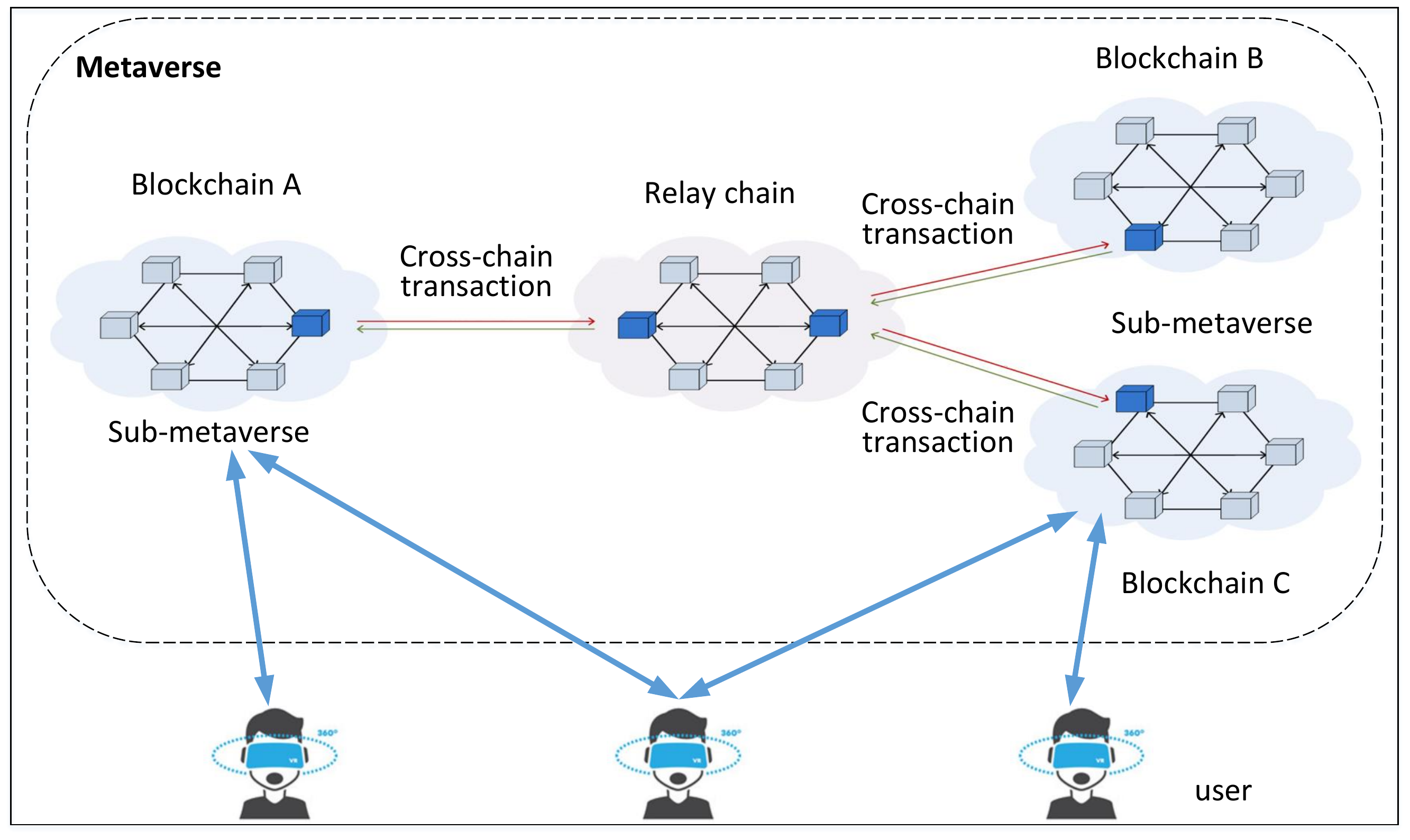}
  \caption{Illustration of cross-chain services among three sub-metaverses which are built on three different blockchains. A relay chain is established to support cross-chain transactions \cite{2020Testimonium}, where the relay chain synchronizes the information of source blockchain \emph{A} to allow destination blockchains \emph{B} and \emph{C} to verify the correctness of transactions on source blockchain \emph{A}.}\label{fig:MetaCrossChain}\vspace{-3mm}
\end{figure}

By getting rid of trusted third parties, blockchain is recognized as the underlying technology to build the future trust-free economy ecosystem in the metaverse. However, distinct sub-metaverses may deploy services on heterogeneous blockchains (e.g., using different transaction formats, block structures, and consensus protocols) to meet QoS requirements, resulting in severe interoperability concerns. As shown in Fig.~\ref{fig:MetaCrossChain}, efficient cross-chain authentication and governance are essential to ensure the security and legitimacy of digital asset-related activities (e.g., asset trading) across different sub-metaverses built on heterogeneous blockchains.
Current cross-chain mechanisms mainly focus on digital asset transfer and rely on the notary scheme, hash-locking, relay chain, and sidechain (details can refer to \cite{9631953}), and few of them consider cross-chain authentication and governance in the metaverse.
The implementation, efficiency, and security of identity authentication across various domains and blockchains in the metaverse need to be further investigated. Moreover, novel decentralized, hierarchical, and penetrating cross-chain governance mechanisms need further research efforts in the metaverse. Besides, efficient metaverse-specific consensus mechanisms, redesigned block structures, as well as well-designed user incentives are required for distinct metaverse applications.
To summarize, open challenges include application-specific governance rule design, programmable and scalable cross-chain governance architecture design, on-chain entity identification and risk assessment, dynamic and collaborative cross-chain governance, etc.

\vspace{-0.3cm}
\subsection{Energy-Efficient and Green Metaverse}\label{subsec:Green}
In the metaverse, on one hand, the wearable XR devices may be resource-constrained and their communication/computation capacities can be highly heterogeneous. On the other hand, the metaverse can be always resource hungry and the computational power demanded in the metaverse will continue to rise, causing increasing environmental concerns (e.g., greenhouse gas emission).
The future metaverse design should be energy-efficient and green to attain sustainability. Users/avatars' cooperation can offer a possible solution for green metaverse in terms of UGC/AIGC dissemination, cooperative networking, and cooperative computation. For example, users' social/locational cooperation can be beneficial to create and distribute high-quality UGC games via the formation of cooperative social groups. Besides, the collaboration among heterogeneous metaverse devices with temporal and spatial correlations can be leveraged to design energy-efficient consensus protocols \cite{9631953} tailored to resource-limited metaverse environments. Apart from user cooperation and new green technology design, other possible solutions include new architecture design, new green edge-cloud computing design, new energy-efficient consensus protocol design, etc., to support green networking and computing in the metaverse.

\vspace{-0.3cm}
\subsection{Content-Centric and Human-Centric Metaverse}\label{subsec:Content-Centric}
In the future metaverse, a surge of UGC is expected to be created, requested, and delivered across various sub-metaverses. Existing IP-based content transmissions can face critical challenges in securing UGC dissemination to massive heterogeneous end devices over the large-scale metaverse across virtual worlds.
Content-centric networking (CCN) stands for a paradigm shift of current Internet architecture. In contrast to current IP-based and host-oriented Internet architecture, contents are addressed and routed directly by their naming information in CCN instead of IP addresses. In CCN-based metaverse, the UGC consumer can request the desired UGC object by sending an interest message to any CCN node that hosts the matched UGC. Besides, CCN embodies a security model which explicitly ensures the security of individual content pieces instead of securing the ``pipe'' or the connection. Therefore, the deployment of CCN can offer a more flexible, scalable, and secure network in the metaverse.
However, CCN also brings new security concerns in the metaverse and one of them is content poisoning, in which adversaries can contaminate the cache space of metaverse nodes by injecting poisoned UGCs and further cause the delay and even failure in retrieving valid UGCs via flooding attacks.
In addition, the design of metaverse should be human-centric. For example, users/avatars' personalized privacy preferences should be ensured in developing privacy-preserving approaches in metaverse environments.

\vspace{-0.2cm}
\section{Conclusion}\label{sec:CONSLUSION}
In this paper, we have presented an in-depth survey of the fundamentals, security, and privacy of metaverse.
Specifically, we have introduced a novel distributed metaverse architecture and discussed its key characteristics, enabling technologies, and modern prototypes.
Afterward, the security and privacy threats, as well as the critical challenges in security defenses and privacy preservation, have been investigated under the distributed metaverse architecture.
Furthermore, we have reviewed the existing/potential solutions in designing tailored security and privacy countermeasures for the metaverse.
We expect that this survey can shed light on the security and privacy provision in metaverse applications, and inspire more pioneering research in this emerging area.

\vspace{-0.2cm}
\section*{Acknowledgment}
This work was supported in part by NSFC (nos. U20A20175, U1808207), and the Fundamental Research Funds for the Central Universities.\vspace{-0.2cm}

\bibliographystyle{IEEETran}
\bibliography{ref_metaverse}

\begin{thebibliography}{100}
\providecommand{\url}[1]{#1}
\csname url@samestyle\endcsname
\providecommand{\newblock}{\relax}
\providecommand{\bibinfo}[2]{#2}
\providecommand{\BIBentrySTDinterwordspacing}{\spaceskip=0pt\relax}
\providecommand{\BIBentryALTinterwordstretchfactor}{4}
\providecommand{\BIBentryALTinterwordspacing}{\spaceskip=\fontdimen2\font plus
\BIBentryALTinterwordstretchfactor\fontdimen3\font minus
  \fontdimen4\font\relax}
\providecommand{\BIBforeignlanguage}[2]{{%
\expandafter\ifx\csname l@#1\endcsname\relax
\typeout{** WARNING: IEEEtran.bst: No hyphenation pattern has been}%
\typeout{** loaded for the language `#1'. Using the pattern for}%
\typeout{** the default language instead.}%
\else
\language=\csname l@#1\endcsname
\fi
#2}}
\providecommand{\BIBdecl}{\relax}
\BIBdecl

\bibitem{sanchez2007second}
J.~Sanchez, ``Second life: An interactive qualitative analysis,'' in
  \emph{Society for Information Technology \& Teacher Education International
  Conference}, Mar. 2007, pp. 1240--1243.

\bibitem{dionisio20133d}
J.~D.~N. Dionisio, W.~G.~B. III, and R.~Gilbert, ``{3D} virtual worlds and the
  metaverse: Current status and future possibilities,'' \emph{ACM Computing
  Surveys (CSUR)}, vol.~45, no.~3, pp. 1--38, Jul. 2013.

\bibitem{2019Lifelogging}
A.~Bruun and M.~L. Stentoft, ``Lifelogging in the wild: Participant experiences
  of using lifelogging as a research tool,'' in \emph{IFIP Conference on
  Human-Computer Interaction}, Aug. 2019, pp. 431--451.

\bibitem{ning2021survey}
H.~Ning, H.~Wang, Y.~Lin, W.~Wang, S.~Dhelim, F.~Farha, J.~Ding, and
  M.~Daneshmand, ``A survey on metaverse: the state-of-the-art, technologies,
  applications, and challenges,'' \emph{arXiv preprint arXiv:2111.09673}, 2021.

\bibitem{MetaverseReport}
\BIBentryALTinterwordspacing
D.~Grider and M.~Maximo. The metaverse: Web3.0 virtual cloud economies.
  Accessed: Nov. 1, 2021. [Online]. Available:
  \url{https://grayscale.com/wp-content/uploads/2021/11/Grayscale_Metaverse_Report_Nov2021.pdf}
\BIBentrySTDinterwordspacing

\bibitem{lee2021all}
L.-H. Lee, T.~Braud, P.~Zhou, L.~Wang, D.~Xu, Z.~Lin, A.~Kumar, C.~Bermejo, and
  P.~Hui, ``All one needs to know about metaverse: A complete survey on
  technological singularity, virtual ecosystem, and research agenda,''
  \emph{arXiv preprint arXiv:2110.05352}, 2021.

\bibitem{yang2022fusing}
Q.~Yang, Y.~Zhao, H.~Huang, and Z.~Zheng, ``Fusing blockchain and {AI} with
  metaverse: A survey,'' \emph{arXiv preprint arXiv:2201.03201}, 2022.

\bibitem{Metaverse2021Duan}
H.~Duan, J.~Li, S.~Fan, Z.~Lin, X.~Wu, and W.~Cai, ``Metaverse for social good:
  A university campus prototype,'' in \emph{ACM International Conference on
  Multimedia (MM)}, Oct. 2021, pp. 153--161.

\bibitem{Wei2022Realizing}
W.~Y.~B. Lim, Z.~Xiong, D.~Niyato, X.~Cao, C.~Miao, S.~Sun, and Q.~Yang,
  ``Realizing the metaverse with edge intelligence: A match made in heaven,''
  \emph{arXiv preprint arXiv:2203.05471}, 2022.

\bibitem{facebookrename}
\BIBentryALTinterwordspacing
Facebook {Inc.} rebrands as {Meta} to stress 'metaverse' plan. Accessed:
  October 28, 2021. [Online]. Available:
  \url{https://machinaresearch.com/news/press-release-global-internet-of-things-market-to-grow-to-27-billion-devices-generating-usd3-trillion-revenue-in-2025/}
\BIBentrySTDinterwordspacing

\bibitem{2008Privacy}
R.~Leenes, ``Privacy in the metaverse: Regulating a complex social construct in
  a virtual world,'' \emph{The Future of Identity in the Information Society},
  pp. 95--112, Jul. 2008.

\bibitem{Falchuk8371577}
B.~Falchuk, S.~Loeb, and R.~Neff, ``The social metaverse: Battle for privacy,''
  \emph{IEEE Technology and Society Magazine}, vol.~37, no.~2, pp. 52--61, Jun.
  2018.

\bibitem{9134928}
J.~Shang, S.~Chen, J.~Wu, and S.~Yin, ``{ARSpy}: Breaking location-based
  multi-player augmented reality application for user location tracking,''
  \emph{IEEE Transactions on Mobile Computing}, vol.~21, no.~2, pp. 433--447,
  Feb. 2022.

\bibitem{7218444}
P.~Hu, H.~Li, H.~Fu, D.~Cansever, and P.~Mohapatra, ``Dynamic defense strategy
  against advanced persistent threat with insiders,'' in \emph{IEEE Conference
  on Computer Communications (INFOCOM)}, 2015, pp. 747--755.

\bibitem{nevelsteen2018virtual}
K.~J. Nevelsteen, ``Virtual world, defined from a technological perspective and
  applied to video games, mixed reality, and the metaverse,'' \emph{Computer
  Animation and Virtual Worlds}, vol.~29, no.~1, pp. 1--22, 2018.

\bibitem{nguyen2021metachain}
C.~T. Nguyen, D.~T. Hoang, D.~N. Nguyen, and E.~Dutkiewicz, ``Metachain: A
  novel blockchain-based framework for metaverse applications,'' \emph{arXiv
  preprint arXiv:2201.00759}, 2021.

\bibitem{Thien2022AI}
T.~Huynh-The, Q.-V. Pham, X.-Q. Pham, T.~T. Nguyen, Z.~Han, and D.-S. Kim,
  ``Artificial intelligence for the metaverse: A survey,'' \emph{arXiv preprint
  arXiv:2202.10336}, 2022.

\bibitem{9667507}
S.-M. Park and Y.-G. Kim, ``A metaverse: Taxonomy, components, applications,
  and open challenges,'' \emph{IEEE Access}, vol.~10, pp. 4209--4251, Jan.
  2022.

\bibitem{Xu2022EdgeMeta}
M.~Xu, W.~C. Ng, W.~Y.~B. Lim, J.~Kang, Z.~Xiong, D.~Niyato, Q.~Yang, X.~Shen,
  and C.~Miao, ``A full dive into realizing the edge-enabled metaverse:
  Visions, enabling technologies,and challenges,'' \emph{arXiv preprint
  arXiv:2203.05471}, 2022.

\bibitem{Bourlakis2009Retail}
M.~Bourlakis, S.~Papagiannidis, and F.~Li, ``Retail spatial evolution: Paving
  the way from traditional to metaverse retailing,'' \emph{Electronic Commerce
  Research}, vol.~9, no. 1–2, pp. 135--148, Jun. 2009.

\bibitem{JEM2020Virtual}
J.~D{\'i}az, C.~Andr{\'e}s, D.~Saldaa, C.~Alberto, and R.~Avila, ``Virtual
  world as a resource for hybrid education,'' \emph{International Journal of
  Emerging Technologies in Learning (iJET)}, vol.~15, no.~15, pp. 94--109,
  2020.

\bibitem{Lee2021When}
L.~Lee, Z.~Lin, R.~Hu, Z.~Gong, A.~Kumar, T.~Li, S.~Li, and P.~Hui, ``When
  creators meet the metaverse: {A} survey on computational arts,'' \emph{CoRR},
  vol. abs/2111.13486, 2021.

\bibitem{MPEG-V}
\BIBentryALTinterwordspacing
{ISO/IEC 23005 (MPEG-V)} standards. Accessed: Sep. 20, 2021. [Online].
  Available: \url{https://mpeg.chiariglione.org/standards/mpeg-v}
\BIBentrySTDinterwordspacing

\bibitem{IEEE-2888}
\BIBentryALTinterwordspacing
{IEEE 2888} standards. Accessed: Dec. 20, 2021. [Online]. Available:
  \url{https://sagroups.ieee.org/2888/}
\BIBentrySTDinterwordspacing

\bibitem{7863165}
L.~Heller and L.~Goodman, ``What do avatars want now? posthuman embodiment and
  the technological sublime,'' in \emph{International Conference on Virtual
  System Multimedia (VSMM)}, Oct. 2016, pp. 1--4.

\bibitem{9495125}
A.~C.~S. Genay, A.~Lecuyer, and M.~Hachet, ``Being an avatar ``for real'': a
  survey on virtual embodiment in augmented reality,'' \emph{IEEE Transactions
  on Visualization and Computer Graphics}, Fourthquarter 2021, doi:
  10.1109/TVCG.2021.3099290.

\bibitem{9171865}
C.~Kai, H.~Zhou, Y.~Yi, and W.~Huang, ``Collaborative cloud-edge-end task
  offloading in mobile-edge computing networks with limited communication
  capability,'' \emph{IEEE Transactions on Cognitive Communications and
  Networking}, vol.~7, no.~2, pp. 624--634, Aug. 2021.

\bibitem{4623222}
S.~Kumar, J.~Chhugani, C.~Kim, D.~Kim, A.~Nguyen, P.~Dubey, C.~Bienia, and
  Y.~Kim, ``Second life and the new generation of virtual worlds,''
  \emph{Computer}, vol.~41, no.~9, pp. 46--53, Sept. 2008.

\bibitem{8364607}
U.~Jayasinghe, G.~M. Lee, T.-W. Um, and Q.~Shi, ``Machine learning based trust
  computational model for {IoT} services,'' \emph{IEEE Transactions on
  Sustainable Computing}, vol.~4, no.~1, pp. 39--52, Jan.-Mar. 2019.

\bibitem{5606335}
J.~Han, J.~Yun, J.~Jang, and K.-r. Park, ``User-friendly home automation based
  on {3D} virtual world,'' \emph{IEEE Transactions on Consumer Electronics},
  vol.~56, no.~3, pp. 1843--1847, Aug. 2010.

\bibitem{sugimoto2021extended}
M.~Sugimoto, ``Extended reality {(XR: VR/AR/MR)}, {3D} printing, holography,
  {AI}, radiomics, and online {VR} {Tele}-medicine for precision surgery,'' in
  \emph{Surgery and Operating Room Innovation}.\hskip 1em plus 0.5em minus
  0.4em\relax Springer, Nov. 2021, pp. 65--70.

\bibitem{10.1145/769953.769967}
C.~Jaynes, W.~B. Seales, K.~Calvert, Z.~Fei, and J.~Griffioen, ``The metaverse:
  A networked collection of inexpensive, self-configuring, immersive
  environments,'' in \emph{Proceedings of the Workshop on Virtual
  Environments}, ser. EGVE '03, 2003, pp. 115--124.

\bibitem{9429703}
Y.~Wu, K.~Zhang, and Y.~Zhang, ``Digital twin networks: A survey,'' \emph{IEEE
  Internet of Things Journal}, vol.~8, no.~18, pp. 13\,789--13\,804, Sept.
  2021.

\bibitem{du2021optimal}
H.~Du, D.~Niyato, J.~Kang, D.~I. Kim, and C.~Miao, ``Optimal targeted
  advertising strategy for secure wireless edge metaverse,'' \emph{arXiv
  preprint arXiv:2111.00511}, 2021.

\bibitem{9631953}
Y.~Wang, Z.~Su, J.~Ni, N.~Zhang, and X.~Shen, ``Blockchain-empowered
  space-air-ground integrated networks: Opportunities, challenges, and
  solutions,'' \emph{IEEE Communications Surveys \& Tutorials}, vol.~24, no.~1,
  pp. 160--209, Firstquarter 2022.

\bibitem{9258002}
E.~H.-K. Wu, C.-S. Chen, T.-K. Yeh, and S.-C. Yeh, ``Interactive medical {VR}
  streaming service based on software-defined network: Design and
  implementation,'' in \emph{IEEE International Conference on Consumer
  Electronics - Taiwan (ICCE-Taiwan)}, Sept. 2020, pp. 1--2.

\bibitem{6157577}
S.~Vural, D.~Wei, and K.~Moessner, ``Survey of experimental evaluation studies
  for wireless mesh network deployments in urban areas towards ubiquitous
  {Internet},'' \emph{IEEE Communications Surveys \& Tutorials}, vol.~15,
  no.~1, pp. 223--239, First Quarter 2013.

\bibitem{wang2021NFT}
Q.~Wang, R.~Li, Q.~Wang, and S.~Chen, ``Non-fungible token ({NFT}): Overview,
  evaluation, opportunities and challenges,'' \emph{arXiv preprint
  arXiv:2105.07447}, 2021.

\bibitem{han2021analysis}
J.~Han, J.~Heo, and E.~You, ``Analysis of metaverse platform as a new play
  culture: Focusing on {Roblox} and {ZEPETO},'' in \emph{International
  Conference on Human-centered Artificial Intelligence}, Oct. 2021, pp. 27--36.

\bibitem{10.1145/3161570}
V.~Kasapakis and D.~Gavalas, ``User-generated content in pervasive games,''
  \emph{ACM Computers in Entertainment}, vol.~16, no.~1, pp. 1--23, Spring
  2018.

\bibitem{MetaHuman}
\BIBentryALTinterwordspacing
Meet the {MetaHuman}. Accessed: Jan. 20, 2022. [Online]. Available:
  \url{https://www.unrealengine.com/en-US/digital-humans}
\BIBentrySTDinterwordspacing

\bibitem{9163078}
A.~Alwarafy, K.~A. Al-Thelaya, M.~Abdallah, J.~Schneider, and M.~Hamdi, ``A
  survey on security and privacy issues in edge-computing-assisted internet of
  things,'' \emph{IEEE Internet of Things Journal}, vol.~8, no.~6, pp.
  4004--4022, Mar. 2021.

\bibitem{9146364}
K.~Tange, M.~De~Donno, X.~Fafoutis, and N.~Dragoni, ``A systematic survey of
  industrial internet of things security: Requirements and fog computing
  opportunities,'' \emph{IEEE Communications Surveys \& Tutorials}, vol.~22,
  no.~4, pp. 2489--2520, Fourthquarter 2020.

\bibitem{opensea2022nft}
\BIBentryALTinterwordspacing
{NFT} investors lose {\$1.7m} in {OpenSea} phishing attack. Accessed: Mar. 1,
  2022. [Online]. Available:
  \url{https://threatpost.com/nft-investors-lose-1-7m-in-opensea-phishing-attack/178558/}
\BIBentrySTDinterwordspacing

\bibitem{9152758}
D.~Antonioli, N.~Tippenhauer, and K.~Rasmussen, ``{BIAS}: {Bluetooth}
  impersonation attacks,'' in \emph{IEEE Symposium on Security and Privacy
  (SP)}, May 2020, pp. 549--562.

\bibitem{Xu2021Wireless}
M.~Xu, D.~Niyato, J.~Kang, Z.~Xiong, C.~Miao, and D.~I. Kim, ``Wireless
  edge-empowered metaverse: A learning-based incentive mechanism for virtual
  reality,'' \emph{arXiv preprint arXiv:2111.03776}, 2021.

\bibitem{8249924}
J.~Yu, Z.~Kuang, B.~Zhang, W.~Zhang, D.~Lin, and J.~Fan, ``Leveraging content
  sensitiveness and user trustworthiness to recommend fine-grained privacy
  settings for social image sharing,'' \emph{IEEE Transactions on Information
  Forensics and Security}, vol.~13, no.~5, pp. 1317--1332, May 2018.

\bibitem{6336740}
J.~Jensen and M.~G. Jaatun, ``Federated identity management - we built it; why
  won't they come?'' \emph{IEEE Security \& Privacy}, vol.~11, no.~2, pp.
  34--41, Mar.-Apr. 2013.

\bibitem{9536956}
E.~Samir, H.~Wu, M.~Azab, C.~S. Xin, and Q.~Zhang, ``{DT-SSIM}: A decentralized
  trustworthy self-sovereign identity management framework,'' \emph{IEEE
  Internet of Things Journal}, 2021, doi: 10.1109/JIOT.2021.3112537.

\bibitem{de2019key}
M.~De~Ree, G.~Mantas, A.~Radwan, S.~Mumtaz, J.~Rodriguez, and I.~E. Otung,
  ``Key management for beyond {5G} mobile small cells: A survey,'' \emph{IEEE
  Access}, vol.~7, pp. 59\,200--59\,236, May. 2019.

\bibitem{8089389}
Z.~Li, Q.~Pei, I.~Markwood, Y.~Liu, and H.~Zhu, ``Secret key establishment via
  {RSS} trajectory matching between wearable devices,'' \emph{IEEE Transactions
  on Information Forensics and Security}, vol.~13, no.~3, pp. 802--817, Mar.
  2018.

\bibitem{sun2020accelerometer}
F.~Sun, W.~Zang, H.~Huang, I.~Farkhatdinov, and Y.~Li, ``Accelerometer-based
  key generation and distribution method for wearable {IoT} devices,''
  \emph{IEEE Internet of Things Journal}, vol.~8, no.~3, pp. 1636--1650, Feb.
  2020.

\bibitem{chen2018lirek}
Z.~Chen, W.~Ren, Y.~Ren, and K.-K.~R. Choo, ``{LiReK}: A lightweight and
  real-time key establishment scheme for wearable embedded devices by gestures
  or motions,'' \emph{Future Generation Computer Systems}, vol.~84, pp.
  126--138, Jul. 2018.

\bibitem{zheng2018critical}
G.~Zheng, R.~Shankaran, W.~Yang, C.~Valli, L.~Qiao, M.~A. Orgun, and S.~C.
  Mukhopadhyay, ``A critical analysis of {ECG}-based key distribution for
  securing wearable and implantable medical devices,'' \emph{IEEE Sensors
  Journal}, vol.~19, no.~3, pp. 1186--1198, Feb. 2018.

\bibitem{srinivas2018cloud}
J.~Srinivas, A.~K. Das, N.~Kumar, and J.~J. Rodrigues, ``Cloud centric
  authentication for wearable healthcare monitoring system,'' \emph{IEEE
  Transactions on Dependable and Secure Computing}, vol.~17, no.~5, pp.
  942--956, Sept.-Oct. 2018.

\bibitem{zhao2020trueheart}
T.~Zhao, Y.~Wang, J.~Liu, Y.~Chen, J.~Cheng, and J.~Yu, ``Trueheart: Continuous
  authentication on wrist-worn wearables using {PPG}-based biometrics,'' in
  \emph{IEEE Conference on Computer Communications (INFOCOM)}, Jul. 2020, pp.
  30--39.

\bibitem{9290438}
M.~A. Jan, F.~Khan, R.~Khan, S.~Mastorakis, V.~G. Menon, M.~Alazab, and
  P.~Watters, ``Lightweight mutual authentication and privacy-preservation
  scheme for intelligent wearable devices in industrial-{CPS},'' \emph{IEEE
  Transactions on Industrial Informatics}, vol.~17, no.~8, pp. 5829--5839, Aug.
  2021.

\bibitem{8299447}
H.~Aksu, A.~S. Uluagac, and E.~S. Bentley, ``Identification of wearable devices
  with {Bluetooth},'' \emph{IEEE Transactions on Sustainable Computing},
  vol.~6, no.~2, pp. 221--230, Apr.-Jun. 2021.

\bibitem{7321811}
O.~Arias, J.~Wurm, K.~Hoang, and Y.~Jin, ``Privacy and security in internet of
  things and wearable devices,'' \emph{IEEE Transactions on Multi-Scale
  Computing Systems}, vol.~1, no.~2, pp. 99--109, Apr.-Jun. 2015.

\bibitem{9036971}
M.~Shen, H.~Liu, L.~Zhu, K.~Xu, H.~Yu, X.~Du, and M.~Guizani,
  ``Blockchain-assisted secure device authentication for cross-domain
  industrial {IoT},'' \emph{IEEE Journal on Selected Areas in Communications},
  vol.~38, no.~5, pp. 942--954, May 2020.

\bibitem{9465710}
J.~Chen, Z.~Zhan, K.~He, R.~Du, D.~Wang, and F.~Liu, ``{XAuth}: Efficient
  privacy-preserving cross-domain authentication,'' \emph{IEEE Transactions on
  Dependable and Secure Computing}, 2021, doi: 10.1109/TDSC.2021.3092375.

\bibitem{7422115}
K.~Yang, Z.~Liu, X.~Jia, and X.~Shen, ``Time-domain attribute-based access
  control for cloud-based video content sharing: A cryptographic approach,''
  \emph{IEEE Transactions on Multimedia}, vol.~18, no.~5, pp. 940--950, May
  2016.

\bibitem{8432128}
L.~Y. Zhang, Y.~Zheng, J.~Weng, C.~Wang, Z.~Shan, and K.~Ren, ``You can access
  but you cannot leak: Defending against illegal content redistribution in
  encrypted cloud media center,'' \emph{IEEE Transactions on Dependable and
  Secure Computing}, vol.~17, no.~6, pp. 1218--1231, Nov.-Dec. 2020.

\bibitem{9268472}
Y.~Wang, Z.~Su, N.~Zhang, J.~Chen, X.~Sun, Z.~Ye, and Z.~Zhou, ``{SPDS}: A
  secure and auditable private data sharing scheme for smart grid based on
  blockchain,'' \emph{IEEE Transactions on Industrial Informatics}, vol.~17,
  no.~11, pp. 7688--7699, Nov. 2021.

\bibitem{7518650}
A.~Ometov, S.~V. Bezzateev, J.~Kannisto, J.~Harju, S.~Andreev, and
  Y.~Koucheryavy, ``Facilitating the delegation of use for private devices in
  the era of the internet of wearable things,'' \emph{IEEE Internet of Things
  Journal}, vol.~4, no.~4, pp. 843--854, Aug 2017.

\bibitem{8399560}
C.~Ma, Z.~Yan, and C.~W. Chen, ``Scalable access control for privacy-aware
  media sharing,'' \emph{IEEE Transactions on Multimedia}, vol.~21, no.~1, pp.
  173--183, Jan. 2019.

\bibitem{9035635}
Z.~Su, Y.~Wang, Q.~Xu, and N.~Zhang, ``{LVBS}: Lightweight vehicular blockchain
  for secure data sharing in disaster rescue,'' \emph{IEEE Transactions on
  Dependable and Secure Computing}, vol.~19, no.~1, pp. 19--32, Jan.-Feb. 2022.

\bibitem{7752958}
G.~{Liang}, S.~R. {Weller}, J.~{Zhao}, F.~{Luo}, and Z.~Y. {Dong}, ``The 2015
  {Ukraine} blackout: Implications for false data injection attacks,''
  \emph{IEEE Transactions on Power Systems}, vol.~32, no.~4, pp. 3317--3318,
  Jul. 2017.

\bibitem{liang2017provchain}
X.~Liang, S.~Shetty, D.~Tosh, C.~Kamhoua, K.~Kwiat, and L.~Njilla, ``Provchain:
  A blockchain-based data provenance architecture in cloud environment with
  enhanced privacy and availability,'' in \emph{IEEE/ACM International
  Symposium on Cluster, Cloud and Grid Computing (CCGRID)}, May 2017, pp.
  468--477.

\bibitem{4428344}
A.~Hendaoui, M.~Limayem, and C.~W. Thompson, ``{3D} social virtual worlds:
  Research issues and challenges,'' \emph{IEEE Internet Computing}, vol.~12,
  no.~1, pp. 88--92, Jan.-Feb. 2008.

\bibitem{btlj2014Lawsuits}
\BIBentryALTinterwordspacing
The right of publicity: Likeness lawsuits against video game companies.
  Accessed: Feb. 2, 2020. [Online]. Available:
  \url{https://btlj.org/2014/12/the-right-of-publicity-likeness-lawsuits-against-video-game-companies/}
\BIBentrySTDinterwordspacing

\bibitem{9660773}
S.~Liao, J.~Wu, A.~K. Bashir, W.~Yang, J.~Li, and U.~Tariq, ``Digital twin
  consensus for blockchain-enabled intelligent transportation systems in smart
  cities,'' \emph{IEEE Transactions on Intelligent Transportation Systems},
  2021, doi: 10.1109/TITS.2021.3134002.

\bibitem{8417973}
T.~Miyato, S.-I. Maeda, M.~Koyama, and S.~Ishii, ``Virtual adversarial
  training: A regularization method for supervised and semi-supervised
  learning,'' \emph{IEEE Transactions on Pattern Analysis and Machine
  Intelligence}, vol.~41, no.~8, pp. 1979--1993, Aug. 2019.

\bibitem{2019Modality}
S.~Mai, H.~Hu, and S.~Xing, ``Modality to modality translation: An adversarial
  representation learning and graph fusion network for multimodal fusion,'' in
  \emph{AAAI}, 2019, pp. 1--9.

\bibitem{2020Stealthy}
J.~Sun, T.~Zhang, X.~Xie, L.~Ma, and Y.~Liu, ``Stealthy and efficient
  adversarial attacks against deep reinforcement learning,'' in \emph{AAAI},
  2020, pp. 1--9.

\bibitem{2020Efficient}
H.~Zheng, Z.~Zhang, J.~Gu, H.~Lee, and A.~Prakash, ``Efficient adversarial
  training with transferable adversarial examples,'' in \emph{IEEE/CVF
  Conference on Computer Vision and Pattern Recognition (CVPR)}, Jun. 2020, pp.
  1--10.

\bibitem{8822494}
C.~Gehrmann and M.~Gunnarsson, ``A digital twin based industrial automation and
  control system security architecture,'' \emph{IEEE Transactions on Industrial
  Informatics}, vol.~16, no.~1, pp. 669--680, Jan. 2020.

\bibitem{2008Spatialized}
R.~Zimmermann and K.~Liang, ``Spatialized audio streaming for networked virtual
  environments,'' in \emph{ACM International Conference on Multimedia (MM)},
  Oct. 2008, pp. 299--308.

\bibitem{Jean2019Rendering}
J.-M. Jot, R.~Audfray, M.~Hertensteiner, and B.~Schmidt, ``Rendering spatial
  sound for interoperable experiences in the audio metaverse,'' in
  \emph{International Conference on Immersive and 3D Audio (i3DA)}, Sep. 2021,
  pp. 1--15.

\bibitem{Patrick2021Experiencing}
P.~Dickinson, A.~Jones, W.~Christian, A.~Westerside, and A.~Parke,
  ``Experiencing simulated confrontations in virtual reality,'' in \emph{ACM
  CHI Conference on Human Factors in Computing Systems (CHI)}, May 2021, pp.
  1--10.

\bibitem{9036917}
Q.~Xu, Z.~Su, and R.~Lu, ``Game theory and reinforcement learning based secure
  edge caching in mobile social networks,'' \emph{IEEE Transactions on
  Information Forensics and Security}, vol.~15, pp. 3415--3429, Mar. 2020.

\bibitem{9478223}
Z.~Su, Y.~Wang, T.~H. Luan, N.~Zhang, F.~Li, T.~Chen, and H.~Cao, ``Secure and
  efficient federated learning for smart grid with edge-cloud collaboration,''
  \emph{IEEE Transactions on Industrial Informatics}, vol.~18, no.~2, pp.
  1333--1344, Feb. 2022.

\bibitem{Han2022ADynamic}
Y.~Han, D.~Niyato, C.~Leung, D.~I. Kim, K.~Zhu, S.~Feng, X.~Shen, and C.~Miao,
  ``A dynamic hierarchical framework for iot-assisted metaverse
  synchronization,'' \emph{arXiv preprint arXiv:2203.03969}, 2022.

\bibitem{Kimberly2019Secure}
K.~Ruth, T.~Kohno, and F.~Roesner, ``Secure multi-user content sharing for
  augmented reality applications,'' in \emph{28th USENIX Security Symposium
  (USENIX Security 19)}, Aug. 2019, pp. 141--158.

\bibitem{Hyunjoo2021AdCube}
H.~Lee, J.~Lee, D.~Kim, S.~Jana, I.~Shin, and S.~Son, ``{AdCube}: {WebVR} ad
  fraud and practical confinement of {Third-Party} ads,'' in \emph{30th USENIX
  Security Symposium (USENIX Security 21)}, Aug. 2021, pp. 2543--2560.

\bibitem{8070948}
Z.~Ning, X.~Hu, Z.~Chen, M.~Zhou, B.~Hu, J.~Cheng, and M.~S. Obaidat, ``A
  cooperative quality-aware service access system for social internet of
  vehicles,'' \emph{IEEE Internet of Things Journal}, vol.~5, no.~4, pp.
  2506--2517, Aug. 2018.

\bibitem{7738534}
B.~Satchidanandan and P.~R. Kumar, ``Dynamic watermarking: Active defense of
  networked cyber–physical systems,'' \emph{Proceedings of the IEEE}, vol.
  105, no.~2, pp. 219--240, Feb. 2017.

\bibitem{kamal2018light}
M.~Kamal and s.~Tariq, ``Light-weight security and data provenance for
  multi-hop internet of things,'' \emph{IEEE Access}, vol.~6, pp.
  34\,439--34\,448, 2018.

\bibitem{9205204}
J.~Wei, J.~Li, Y.~Lin, and J.~Zhang, ``{LDP}-based social content protection
  for trending topic recommendation,'' \emph{IEEE Internet of Things Journal},
  vol.~8, no.~6, pp. 4353--4372, Mar. 2021.

\bibitem{4492778}
S.~Wasserkrug, A.~Gal, and O.~Etzion, ``Inference of security hazards from
  event composition based on incomplete or uncertain information,'' \emph{IEEE
  Transactions on Knowledge and Data Engineering}, vol.~20, no.~8, pp.
  1111--1114, Aug. 2008.

\bibitem{9531392}
X.~Li, J.~He, P.~Vijayakumar, X.~Zhang, and V.~Chang, ``A verifiable
  privacy-preserving machine learning prediction scheme for edge-enhanced
  {HCPSs},'' \emph{IEEE Transactions on Industrial Informatics}, Aug. 2021,
  doi: 10.1109/TII.2021.3110808.

\bibitem{GDPR}
\BIBentryALTinterwordspacing
General data protection regulation {(GDPR)}. Accessed: Mar. 2, 2022. [Online].
  Available: \url{https://gdpr-info.eu/}
\BIBentrySTDinterwordspacing

\bibitem{7842850}
E.~{Bertino} and N.~{Islam}, ``Botnets and internet of things security,''
  \emph{Computer}, vol.~50, no.~2, pp. 76--79, Feb. 2017.

\bibitem{SL2006breached}
\BIBentryALTinterwordspacing
Metaverse breached: {Second Life} customer database hacked. Accessed: Jan. 15,
  2021. [Online]. Available:
  \url{https://techcrunch.com/2006/09/08/metaverse-breached-second-life-customer-database-hacked/}
\BIBentrySTDinterwordspacing

\bibitem{8933081}
H.~Song, T.~Luo, X.~Wang, and J.~Li, ``Multiple sensitive values-oriented
  personalized privacy preservation based on randomized response,'' \emph{IEEE
  Transactions on Information Forensics and Security}, vol.~15, pp. 2209--2224,
  Dec. 2020.

\bibitem{7486070}
Z.~Wu, G.~Li, Q.~Liu, G.~Xu, and E.~Chen, ``Covering the sensitive subjects to
  protect personal privacy in personalized recommendation,'' \emph{IEEE
  Transactions on Services Computing}, vol.~11, no.~3, pp. 493--506, May-Jun.
  2018.

\bibitem{5054904}
S.~Bono, D.~Caselden, G.~Landau, and C.~Miller, ``Reducing the attack surface
  in massively multiplayer online role-playing games,'' \emph{IEEE Security \&
  Privacy}, vol.~7, no.~3, pp. 13--19, May-Jun. 2009.

\bibitem{8418615}
K.~Lebeck, K.~Ruth, T.~Kohno, and F.~Roesner, ``Towards security and privacy
  for multi-user augmented reality: Foundations with end users,'' in \emph{IEEE
  Symposium on Security and Privacy (SP)}, 2018, pp. 392--408.

\bibitem{7535117}
J.~Laakkonen, J.~Parkkila, P.~Jäppinen, J.~Ikonen, and A.~Seffah,
  ``Incorporating privacy into digital game platform design: The what, why, and
  how,'' \emph{IEEE Security \& Privacy}, vol.~14, no.~4, pp. 22--32, July-Aug.
  2016.

\bibitem{8516484}
P.~M. Corcoran and C.~Costache, ``A privacy framework for games \& interactive
  media,'' in \emph{IEEE Games, Entertainment, Media Conference (GEM)}, Aug.
  2018, pp. 1--9.

\bibitem{9488776}
D.~Y. Zhang, Z.~Kou, and D.~Wang, ``{FedSens}: A federated learning approach
  for smart health sensing with class imbalance in resource constrained edge
  computing,'' in \emph{IEEE Conference on Computer Communications (INFOCOM)},
  May 2021, pp. 1--10.

\bibitem{9200775}
Z.~Guan, Z.~Wan, Y.~Yang, Y.~Zhou, and B.~Huang, ``{BlockMaze}: An efficient
  privacy-preserving account-model blockchain based on {zk-SNARKs},''
  \emph{IEEE Transactions on Dependable and Secure Computing}, May-Jun. 2020,
  doi: 10.1109/TDSC.2020.3025129.

\bibitem{7120947}
K.~Xu, Y.~Guo, L.~Guo, Y.~Fang, and X.~Li, ``My privacy my decision: Control of
  photo sharing on online social networks,'' \emph{IEEE Transactions on
  Dependable and Secure Computing}, vol.~14, no.~2, pp. 199--210, Mar.-Apr.
  2017.

\bibitem{6509878}
R.~Raguram, A.~M. White, Y.~Xu, J.-M. Frahm, P.~Georgel, and F.~Monrose, ``On
  the privacy risks of virtual keyboards: Automatic reconstruction of typed
  input from compromising reflections,'' \emph{IEEE Transactions on Dependable
  and Secure Computing}, vol.~10, no.~3, pp. 154--167, May-Jun. 2013.

\bibitem{9492755}
J.~Mills, J.~Hu, and G.~Min, ``Multi-task federated learning for personalised
  deep neural networks in edge computing,'' \emph{IEEE Transactions on Parallel
  and Distributed Systems}, vol.~33, no.~3, pp. 630--641, Mar. 2022.

\bibitem{groping2021MIT}
\BIBentryALTinterwordspacing
The metaverse has a groping problem already ({MIT} technology review).
  Accessed: Dec. 17, 2021. [Online]. Available:
  \url{https://www.technologyreview.com/2021/12/16/1042516/the-metaverse-has-a-groping-problem/}
\BIBentrySTDinterwordspacing

\bibitem{distance2022Meta}
\BIBentryALTinterwordspacing
Meta establishes 4-foot ``personal boundary'' to deter {VR} groping. Accessed:
  Feb. 9, 2022. [Online]. Available:
  \url{https://arstechnica.com/gaming/2022/02/meta-establishes-four-foot-personal-boundary-to-deter-vr-groping/}
\BIBentrySTDinterwordspacing

\bibitem{privacysandbox2022}
\BIBentryALTinterwordspacing
{The Privacy Sandbox}. Accessed: Mar. 20, 2022. [Online]. Available:
  \url{https://privacysandbox.com/}
\BIBentrySTDinterwordspacing

\bibitem{DPpaperApple}
\BIBentryALTinterwordspacing
Learning with privacy at scale. Accessed: Mar. 9, 2022. [Online]. Available:
  \url{https://docs-assets.developer.apple.com/ml-research/papers/learning-with-privacy-at-scale.pdf}
\BIBentrySTDinterwordspacing

\bibitem{NFT2022hashkey}
\BIBentryALTinterwordspacing
Key infrastructure of the metaverse: status, opportunities, and challenges of
  {NFT} data storage. Accessed: Feb. 2, 2022. [Online]. Available:
  \url{https://www.hashkey.com/key-infrastructure-of-the-metaverse-status-opportunities-and-challenges-of-nft-data-storage/}
\BIBentrySTDinterwordspacing

\bibitem{9676467}
J.~Woodward and J.~Ruiz, ``Analytic review of using augmented reality for
  situational awareness,'' \emph{IEEE Transactions on Visualization and
  Computer Graphics}, 2022, doi: 10.1109/TVCG.2022.3141585.

\bibitem{9261134}
U.~Ju, L.~L. Chuang, and C.~Wallraven, ``Acoustic cues increase situational
  awareness in accident situations: A {VR} car-driving study,'' \emph{IEEE
  Transactions on Intelligent Transportation Systems}, pp. 1--11, Apr. 2020.

\bibitem{9130693}
Z.~Lv, D.~Chen, R.~Lou, and H.~Song, ``Industrial security solution for virtual
  reality,'' \emph{IEEE Internet of Things Journal}, vol.~8, no.~8, pp.
  6273--6281, Apr. 2021.

\bibitem{9199886}
L.~Vu, V.~L. Cao, Q.~U. Nguyen, D.~N. Nguyen, D.~T. Hoang, and E.~Dutkiewicz,
  ``Learning latent representation for {IoT} anomaly detection,'' \emph{IEEE
  Transactions on Cybernetics}, pp. 1--14, May. 2020.

\bibitem{6507636}
M.~Zhang, A.~Raghunathan, and N.~K. Jha, ``{MedMon}: Securing medical devices
  through wireless monitoring and anomaly detection,'' \emph{IEEE Transactions
  on Biomedical Circuits and Systems}, vol.~7, no.~6, pp. 871--881, Dec. 2013.

\bibitem{9277640}
R.~Heartfield, G.~Loukas, A.~Bezemskij, and E.~Panaousis, ``Self-configurable
  cyber-physical intrusion detection for smart homes using reinforcement
  learning,'' \emph{IEEE Transactions on Information Forensics and Security},
  vol.~16, pp. 1720--1735, Dec. 2021.

\bibitem{9311786}
X.~Zhou, W.~Liang, S.~Shimizu, J.~Ma, and Q.~Jin, ``Siamese neural network
  based few-shot learning for anomaly detection in industrial cyber-physical
  systems,'' \emph{IEEE Transactions on Industrial Informatics}, vol.~17,
  no.~8, pp. 5790--5798, Aug. 2021.

\bibitem{2017Wild}
B.~Biggio and F.~Roli, ``Wild patterns: Ten years after the rise of adversarial
  machine learning,'' \emph{Pattern Recognition}, vol.~84, pp. 317--331, Dec.
  2018.

\bibitem{9060972}
A.~M. Zarca, J.~B. Bernabe, A.~Skarmeta, and J.~M. Alcaraz~Calero, ``Virtual
  {IoT} honeynets to mitigate cyberattacks in {SDN/NFV}-enabled {IoT}
  networks,'' \emph{IEEE Journal on Selected Areas in Communications}, vol.~38,
  no.~6, pp. 1262--1277, Jun. 2020.

\bibitem{8640020}
A.~Shahsavari, M.~Farajollahi, E.~M. Stewart, E.~Cortez, and H.~Mohsenian-Rad,
  ``Situational awareness in distribution grid using micro-{PMU} data: A
  machine learning approach,'' \emph{IEEE Transactions on Smart Grid}, vol.~10,
  no.~6, pp. 6167--6177, 2019.

\bibitem{9540743}
P.~Krishnan, K.~Jain, R.~Buyya, P.~Vijayakumar, A.~Nayyar, M.~Bilal, and
  H.~Song, ``{MUD}-based behavioral profiling security framework for
  software-defined {IoT} networks,'' \emph{IEEE Internet of Things Journal},
  May 2021, doi: 10.1109/JIOT.2021.3113577.

\bibitem{8915712}
W.~Zhang, B.~Zhang, Y.~Zhou, H.~He, and Z.~Ding, ``An {IoT} honeynet based on
  multiport honeypots for capturing {IoT} attacks,'' \emph{IEEE Internet of
  Things Journal}, vol.~7, no.~5, pp. 3991--3999, May 2020.

\bibitem{7587350}
J.~Wu, K.~Ota, M.~Dong, J.~Li, and H.~Wang, ``Big data analysis-based security
  situational awareness for smart grid,'' \emph{IEEE Transactions on Big Data},
  vol.~4, no.~3, pp. 408--417, Sept. 2018.

\bibitem{Paraluni2022Reen}
\BIBentryALTinterwordspacing
Hackers exploited reentrancy vulnerability to attack {Paraluni}, making more
  than \$1.7 million. Accessed: Mar. 14, 2022. [Online]. Available:
  \url{https://webscrypto.com/hackers-exploited-reentrancy-vulnerability-to-attack-paraluni-making-more-than-1-7-million-about-1-3-of-which-has-flowed-into-tornado/}
\BIBentrySTDinterwordspacing

\bibitem{8360480}
H.~Ritzdorf, C.~Soriente, G.~O. Karame, S.~Marinovic, D.~Gruber, and S.~Capkun,
  ``Toward shared ownership in the cloud,'' \emph{IEEE Transactions on
  Information Forensics and Security}, vol.~13, no.~12, pp. 3019--3034, Dec.
  2018.

\bibitem{NFT2021lost}
\BIBentryALTinterwordspacing
Don't get rugged: {DeFi} scams go from zero to \$129 million in a year to
  become top financial hack. Accessed: Aug. 25, 2021. [Online]. Available:
  \url{https://www.techrepublic.com/article/dont-get-rugged-defi-scams-go-from-zero-to-129-million-in-a-year-to-become-top-financial-hack/}
\BIBentrySTDinterwordspacing

\bibitem{9354053}
M.~Zhang, L.~Yang, S.~He, M.~Li, and J.~Zhang, ``Privacy-preserving data
  aggregation for mobile crowdsensing with externality: An auction approach,''
  \emph{IEEE/ACM Transactions on Networking}, vol.~29, no.~3, pp. 1046--1059,
  Jun. 2021.

\bibitem{6725569}
Z.~Xu and W.~Liang, ``Collusion-resistant repeated double auctions for relay
  assignment in cooperative networks,'' \emph{IEEE Transactions on Wireless
  Communications}, vol.~13, no.~3, pp. 1196--1207, Mar. 2014.

\bibitem{4359459}
M.~Li, J.~Yu, and J.~Wu, ``Free-riding on {BitTorrent}-like peer-to-peer file
  sharing systems: Modeling analysis and improvement,'' \emph{IEEE Transactions
  on Parallel and Distributed Systems}, vol.~19, no.~7, pp. 954--966, Jul.
  2008.

\bibitem{8892660}
M.~H.~u. Rehman, K.~Salah, E.~Damiani, and D.~Svetinovic, ``Trust in blockchain
  cryptocurrency ecosystem,'' \emph{IEEE Transactions on Engineering
  Management}, vol.~67, no.~4, pp. 1196--1212, Nov. 2020.

\bibitem{8572758}
L.~C.~C. De~Biase, P.~C. Calcina-Ccori, G.~Fedrecheski, G.~M. Duarte, P.~S.~S.
  Rangel, and M.~K. Zuffo, ``Swarm economy: A model for transactions in a
  distributed and organic {IoT} platform,'' \emph{IEEE Internet of Things
  Journal}, vol.~6, no.~3, pp. 4561--4572, Jun. 2019.

\bibitem{8502858}
C.~Liu, Y.~Xiao, V.~Javangula, Q.~Hu, S.~Wang, and X.~Cheng, ``{NormaChain}: A
  blockchain-based normalized autonomous transaction settlement system for
  {IoT}-based {E}-commerce,'' \emph{IEEE Internet of Things Journal}, vol.~6,
  no.~3, pp. 4680--4693, Jun. 2019.

\bibitem{9606227}
L.~Jiang, H.~Zheng, H.~Tian, S.~Xie, and Y.~Zhang, ``Cooperative federated
  learning and model update verification in blockchain empowered digital twin
  edge networks,'' \emph{IEEE Internet of Things Journal}, 2021, doi:
  10.1109/JIOT.2021.3126207.

\bibitem{6081879}
A.~Das and M.~M. Islam, ``{SecuredTrust}: A dynamic trust computation model for
  secured communication in multiagent systems,'' \emph{IEEE Transactions on
  Dependable and Secure Computing}, vol.~9, no.~2, pp. 261--274, Mar.-Apr.
  2012.

\bibitem{wang2013enabling}
X.~Wang, W.~Cheng, P.~Mohapatra, and T.~Abdelzaher, ``Enabling reputation and
  trust in privacy-preserving mobile sensing,'' \emph{IEEE Transactions on
  Mobile Computing}, vol.~13, no.~12, pp. 2777--2790, Dec. 2013.

\bibitem{7559774}
F.~Wu, T.~Zhang, C.~Qiao, and G.~Chen, ``A strategy-proof auction mechanism for
  adaptive-width channel allocation in wireless networks,'' \emph{IEEE Journal
  on Selected Areas in Communications}, vol.~34, no.~10, pp. 2678--2689, Oct.
  2016.

\bibitem{9664267}
Y.~Wang, Z.~Su, T.~Luan, R.~Li, and K.~Zhang, ``Federated learning with fair
  incentives and robust aggregation for {UAV}-aided crowdsensing,'' \emph{IEEE
  Transactions on Network Science and Engineering}, 2021, doi:
  10.1109/TNSE.2021.3138928.

\bibitem{9354536}
Z.~Wan, T.~Zhang, W.~Liu, M.~Wang, and L.~Zhu, ``Decentralized
  privacy-preserving fair exchange scheme for {V2G} based on blockchain,''
  \emph{IEEE Transactions on Dependable and Secure Computing}, 2021, doi:
  10.1109/TDSC.2021.3059345.

\bibitem{8493354}
Y.~Chen, X.~Tian, Q.~Wang, M.~Li, M.~Du, and Q.~Li, ``{ARMOR}: A secure
  combinatorial auction for heterogeneous spectrum,'' \emph{IEEE Transactions
  on Mobile Computing}, vol.~18, no.~10, pp. 2270--2284, Oct. 2019.

\bibitem{7894274}
K.~Shin, C.~Joe-Wong, S.~Ha, Y.~Yi, I.~Rhee, and D.~S. Reeves, ``{T-Chain}: A
  general incentive scheme for cooperative computing,'' \emph{IEEE/ACM
  Transactions on Networking}, vol.~25, no.~4, pp. 2122--2137, Aug. 2017.

\bibitem{8827301}
M.~Li, J.~Weng, A.~Yang, J.-N. Liu, and X.~Lin, ``Toward blockchain-based fair
  and anonymous ad dissemination in vehicular networks,'' \emph{IEEE
  Transactions on Vehicular Technology}, vol.~68, no.~11, pp. 11\,248--11\,259,
  Nov. 2019.

\bibitem{1709951}
R.~Ma, S.~Lee, J.~Lui, and D.~Yau, ``Incentive and service differentiation in
  {P2P} networks: A game theoretic approach,'' \emph{IEEE/ACM Transactions on
  Networking}, vol.~14, no.~5, pp. 978--991, Oct. 2006.

\bibitem{7295625}
H.~Shen, Y.~Lin, K.~Sapra, and Z.~Li, ``Enhancing collusion resilience in
  reputation systems,'' \emph{IEEE Transactions on Parallel and Distributed
  Systems}, vol.~27, no.~8, pp. 2274--2287, Aug. 2016.

\bibitem{5752848}
J.~Liu and B.~Yang, ``Collusion-resistant multicast key distribution based on
  homomorphic one-way function trees,'' \emph{IEEE Transactions on Information
  Forensics and Security}, vol.~6, no.~3, pp. 980--991, Sept. 2011.

\bibitem{9328528}
K.~Li, S.~Wang, X.~Cheng, and Q.~Hu, ``A misreport- and collusion-proof
  crowdsourcing mechanism without quality verification,'' \emph{IEEE
  Transactions on Mobile Computing}, 2021, doi: 10.1109/TMC.2021.3052873.

\bibitem{Wei2021Unified}
W.~C. Ng, W.~Y.~B. Lim, J.~S. Ng, Z.~Xiong, D.~Niyato, and C.~Miao, ``Unified
  resource allocation framework for the edge intelligence-enabled metaverse,''
  \emph{arXiv preprint arXiv:2110.14325}, 2021.

\bibitem{Yuna2021Reliable}
Y.~Jiang, J.~Kang, D.~Niyato, X.~Ge, Z.~Xiong, and C.~Miao, ``Reliable coded
  distributed computing for metaverse services: Coalition formation and
  incentive mechanism design,'' \emph{arXiv preprint arXiv:2111.10548}, 2021.

\bibitem{8674550}
S.~Mondal, K.~P. Wijewardena, S.~Karuppuswami, N.~Kriti, D.~Kumar, and
  P.~Chahal, ``Blockchain inspired {RFID}-based information architecture for
  food supply chain,'' \emph{IEEE Internet of Things Journal}, vol.~6, no.~3,
  pp. 5803--5813, Jun. 2019.

\bibitem{9632411}
Y.~Wang, Z.~Su, J.~Li, N.~Zhang, K.~Zhang, K.-K.~R. Choo, and Y.~Liu,
  ``Blockchain-based secure and cooperative private charging pile sharing
  services for vehicular networks,'' \emph{IEEE Transactions on Vehicular
  Technology}, vol.~71, no.~2, pp. 1857--1874, Feb. 2022.

\bibitem{8939473}
M.~Baza, N.~Lasla, M.~M. E.~A. Mahmoud, G.~Srivastava, and M.~Abdallah,
  ``{B-Ride}: Ride sharing with privacy-preservation, trust and fair payment
  atop public blockchain,'' \emph{IEEE Transactions on Network Science and
  Engineering}, vol.~8, no.~2, pp. 1214--1229, Apr.-Jun. 2021.

\bibitem{8931796}
Y.~Zhou, F.~R. Yu, J.~Chen, and Y.~Kuo, ``Cyber-physical-social systems: A
  state-of-the-art survey, challenges and opportunities,'' \emph{IEEE
  Communications Surveys \& Tutorials}, vol.~22, no.~1, pp. 389--425,
  Firstquarter 2020.

\bibitem{8675340}
P.~Casey, I.~Baggili, and A.~Yarramreddy, ``Immersive virtual reality attacks
  and the human joystick,'' \emph{IEEE Transactions on Dependable and Secure
  Computing}, vol.~18, no.~2, pp. 550--562, Mar.-Apr. 2021.

\bibitem{XRSI2022Meta}
\BIBentryALTinterwordspacing
Metaverse rollout brings new security risks, challenges. Accessed: Feb. 8,
  2022. [Online]. Available:
  \url{https://www.techtarget.com/searchsecurity/news/252513072/Metaverse-rollout-brings-new-security-risks-challenges}
\BIBentrySTDinterwordspacing

\bibitem{6979242}
C.~Vellaithurai, A.~Srivastava, S.~Zonouz, and R.~Berthier, ``{CPIndex}:
  Cyber-physical vulnerability assessment for power-grid infrastructures,''
  \emph{IEEE Transactions on Smart Grid}, vol.~6, no.~2, pp. 566--575, Mar.
  2015.

\bibitem{9580681}
S.~Valluripally, A.~Gulhane, K.~A. Hoque, and P.~Calyam, ``Modeling and defense
  of social virtual reality attacks inducing cybersickness,'' \emph{IEEE
  Transactions on Dependable and Secure Computing}, 2021, doi:
  10.1109/TDSC.2021.3121216.

\bibitem{9107084}
J.~Zhu, P.~Ni, and G.~Wang, ``Activity minimization of misinformation influence
  in online social networks,'' \emph{IEEE Transactions on Computational Social
  Systems}, vol.~7, no.~4, pp. 897--906, Aug. 2020.

\bibitem{meta2022terrorist}
\BIBentryALTinterwordspacing
The metaverse offers a future full of potential – for terrorists and
  extremists, too. Accessed: Jan. 7, 2022. [Online]. Available:
  \url{https://theconversation.com/the-metaverse-offers-a-future-full-of-potential-for-terrorists-and-extremists-too-173622}
\BIBentrySTDinterwordspacing

\bibitem{9428549}
P.~Lau, L.~Wang, Z.~Liu, W.~Wei, and C.-W. Ten, ``A coalitional cyber-insurance
  design considering power system reliability and cyber vulnerability,''
  \emph{IEEE Transactions on Power Systems}, vol.~36, no.~6, pp. 5512--5524,
  Nov. 2021.

\bibitem{8496785}
S.~Feng, W.~Wang, Z.~Xiong, D.~Niyato, P.~Wang, and S.~S. Wang, ``On cyber risk
  management of blockchain networks: A game theoretic approach,'' \emph{IEEE
  Transactions on Services Computing}, vol.~14, no.~5, pp. 1492--1504,
  Sept.-Oct. 2021.

\bibitem{8051056}
M.~U. Tariq, J.~Florence, and M.~Wolf, ``Improving the safety and security of
  wide-area cyber–physical systems through a resource-aware, service-oriented
  development methodology,'' \emph{Proceedings of the IEEE}, vol. 106, no.~1,
  pp. 144--159, Jan. 2018.

\bibitem{9697993}
F.-Y. Wang, R.~Qin, X.~Wang, and B.~Hu, ``{MetaSocieties in Metaverse:
  MetaEconomics and MetaManagement for MetaEnterprises and MetaCities},''
  \emph{IEEE Transactions on Computational Social Systems}, vol.~9, no.~1, pp.
  2--7, Feb. 2022.

\bibitem{FYI2021Facebook}
\BIBentryALTinterwordspacing
Improving user experience in our transfer your information tool. Accessed: Oct.
  1, 2021. [Online]. Available:
  \url{https://about.fb.com/news/2021/08/improving-user-experience-in-our-transfer-your-information-tool/}
\BIBentrySTDinterwordspacing

\bibitem{9351745}
V.~Almeida, F.~Filgueiras, and D.~Doneda, ``The ecosystem of digital content
  governance,'' \emph{IEEE Internet Computing}, vol.~25, no.~3, pp. 13--17,
  May-June 2021.

\bibitem{9462451}
Y.~Bai, Q.~Hu, S.-H. Seo, K.~Kang, and J.~J. Lee, ``Public participation
  consortium blockchain for smart city governance,'' \emph{IEEE Internet of
  Things Journal}, vol.~9, no.~3, pp. 2094--2108, Feb. 2022.

\bibitem{8976179}
S.~{Sayeed}, H.~{Marco-Gisbert}, and T.~{Caira}, ``Smart contract: Attacks and
  protections,'' \emph{IEEE Access}, vol.~8, pp. 24\,416--24\,427, Jan. 2020.

\bibitem{8885317}
G.~Huang, C.~Luo, K.~Wu, Y.~Ma, Y.~Zhang, and X.~Liu, ``Software-defined
  infrastructure for decentralized data lifecycle governance: Principled design
  and open challenges,'' in \emph{IEEE International Conference on Distributed
  Computing Systems (ICDCS)}, Jul. 2019, pp. 1674--1683.

\bibitem{9555823}
M.~Li, J.~Weng, J.-N. Liu, X.~Lin, and C.~Obimbo, ``Towards vehicular digital
  forensics from decentralized trust: An accountable, privacy-preserving, and
  secure realization,'' \emph{IEEE Internet of Things Journal}, May. 2021, doi:
  10.1109/JIOT.2021.3116957.

\bibitem{9384217}
X.~He, Q.~Gong, Y.~Chen, Y.~Zhang, X.~Wang, and X.~Fu, ``{DatingSec}: Detecting
  malicious accounts in dating apps using a content-based attention network,''
  \emph{IEEE Transactions on Dependable and Secure Computing}, vol.~18, no.~5,
  pp. 2193--2208, Sept.-Oct. 2021.

\bibitem{8114684}
U.~Gasser and V.~A. Almeida, ``A layered model for {AI} governance,''
  \emph{IEEE Internet Computing}, vol.~21, no.~6, pp. 58--62, Nov./Dec. 2017.

\bibitem{8371566}
F.~Zambonelli, F.~Salim, S.~W. Loke, W.~De~Meuter, and S.~Kanhere,
  ``Algorithmic governance in smart cities: The conundrum and the potential of
  pervasive computing solutions,'' \emph{IEEE Technology and Society Magazine},
  vol.~37, no.~2, pp. 80--87, June 2018.

\bibitem{7484666}
B.~Li, R.~Lu, W.~Wang, and K.-K.~R. Choo, ``{DDOA}: A {Dirichlet}-based
  detection scheme for opportunistic attacks in smart grid cyber-physical
  system,'' \emph{IEEE Transactions on Information Forensics and Security},
  vol.~11, no.~11, pp. 2415--2425, Nov. 2016.

\bibitem{4451097}
A.~Swaminathan, M.~Wu, and K.~R. Liu, ``Digital image forensics via intrinsic
  fingerprints,'' \emph{IEEE Transactions on Information Forensics and
  Security}, vol.~3, no.~1, pp. 101--117, Mar. 2008.

\bibitem{6222325}
M.~C. Stamm, W.~S. Lin, and K.~J.~R. Liu, ``Temporal forensics and
  anti-forensics for motion compensated video,'' \emph{IEEE Transactions on
  Information Forensics and Security}, vol.~7, no.~4, pp. 1315--1329, Aug.
  2012.

\bibitem{8361414}
D.~Zou, J.~Zhao, W.~Li, Y.~Wu, W.~Qiang, H.~Jin, Y.~Wu, and Y.~Yang, ``A
  multigranularity forensics and analysis method on privacy leakage in cloud
  environment,'' \emph{IEEE Internet of Things Journal}, vol.~6, no.~2, pp.
  1484--1494, Apr. 2019.

\bibitem{9141216}
Z.~Zhou, X.~Kuang, L.~Sun, L.~Zhong, and C.~Xu, ``Endogenous security defense
  against deductive attack: When artificial intelligence meets active defense
  for online service,'' \emph{IEEE Communications Magazine}, vol.~58, no.~6,
  pp. 58--64, Jun. 2020.

\bibitem{9684555}
Y.~Cao, Y.~Zhao, Q.~Wang, J.~Zhang, S.~X. Ng, and L.~Hanzo, ``The evolution of
  quantum key distribution networks: On the road to the qinternet,'' \emph{IEEE
  Communications Surveys \& Tutorials}, vol.~24, no.~2, pp. 839--894,
  Secondquarter 2022.

\bibitem{2020Testimonium}
P.~Frauenthaler, M.~Sigwart, C.~Spanring, and S.~Schulte, ``Testimonium: A
  cost-efficient blockchain relay,'' \emph{arXiv preprint arXiv:2002.12837},
  2020.

\end{thebibliography}

\begin{IEEEbiography}[{\includegraphics[width=1in,height=1.25in,clip,keepaspectratio]{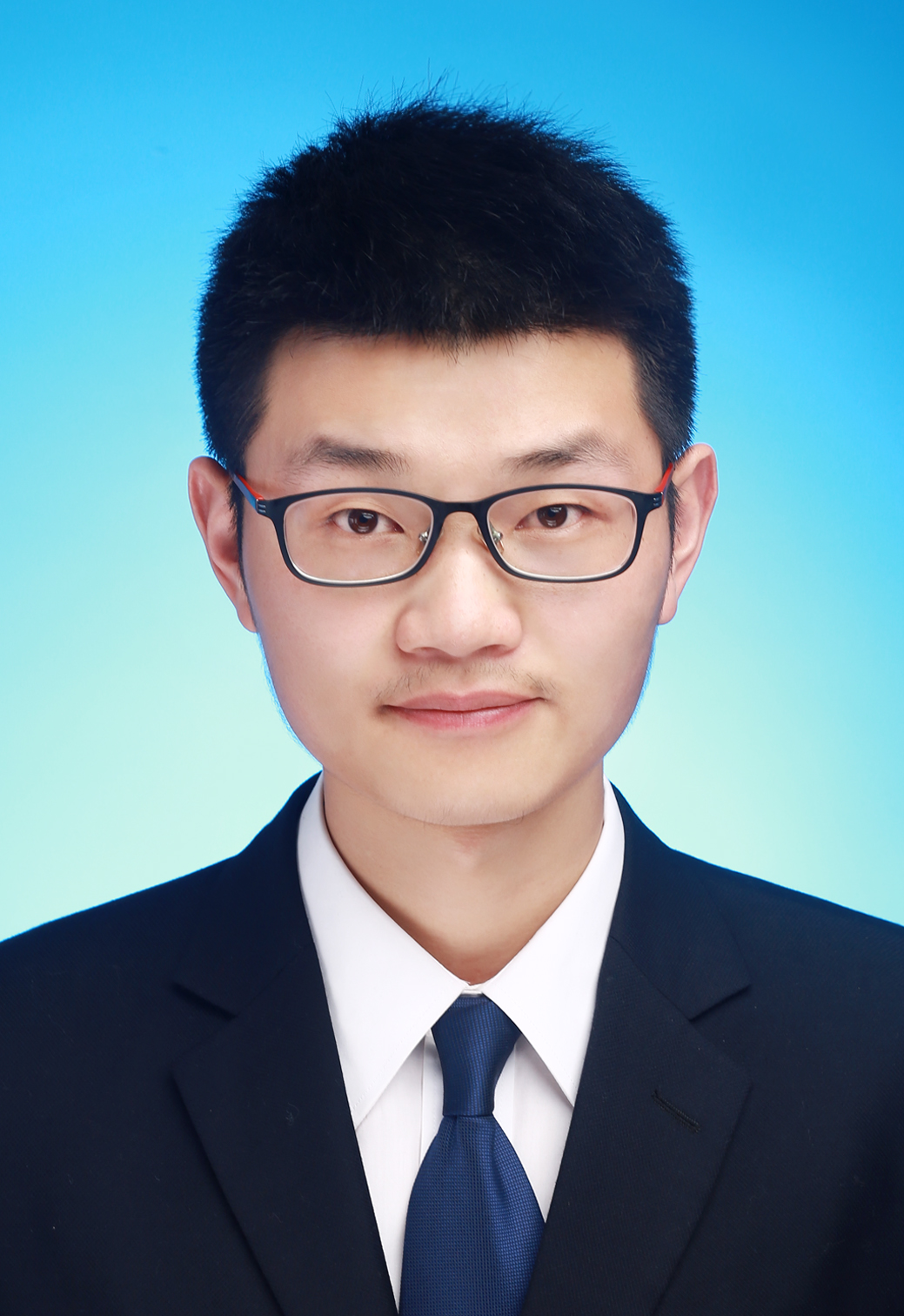}}]{Yuntao Wang}
is working on his Ph.D degree with the School of Cyber Science and Engineering of Xi'an Jiaotong University, Xi'an, China. His research interests include security and privacy protection in general wireless networks and vehicular networks.
\end{IEEEbiography}

\begin{IEEEbiography}[{\includegraphics[width=1in,height=1.25in,clip,keepaspectratio]{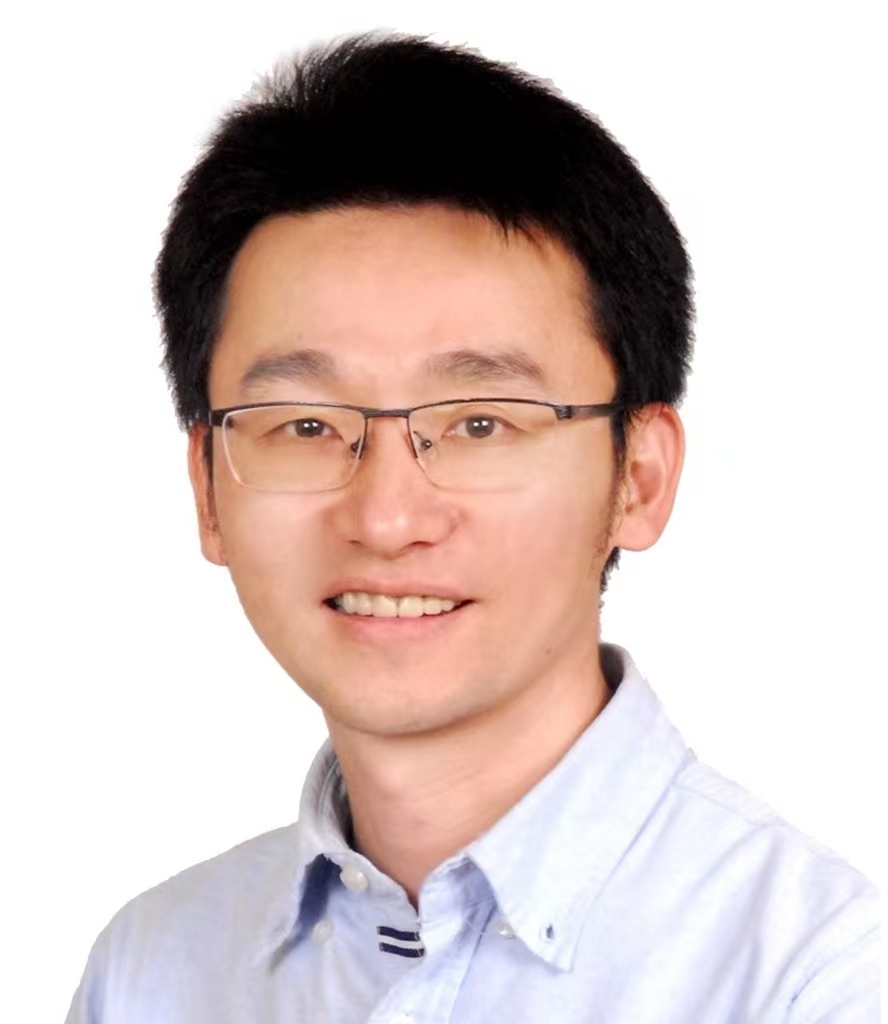}}]{Zhou Su}
has published technical papers, including top journals and top conferences, such as IEEE JOURNAL ON SELECTED AREAS IN COMMUNICATIONS, IEEE TRANSACTIONS ON INFORMATION FORENSICS AND SECURITY, IEEE TRANSACTIONS ON DEPENDABLE AND SECURE COMPUTING, IEEE TRANSACTIONS ON MOBILE COMPUTING, IEEE/ACM TRANSACTIONS ON NETWORKING, and INFOCOM. His research interests include multimedia communication, wireless communication, and network traffic. Dr. Su received the Best Paper Award of International Conference IEEE ICC2020, IEEE BigdataSE2019, and IEEE CyberSciTech2017. He is an Associate Editor of IEEE INTERNET OF THINGS JOURNAL, IEEE OPEN JOURNAL OF COMPUTER SOCIETY, and IET Communications.
\end{IEEEbiography}

\begin{IEEEbiography}[{\includegraphics[width=1in,height=1.25in,clip,keepaspectratio]{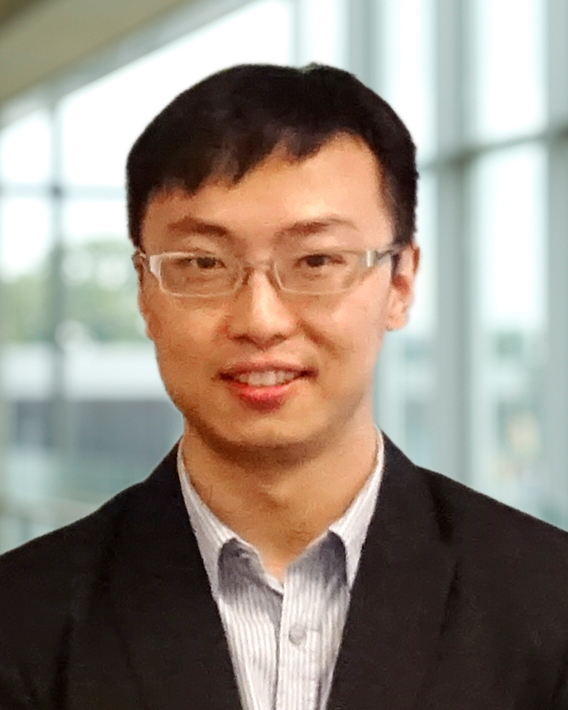}}]{Ning Zhang}
received the Ph.D degree from University of Waterloo, Canada, in 2015. He is an Associate Professor at University of Windsor, Canada. He serves as an Associate Editor of IEEE Internet of Things Journal, IEEE Transactions on Cognitive Communications and Networking, IEEE Access, and IET Communications, and Vehicular Communications (Elsevier); and a Guest Editor of several international journals, such as IEEE Wireless Communications, IEEE Transactions on Industrial Informatics, and IEEE Transactions on Cognitive Communications and Networking. He also serves/served as a track chair for several international conferences and a co-chair for several international workshops. He received the Best Paper Awards from IEEE Globecom in 2014, IEEE WCSP in 2015, and Journal of Communications and Information Networks in 2018, IEEE ICC in 2019, IEEE Technical Committee on Transmission Access and Optical Systems in 2019, and IEEE ICCC in 2019, respectively. 
\end{IEEEbiography}

\begin{IEEEbiography}[{\includegraphics[width=1in,height=1.25in,clip,keepaspectratio]{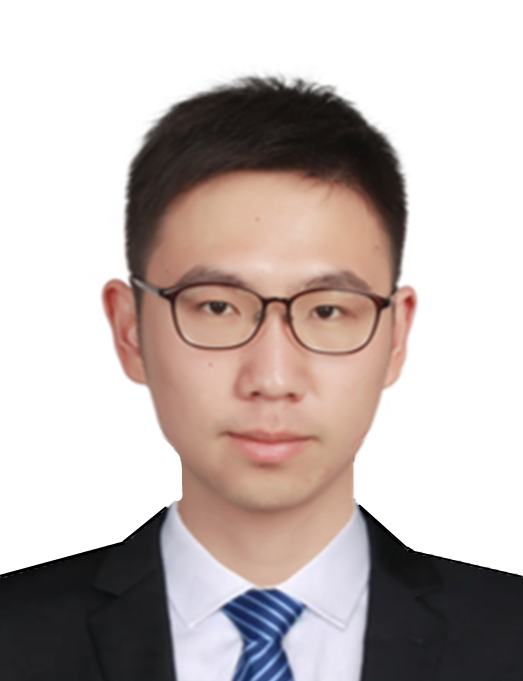}}]{Rui Xing}
is currently working toward the Ph.D degree with the School of Cyber Science and Engineering, Xi'an Jiaotong University, Xi'an, China. His research interests include security protection and system optimization in wireless networks and vehicular networks.
\end{IEEEbiography}

\begin{IEEEbiography}[{\includegraphics[width=1in,height=1.25in,clip,keepaspectratio]{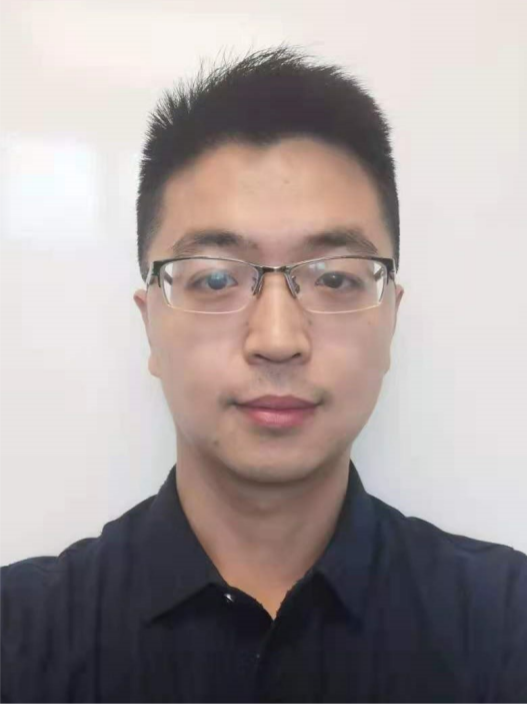}}]{Dongxiao Liu}
received the PhD degree in the Department of Electrical and Computer Engineering, University of Waterloo, Canada in 2020. He is currently a Postdoctoral Fellow in the Department of Electrical and Computer Engineering, University of Waterloo. His research interests include security and privacy in intelligent transportation systems, blockchain, and mobile networks.
\end{IEEEbiography}

\begin{IEEEbiography}[{\includegraphics[width=1in,height=1.25in,clip,keepaspectratio]{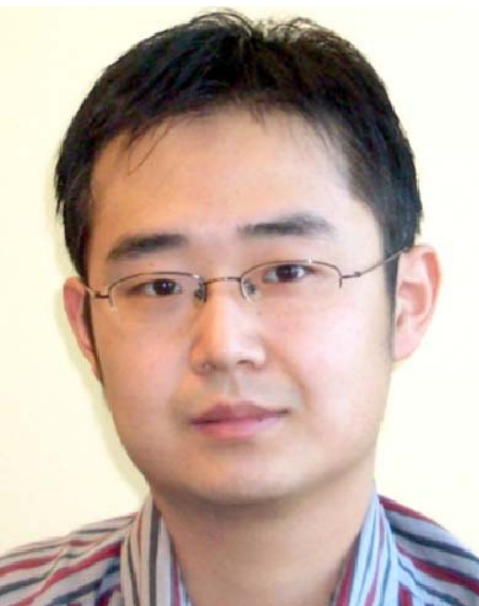}}]{Tom H. Luan}
received the Ph.D. degree from the University of Waterloo, Canada, in 2012. He is currently a Professor with the School of Cyber Science and Engineering, Xi'an Jiaotong University, China. He has authored/coauthored more than 40 journal articles and 30 technical articles in conference proceedings. He awarded one U.S. patent. His research mainly focuses on content distribution and media streaming in vehicular ad hoc networks and peer-to-peer networking and the protocol design and performance evaluation of wireless cloud computing and edge computing. He served as a TPC Member for IEEE Globecom, ICC, and PIMRC.
\end{IEEEbiography}

\begin{IEEEbiography}[{\includegraphics[width=1in,height=1.25in,clip,keepaspectratio]{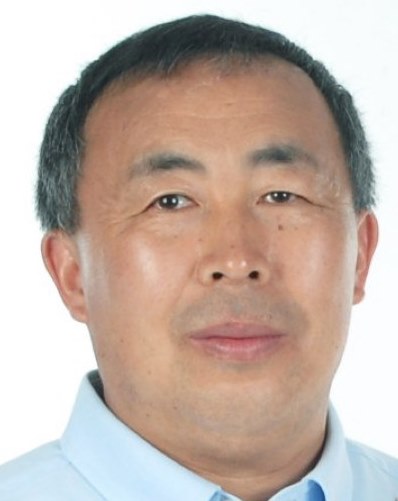}}]{Xuemin (Sherman) Shen}
(Fellow, IEEE) received the Ph.D. degree in electrical engineering from Rutgers University, New Brunswick, NJ, USA, in 1990. He is currently a University Professor with the Department of Electrical and Computer Engineering, University of Waterloo, Canada. He is also a registered Professional Engineer in ON, Canada. 
He is an Engineering Institute of Canada Fellow, a Canadian Academy of Engineering Fellow, a Royal Society of Canada Fellow, a Chinese Academy of Engineering Foreign Member, and a Distinguished Lecturer of the IEEE Vehicular Technology Society and Communications Society. He received the R.A. Fessenden Award from IEEE, Canada, in 2019; the Award of Merit from the Federation of Chinese Canadian Professionals, ON, Canada, in 2019; the Technical Recognition Award from the Wireless Communications Technical Committee in 2019; the James Evans Avant Garde Award from the IEEE Vehicular Technology Society in 2018; the Education Award in 2017 and the Joseph LoCicero Award in 2015 from the IEEE Communications Society; the AHSN Technical Committee in 2013; the Excellent Graduate Supervision Award from the University of Waterloo in 2006; and the Premier's Research Excellence Award (PREA) from the Province of Ontario, Canada, in 2003. He served as the Technical Program Committee Chair/Co-Chair for IEEE Globecom'16, IEEE Infocom'14, IEEE VTC'10 Fall, and IEEE Globecom'07, and the Chair for the IEEE Communications Society Technical Committee on Wireless Communications. He was the Vice President for the Technical and Educational Activities and Publications; a Member-at-Large on the Board of Governors; and the Chair of the Distinguished Lecturer Selection Committee. He is also the President Elect of the IEEE Communications Society. He served as the Editor-in-Chief for IEEE Internet of Things Journal, IEEE Network, and IET Communications.
\end{IEEEbiography}
\end{document}